\documentclass{aa}

\input{./preamble/packages.sty}
\input{./preamble/commands.sty}
\input{./preamble/acronyms.sty}
\input{./preamble/units.sty}

\DTLsetseparator{;}
\DTLloaddb[noheader]{opacityFit}{tables/opacity_fit.csv}
\DTLloaddb[noheader]{bestFit}{tables/best_fit.csv}
\DTLloaddb[noheader]{observations}{tables/observations.csv}

\begin{document}
\begin{filecontents*}{tables/opacity_fit.csv}
$\log(s/\unit{\steradian})$;;\num{17.18(0.07:0.24)}
$w_{\text{PAH}}$;;\num{1.28(1.31:1.20)}
$T_{\text{c}}$;(\unit{\kelvin});\num{420.98(23.36:57.21)}
$w_{\text{cont}}$;(\unit{\percent});\num{69.38(16.85:3.65)}
$w_{\text{large},\,\text{enst}}$;(\unit{\percent});$2.54^{+53.41}_{-0.81}$
$w_{\text{small},\,\text{enst}}$;(\unit{\percent});$5.91^{+16.87}_{-5.56}$
$w_{\text{large},\,\text{forst}}$;(\unit{\percent});$2.10^{+19.92}_{-1.97}$
$w_{\text{small},\,\text{forst}}$;(\unit{\percent});\num{5.64(5.87:5.50)}
$w_{\text{large},\,\text{oliv}}$;(\unit{\percent});\num{6.52(7.64:6.52)}
$w_{\text{small},\,\text{pyrox}}$;(\unit{\percent});\num{1.37(8.13:1.37)}
$w_{\text{large},\,\text{sil}}$;(\unit{\percent});\num{7.73(3.44:7.73)}
$w_{\text{small},\,\text{sil}}$;(\unit{\percent});\num{1.70(1.42:1.70)}
\end{filecontents*}

\begin{filecontents*}{tables/best_fit.csv}
$R_{\text{out},\,1}$;(\unit{\astronomicalunit});\num{5.32(0.07:0.06)}
$w_{\text{cont},\,1}$;(\unit{\percent});\num{79.30(0.98:0.87)}
$\Sigma_{0, 1}$;(\qty{1e-4}{\gram\per\square\centi\metre});\num{1.24(0.03:0.02)}
$p_1$;;\num{0.29(0.01)}
$\log(f/\unit{\steradian})$;;\num{15.30(0.01)}
$\rho_\text{Gauss, t0}$;(\unit{\astronomicalunit});\num{1.21(0.03)}
$\rho_\text{Gauss, t1}$;(\unit{\astronomicalunit});\num{1.10(0.01)}
$\rho_\text{Gauss, t2}$;(\unit{\astronomicalunit});\num{1.06(0.02:0.01)}
$\Phi_\text{Gauss, t0}$;(\unit{\degree});\num{150.11(2.02:1.97)}
$\Phi_\text{Gauss, t1}$;(\unit{\degree});\num{-125.34(1.49:1.09)}
$\Phi_\text{Gauss, t2}$;(\unit{\degree});\num{-154.24(1.62:2.00)}
$\Sigma_\text{Gauss}$;(\qty{1e-4}{\gram\per\square\centi\metre});\num{16.9(0.56:0.68)}
$w_\text{cont, Gauss}$;(\unit{\percent});\num{60.71(36.86:58.19)}
$T_0$;(\unit{\kelvin});\num{599.74(2.61:3.11)}
\end{filecontents*}

\begin{filecontents*}{tables/observations.csv}
2013-02-20T08:14\textsuperscript{\ensuremath{\dagger}};PIONIER;0.8;3.1;M: D0-G1-H0-I1;H;$\sim$$\pm$20;HD142135;0.42;0.7;4.0
2013-06-04T04:53\textsuperscript{\ensuremath{\dagger}}\textsuperscript{\ensuremath{\#}};PIONIER;1.2;3.7;L: K0-A1-G1-J3;H;--;--;--;--;--
2013-06-06T02:16\textsuperscript{\ensuremath{\dagger}}\textsuperscript{\ensuremath{\#}};PIONIER;1.1;3.8;L: K0-A1-G1-J3;H;--;--;--;--;--
2013-06-07T01:39\textsuperscript{\ensuremath{\dagger}}\textsuperscript{\ensuremath{\#}};PIONIER;0.8;7.4;L: K0-A1-G1-J3;H;--;--;--;--;--
2013-06-10T01:08\textsuperscript{\ensuremath{\dagger}}\textsuperscript{\ensuremath{\#}};PIONIER;0.9;5.3;L: K0-A1-G1-J3;H;--;--;--;--;--
2013-06-15T02:57\textsuperscript{\ensuremath{\dagger}};PIONIER;1.4;2.2;M: D0-G1-H0-I1;H;$\sim$$\pm$20;HD 142386;0.24;1.3;1.3
2013-07-03T01:16\textsuperscript{\ensuremath{\dagger}}\textsuperscript{\ensuremath{\#}};PIONIER;1.4;2.7;S: D0-A1-C1-B2;H;--;--;--;--;--
2017-03-19T08:51\textsuperscript{\ensuremath{\dagger\dagger}};GRAVITY;0.7;8.0;L: A0-G1-J2-K0;K;$+$1:18;HD 143118;0.37;0.8;4.9
\textcolor{red}{2019-03-23T09:06\textsuperscript{\ensuremath{\ddagger}}};\textcolor{red}{MATISSE};\textcolor{red}{0.5};\textcolor{red}{10.8};\textcolor{red}{S: A0-B2-D0-C1};\textcolor{red}{L};\textcolor{red}{$+$0:13};\textcolor{red}{* eps Sco};\textcolor{red}{5.8};\textcolor{red}{0.4};\textcolor{red}{11.3}
\textcolor{red}{2019-05-06T04:00\textsuperscript{\ensuremath{\ddagger}}};\textcolor{red}{MATISSE};\textcolor{red}{0.5};\textcolor{red}{5.8};\textcolor{red}{L: K0-G1-D0-J3};\textcolor{red}{L};\textcolor{red}{$-$0:13};\textcolor{red}{HD 138742};\textcolor{red}{1.2};\textcolor{red}{0.7};\textcolor{red}{5.6}
\textcolor{red}{2019-06-30T04:49};\textcolor{red}{MATISSE};\textcolor{red}{0.9};\textcolor{red}{1.8};\textcolor{red}{S: A0-B2-D0-C1};\textcolor{red}{L};\textcolor{red}{$+$0:30};\textcolor{red}{* eps Sco};\textcolor{red}{5.8};\textcolor{red}{1.5};\textcolor{red}{1.7}
\textcolor{red}{2021-03-08T07:52\textsuperscript{\ensuremath{\ddagger}}};\textcolor{red}{MATISSE};\textcolor{red}{1.1};\textcolor{red}{3.8};\textcolor{red}{L: A0-G1-J2-J3};\textcolor{red}{L};\textcolor{red}{$-$0:27};\textcolor{red}{HD139127};\textcolor{red}{3.2};\textcolor{red}{1.8};\textcolor{red}{3.0}
\textcolor{red}{2021-03-11T06:47\textsuperscript{\ensuremath{\ddagger}}};\textcolor{red}{MATISSE};\textcolor{red}{0.7};\textcolor{red}{8.4};\textcolor{red}{M: K0-G2-D0-J3};\textcolor{red}{L};\textcolor{red}{$-$0:26};\textcolor{red}{HD139127};\textcolor{red}{3.2};\textcolor{red}{0.8};\textcolor{red}{6.4}
\textcolor{red}{2021-03-16T06:50\textsuperscript{\ensuremath{\ddagger}}};\textcolor{red}{MATISSE};\textcolor{red}{1.0};\textcolor{red}{2.5};\textcolor{red}{S: A0-B2-D0-C1};\textcolor{red}{L};\textcolor{red}{$+$0:26};\textcolor{red}{* H Sco};\textcolor{red}{4.7};\textcolor{red}{1.4};\textcolor{red}{1.7}
2021-03-27T05:29;MATISSE;1.0;5.7;U: U1-U2-U3-U4;LN;$-$0:31;HD134505;2.5;1.0;7.3
2022-03-14T06:48;MATISSE;0.9;4.1;S: A0-B2-D0-C1;L;$+$0:26;* H Sco;4.7;0.8;4.1
2022-03-23T08:20;MATISSE;0.6;4.1;U: U1-U2-U3-U4;LN;$-$0:31;HD138816;2.1;0.7;3.8
2022-05-11T06:22;MATISSE;1.2;2.9;L: A0-G1-J2-J3;L;$+$0:29;HD149401;2.5;1.0;3.3
2023-05-14T04:25;MATISSE;0.8;4.4;S: A0-B2-D0-C1;L;$-$0:41;* ups Lib;4.6;1.4;2.8
\textcolor{red}{2023-06-12T04:10\textsuperscript{\ensuremath{\ddagger}}};\textcolor{red}{MATISSE};\textcolor{red}{0.7};\textcolor{red}{2.5};\textcolor{red}{S: A0-B2-D0-C1};\textcolor{red}{L};\textcolor{red}{$-$0:27};\textcolor{red}{* ups Lib};\textcolor{red}{4.6};\textcolor{red}{0.8};\textcolor{red}{3.4}
2023-07-01T00:57;MATISSE;1.0;4.8;U: U1-U2-U3-U4;LN;$-$0:24;HD134505;2.5;1.2;4.5
2023-07-09T03:53;MATISSE;0.6;2.8;S: A0-B2-D0-C1;L;$-$0:27;* ups Lib;4.6;0.6;3.3
\textcolor{red}{2023-07-12T00:36};\textcolor{red}{MATISSE};\textcolor{red}{1.6};\textcolor{red}{0.8};\textcolor{red}{S: A0-B2-D0-C1};\textcolor{red}{L};\textcolor{red}{$-$0:34};\textcolor{red}{* ups Lib};\textcolor{red}{4.6};\textcolor{red}{1.8};\textcolor{red}{1.3}
2023-08-12T01:32;MATISSE;0.9;3.2;S: A0-B2-D0-C1;L;$-$0:28;* ups Lib;4.6;1.3;3.1
\end{filecontents*}

\title{The asymmetric structure of the inner disc around HD~142527~A with \acs{vlti}/\acs{matisse}}
\author{M. B. Scheuck\inst{\ref{inst:mpia}, \ref{inst:uniHeidelberg}}
        \and R. van Boekel\inst{\ref{inst:mpia}}
        \and Th. Henning\inst{\ref{inst:mpia}}
        \and P. A. Boley\inst{\ref{inst:mpia}}
        \and J. Varga\inst{\ref{inst:hunRen}}
        \and A. Matter\inst{\ref{inst:cotDAzur}}
        \and A. Penzlin\inst{\ref{inst:lmu}}
        \and J. H. Leftley\inst{\ref{inst:cotDAzur}}
        \and L. van Haastere\inst{\ref{inst:uniLeiden}}
        \and K. Perraut\inst{\ref{inst:uniGrenoble}}
        \and L. Labadie\inst{\ref{inst:uniKöln}}
        \and M. Min\inst{\ref{inst:uniAmsterdam}, \ref{inst:sron}}
        \and J. P. Berger\inst{\ref{inst:uniGrenoble}}
        \and L. B. F. M. Waters\inst{\ref{inst:sron}, \ref{inst:uniRadboud}}
        \and S. Zieba\inst{\ref{inst:uniHarvard}}
        \and B. Lopez\inst{\ref{inst:cotDAzur}}
        \and F. Lykou\inst{\ref{inst:hunRen}}
        \and J.-C. Augereau\inst{\ref{inst:uniGrenoble}}
        \and P. Cruzalèbes\inst{\ref{inst:cotDAzur}}
        \and W. C. Danchi\inst{\ref{inst:nasa}}
        \and V. Gámez Rosas\inst{\ref{inst:star}}
        \and M. Hogerheijde\inst{\ref{inst:uniLeiden}}
        \and M. Letessier\inst{\ref{inst:uniGrenoble}}
        \and J. Scigliuto\inst{\ref{inst:cotDAzur}}
        \and G. Weigelt\inst{\ref{inst:mpifr}}
        \and S. Wolf\inst{\ref{inst:uniKiel}}
        \and the \acs{matisse} and GRAVITY collaborations
       }
\institute{
    Max-Planck-Institut für Astronomie, Königstuhl 17, 69117 Heidelberg, Germany\\\email{scheuck@mpia.de}
    \label{inst:mpia}
    \and
    Fakultät für Physik und Astronomie, Universität Heidelberg, Im Neuenheimer Feld 226, 69120 Heidelberg, Germany
    \label{inst:uniHeidelberg}
    \and
    HUN-REN Research Centre for Astronomy and Earth Sciences, Konkoly Observatory,
    MTA Centre of Excellence, Konkoly\-Thege Miklós út 15-17., H-1121 Budapest, Hungary
    \label{inst:hunRen}
    \and
    Laboratoire Lagrange, Université Côte d’Azur, Observatoire de la Côte d’Azur, CNRS, Boulevard de l’Observatoire,
    CS 34229, 06304, Nice Cedex 4, France
    \label{inst:cotDAzur}
    \and
    Ludwig-Maximilians-Universität München, Universitäts-Sternwarte, Scheinerstr. 1, 81679 München, Germany
    \label{inst:lmu}
    \and
    Leiden Observatory, Leiden University, PO Box 9513, 2300 RA, Leiden, The Netherlands
    \label{inst:uniLeiden}
    \and
    Univ. Grenoble Alpes, CNRS, IPAG, 38000 Grenoble, France
    \label{inst:uniGrenoble}
    \and 
    I. Physikalisches Institut, Universität zu Köln, Zülpicher Str. 77, 50937 Köln, Germany
    \label{inst:uniKöln}
    \and
    Anton Pannekoek Institute for Astronomy, University of Amsterdam, Science Park 904, 1098XH Amsterdam, The Netherlands
    \label{inst:uniAmsterdam}
    \and
    SRON Netherlands Institute for Space Research, Niels Bohrweg 4, 2333 CA Leiden, The Netherlands
    \label{inst:sron}
    \and
    Institute for Mathematics, Astrophysics and Particle Physics, Radboud University,
    P.O. Box 9010, MC 62 NL-6500 GL Nijmegen, The Netherlands
    \label{inst:uniRadboud}
    \and
    Center for Astrophysics Harvard \& Smithsonian: 60 Garden Street, Cambridge, MA 02138, USA
    \label{inst:uniHarvard}
    \and
    NASA Goddard Space Flight Center, Astrophysics Division,  Greenbelt, MD, 20771, USA
    \label{inst:nasa}
    \and
    STAR Institute, University of Liège, Quartier Agora - Bât. B5c Allée du Six Août 19C, B-4000 Liège, Belgium
    \label{inst:star}
    \and
    Max-Planck-Institut für Radioastronomie, Auf dem Hügel 69, 53121, Bonn, Germany
    \label{inst:mpifr}
    \and
    Institute of Theoretical Physics and Astrophysics, University of Kiel, Leibnizstraße 15, 24118, Kiel, Germany
    \label{inst:uniKiel}
}
\date{Received 8 August 2025 / Accepted 2 February 2026}
\abstract
{%
    Circumstellar discs, and especially their inner regions, which cover ranges from less than \qty{1}{\astronomicalunit} to a few \unit{\astronomicalunit}, are the birthplaces of terrestrial planets. The inner regions are thought to be as diverse in structure as the well-observed outer regions probed by ALMA.
}
{%
    By combining data and results from previous studies using the \acs{vlti}/\acs{pionier} and \acs{vlti}/GRAVITY instruments with new, multi-epoch \acs{vlti}/\acs{matisse} observations, we aim to provide a comprehensive picture of the structure of the inner regions of the circumstellar disc around the F-type Herbig Ae/Be star HD~142527~A, the primary of a binary star system.
}
{%
    We modelled the multi-wavelength interferometric data using a parametrised, geometrically thin disc model, allowing for azimuthal asymmetry, and exploring a first-order disc modulation and an off-centre Gaussian component.
}
{%
    We find time-variable structures in the \textit{N}-band observables, which we reproduce with time-dependent models. This variability manifests as azimuthally asymmetric emission, evidenced by strong, non-zero closure phases in the \textit{N}-band data. Fits to individual epochs of the \textit{N}-band observations yield better $\chi^2_\text{r}$ values than fits to all epochs simultaneously. This suggests substantial changes in the geometry of the inner disc emission from \qty{\sim1}{\astronomicalunit} up to a few astronomical-unit scales from one year to the next. Moreover, our models produce a very close-in inner disc rim $R_\text{rim}\approx\qty{0.1}{\astronomicalunit}$. Altogether, we find a very complex, substantially non-point symmetric and temporally variable disc ($r_\text{out}\lesssim\qty{6}{\astronomicalunit}$) around the primary.
}
{%
    The very close-in inner rim indicates the presence of material within the typical wall-like sublimation radius, $R_\text{rim,literature}\approx\qty{0.3}{\astronomicalunit}$. The complex, temporally variable inner-disc geometry is likely affected, or even caused by, the close passage (\qty{\sim5}{\astronomicalunit}) and short orbit ($P\approx\qty{24}{\year}$) of the companion HD~142527~B.
}

\maketitle
\nolinenumbers

\section{Introduction}
\label{sec:introduction}
Herbig Ae/Be and T Tauri stars are often accompanied by planet-forming discs composed of gas and dust within the first \qty{10}{\mega\year} of their existence. These discs are highly complex environments, with properties such as their mass and chemical composition influenced not only by the host star but also by various concurrent processes, including turbulence, winds, the dust grain growth and migration, and stellar accretion \citep{Andrews2018,Andrews2020,Benisty2023,Birnstiel2024}. Observations at millimetre wavelengths and scattered-light images at optical and near-\ac{ir} wavelengths have mapped the outer regions of these discs \citep{Pohl2017,Garufi2017,Garufi2018}, revealing diverse structures such as rings, radial gaps, spiral arms, dust clumps, and other azimuthal asymmetries \citep{ALMAPartnership2015,Rubinstein2018}. These regions also serve as a rich reservoir of gas and dust essential for planetary system formation, as supported by numerous studies examining various disc systems \citep[e.g.][]{Boccaletti2020}. Recent research has also provided additional approaches to identify planets embedded within these discs \citep{Fedele2017,Teague2018,Pinte2019}, and in the case of PDS~70, at least two planets have been detected \citep{Mueller2018,Keppler2018,Haffert2019}.

Similarly, the inner regions likely host smaller-scale structures akin to those observed in the outer disc \citep[e.g.][]{Menu2015,Varga2018,GRAVITYCollaboration2021,Kluska2022,Varga2024,GRAVITYCollaboration2024}. Emerging evidence suggests that some of these substructures, such as radial gaps, arise from young, accreting planets \citep{Haffert2019,Pinte2019}.

We aim to expand knowledge of planet formation within circumstellar discs and to complement previous and concurrent research by studying the innermost regions of the discs (from \qty{<1}{\astronomicalunit} scales up to a few \unit{\astronomicalunit}). Thus, we observed the mid-\ac{ir} wavelength region with the spectro-interferometric \ac{matisse} instrument \citep{Lopez2022} located at the \ac{vlti} on Cerro Paranal in Chile and operated by the \ac{eso}. The \ac{matisse} instrument at the \ac{vlti} offers spatial resolutions superior to those of single telescopes (e.g. the \acl{jwst}) and provides a broad wavelength range. These resolutions facilitate the study of the structure of the innermost disc regions.

In this work, we focus on HD~142527, a binary system composed of an F-type Herbig Ae/Be star \citep{Hunziker2021} accompanied by an M-dwarf companion \citep{Biller2012,Lacour2016,Christiaens2018} on an eccentric orbit \citep[$e\approx$~\qtyrange{0.45}{0.7}{},][]{Claudi2019}. HD~142527~A possesses a circumstellar disc, which was determined to be close to face-on by observations with the \ac{vlti}/\ac{pionier} \citep{Lazareff2017} and \ac{vlti}/GRAVITY \citep{GRAVITYcollaboration2019} instruments and the \ac{alma} \citep{Casassus2013}. These observations showed inclination angles of $i_\text{in}\approx$~\qtyrange{20}{33}{\degree} and position angles of $\theta_\text{in}\approx$~\qtyrange{5}{20}{\degree} (east of north). For the inner disc, fits to the \textit{N}-band emission indicate a radial gap within this region, which is not resolved by either the \ac{sphere} instrument or \ac{alma} \citep{Menu2015, Varga2018}. The companion, HD~142527~B, lies close to the inner disc (periapsis to apoapsis: \qtyrange{\sim5}{15}{\astronomicalunit}) and orbits within a large gap in the surrounding circumbinary disc (\qtyrange{\sim113}{170}{\astronomicalunit} from east to west) \citep{Boehler2017}. A possible explanation for the large gap between the inner and outer discs is one or more unseen planets \citep{Lacour2016}. From the shadows cast from the inner onto the outer disc, \citet{Marino2015} determined the misalignment ($\Delta\vartheta_\text{in-out}=\qty{70}{\degree}$) and the position angle of the inner disc ($\theta_\text{in}=\qty{352(5.)}{\degree}$). The inclination and position angle of the outer disc are estimated to be $i_\text{out}=38\fdg21$ and $\theta_\text{out}=162\fdg72$ from \ac{alma} observations \citep{Bohn2022}. These, along with information on the stellar parameters of HD~142527~A and B, plus the parameters of the companion's orbit and the outer disc, are listed in Table~\ref{tab:stellarParameters}.\\

In this work, we present \ac{matisse} observations of HD~142527, combined with modelling of previous observations from \ac{pionier} and GRAVITY data from \citet{Lazareff2017} and \citet{GRAVITYcollaboration2019}, respectively. Following earlier studies of \acp{yso} \citep[e.g.][]{GRAVITYCollaboration2021,Varga2024}, we analyse these data using geometric disc models to determine the inner-disc structure. In contrast to these previous works, the extensive observations from the \textit{H} to the \textit{N} band also allow us to study the time variability of HD~142527 across different epochs.

The paper is structured as follows. In Sect.~\ref{sec:observations}, we give an overview of the \ac{matisse} observations and provide the necessary background on data reduction and quality assessment. We present first results, obtained directly from the observations, in Sect.~\ref{sec:results}. We explain our modelling approach in Sect.~\ref{sec:modelling}, and discuss it in Sect.~\ref{sec:discussion}, where we place the findings in a wider context. Finally, Sect.~\ref{sec:summaryAndConclusions} summarises the main conclusions.

\section{Observations}
\label{sec:observations}

\begin{table}[ht!]
    \caption{Stellar parameters of HD~142527~A and B, together with outer disc and orbital parameters of the companion.}
    \begin{center}
        \bgroup
        \def\arraystretch{1.2}
        \begin{tabular}{l c c c}
             \toprule\toprule
             Parameter & Unit & Value & Reference \\
             \midrule
             \textbf{HD~142527} & & & \\
             RA\quad(J2016) & (h:\unit{\arcminute}:\unit{\arcsecond}) & 15:56:41.87 & (\ref{itm:gaia}) \\
             DEC \,(J2016) & (\unit{\degree}:\unit{\arcminute}:\unit{\arcsecond}) & -42:19:23.67 & (\ref{itm:gaia}) \\
             $d$ & (pc) & \num{159.3(7)} & (\ref{itm:gaia}) \\
             \midrule
             \textbf{HD~142527~A} & & & \\
             Spectral Type & & F2\,--\,F3  & (\ref{itm:guzman}\tablefootmark{a}) \\
             $T_\star$ & (\unit{\kelvin}) & \num{6500(250.)} & (\ref{itm:guzman}\tablefootmark{a}, \ref{itm:fairlamb}, \ref{itm:vioque}) \\
             $M_\star$ & (\unit{\mass\sun}) & \num{2.2(.05)} & (\ref{itm:guzman}\tablefootmark{a}) \\
             $\log(L_\star/\unit{\lum\solar})$ & & \num{1.35(1)} & (\ref{itm:guzman}\tablefootmark{a}) \\
             $R_\star$ & (\unit{\radius\sun}) & \num{3.46(13)} & (\ref{itm:guzman}\tablefootmark{a}) \\
             $t_\star$ & (\unit{\mega\year}) & \num{4.4(.49:.38)} & (\ref{itm:guzman}\tablefootmark{a}) \\
             \midrule
             \textbf{HD~142527~B} & & & \\
             \textbf{\textsc{Star}} & & & \\
             Spectral Type & & M\num{2.5(1.)} & (\ref{itm:christiaens}) \\
             $T_\star$ & (\unit{\kelvin}) & \num{3500(100.)} & (\ref{itm:christiaens}) \\
             $M_\star$ & (\unit{\mass\sun}) & \num{0.34(6)} & (\ref{itm:christiaens}) \\
             $\log(L_\star/\unit{\lum\solar})$ & & \num{-0.6(.08)} & (\ref{itm:christiaens}\tablefootmark{b}) \\
             $R_\star$ & (\unit{\radius\sun}) & \num{1.37(5)} & (\ref{itm:christiaens}) \\
             $t_\star$ & (\unit{\mega\year}) & \num{1.8(1.2:0.5)} & (\ref{itm:christiaens}) \\
             \cdashline{1-4}
             \textbf{\textsc{Orbit}} & & & \\
             $a$ & (\unit{\milliArcsecond}) & \num{67.8(1.38)} & (\ref{itm:nowak}\tablefootmark{c}) \\
             $e$ & & \num{0.47(1)} & (\ref{itm:nowak}) \\
             $i$ & (\unit{\degree}) & \num{149.47(71)} & (\ref{itm:nowak}) \\
             $\omega$ & (\unit{\degree}) & \num{186.45(48)} & (\ref{itm:nowak}) \\
             $\Omega$ & (\unit{\degree}) & \num{161.51(1.01)} & (\ref{itm:nowak}) \\
             $\tau$ & (\unit{\year}) & \num{2020.42(5)} & (\ref{itm:nowak}) \\
             $P$ & (\unit{\year}) & \num{23.5(.85)} & (\ref{itm:nowak}) \\
             \midrule
             \textbf{Outer disc} & & & \\
             $R_\text{in, east}$ & (\unit{\milliArcsecond}) & $\sim$714.29 & (\ref{itm:fukagawa}\tablefootmark{d}) \\
             $R_\text{in, west}$ & (\unit{\milliArcsecond}) & $\sim$1071.43 & (\ref{itm:fukagawa}\tablefootmark{d}) \\
             $R_\text{out}$ & (\unit{\milliArcsecond}) & $\sim$3500 & (\ref{itm:fukagawa}\tablefootmark{d}) \\
             $\Delta\vartheta_\text{in-out}$ & (\unit{\degree}) & 70 & (\ref{itm:marino}) \\
             $i_\text{out}$ & (\unit{\degree}) & 38.21 & (\ref{itm:bohn}) \\
             $\theta_\text{out}$ & (\unit{\degree}) & 162.72 & (\ref{itm:bohn}) \\
             \bottomrule
        \end{tabular}
        \egroup
    \end{center}
    \tablebib{%
        \counterlabel{stellarParamsBib}{itm:gaia} \citet{GaiaCollaboration2023};
        \counterlabel{stellarParamsBib}{itm:guzman} \citet{GuzmanDiaz2021};
        \counterlabel{stellarParamsBib}{itm:fairlamb} \citet{Fairlamb2015};
        \counterlabel{stellarParamsBib}{itm:vioque} \citet{Vioque2018};
        \counterlabel{stellarParamsBib}{itm:christiaens} \citet{Christiaens2018};
        \counterlabel{stellarParamsBib}{itm:nowak} \citet{Nowak2024};
        \counterlabel{stellarParamsBib}{itm:fukagawa} \citet{Fukagawa2006};
        \counterlabel{stellarParamsBib}{itm:marino} \citet{Marino2015}.
        \counterlabel{stellarParamsBib}{itm:bohn} \citet{Bohn2022};
    }
    \tablefoot{%
        \tablefoottext{a, c, d}{Converted to milliarsecond with \acs{gaiaedr3} $d=\qty{158.51}{\parsec}$ \citep{GaiaCollaboration2021}, \acs{gaiadr3}, and \acs{gaiadr2} $d\approx\qty{140}{\parsec}$ distance \citep{GaiaCollaboration2018}.}
        \tablefoottext{b}{Computed with $L=4\pi R_\star^2\sigma T_\star^4$ \citep{Dullemond2010}.}
    }
    \label{tab:stellarParameters}
\end{table}

First, we provide a brief overview of the observations used, as well as the data treatment. Our work focuses on eight observations of HD~142527 on the \ac{matisse} instrument taken from 2021 to 2023, as presented in Table~\ref{tab:observations}. These observations were obtained as part of the \ac{gto} survey `Initial conditions of planet formation in protoplanetary discs with GRA4MAT and \ac{matisse}'. We excluded any observation with a coherence time $\tau_0<\qty{2}{\milli\second}$, seeing \qty{>1.5}{\arcsecond}, or obvious artefacts (such as glitches during observations).

Each \ac{matisse} dataset covers a wide wavelength range from the \textit{L}/\textit{M} (\qtyrange{2.8}{4.2}{\micro\metre}/\qtyrange{4.5}{5}{\micro\meter}) to the \textit{N} band (\qtyrange{8}{13}{\micro\metre}). We obtained three of the eight datasets in the quadruplet using the \qty{8.2}{\metre} \acp{ut}. We obtained the remaining datasets with the \qty{1.8}{\metre} \ac{at} in the small, medium, large, or extended configurations. With the associated baseline lengths $B\approx$~\qtyrange{10}{130}{\metre}, we reach spatial resolutions of $\theta=\frac{\lambda}{2B}\approx$~\qtyrange{30.9}{2.4}{\milliArcsecond} at \qty{3}{\micro\metre} and $\theta\approx$~\qtyrange{123.8}{9.5}{\milliArcsecond} at \qty{12}{\micro\metre}. We were able to combine the \ac{at} with the \ac{ut} observations for HD~142527 because the inner disc is fully contained within the \ac{fov} for all configurations, whereas the outer disc lies outside it, with any scattered light from it most likely resolved out. For the \textit{N} band, we used only \ac{ut} observations, as the \ac{at}s have a poor \ac{sn}.

The bulk of the \ac{matisse} data were reduced using a modification\footnote{Expected to be included in the official 2.3.0 release by \acs{eso}.} to the 2.0.2 version of the \ac{matisse} \ac{drs}\footnote{Provided by \ac{eso} at \url{https://www.eso.org/sci/software/pipelines/matisse}.} \citep{Millour2016} in combination with the \ac{matisse} tools\footnote{Additional data-reduction tools are available at the \ac{matisse} GitHub organisation: \url{https://github.com/Matisse-Consortium/tools}.}. This modified version implements an improved coherent processing that uses the 2D Fourier transform of the interferograms (previously, it used the 1D Fourier transform), which provides more accurate estimates of the correlated flux and differential phase. To calibrate the data, we observed a bright, unresolved star with well-known properties (e.g. the \ac{ldd}) and time closely matching HD~142527. \citet{Varga2021} provide a detailed description of this calibration process.

From a standard reduction of a \ac{matisse} observation, we obtain  a total spectrum $F_\nu$ (averaged from the four individual \ac{vlti} telescopes), six correlated fluxes $F_{\nu,\text{corr}}$, and four closure phases $\Phi_{\nu,\text{cp}}$. We produced the single-dish spectra from the incoherently reduced and chopped data, whereas the correlated fluxes and the closure phases stem from the coherently reduced, non-chopped data. The studies by \citet{Lopez2022} and \citet{Varga2021, Varga2024} provide detailed descriptions of a typical \ac{matisse} observation and the data reduction. In this work, we provide a brief overview of the differences between the chopped and non-chopped observing modes, as well as the incoherent and coherent reduction methods. When chopping is active, the pointing of the telescopes intermittently switches from the object to the sky background. This improves the \ac{sn} of the total spectrum by enabling background subtraction, but interferes with the stability of the interferometric fringe tracking \citep{Woillez2024}. This worsens the quality of the correlated flux and differential phase measurements. We therefore used the non-chopping mode for these quantities. In this mode, the background for the interferometric observables is removed via \ac{opd} modulation. During reduction in the coherent mode, the individual frames of the observation are aligned, preserving the phase information \citep{Lopez2022}. We then computed the noisy correlated flux estimator over the entire exposure, from which we extracted the correlated flux and the differential phase. In incoherent mode, we averaged the measured intensities without preserving the phase information of the individual frames \citep{Petrov2020}. We obtained all observations in this work either in low, medium, or high spectral resolution and then binned them to low resolution during the reduction and post-processing. This yielded data with spectral resolutions of $R=\frac{\lambda}{\Delta\lambda}\approx34$ in the \textit{L} and \textit{M} band, and of $R\approx30$ in the \textit{N} band.

In addition to the \ac{matisse} data, we used archival \textit{H} band (\qtyrange{1.5}{1.85}{\micro\metre}) observations from the \ac{oidb} at the \ac{jmmc}, obtained with the \ac{pionier} instrument \citep{Lazareff2017}. Moreover, we used \textit{K} band observations (\qtyrange{1.95}{2.45}{\micro\metre}) provided by the GRAVITY collaboration \citep{GRAVITYCollaboration2017}.

\section{Results}
\label{sec:results}
Before modelling, we interpret the calibrated data from the Figs. presented in Appendix~\ref{app:observations}. Within the \textit{H}, \textit{K}, \textit{L}, and \textit{N} bands, the correlated fluxes reach values of $3.46,3.02,2.03$, and \qty{0.63}{\jansky} for the longest baselines of $139.93,129.23,130.19$, and \qty{130.19}{\metre}, across all observations. For the same observations and baselines, the closure phases reach values of $2\fdg67,2\fdg52,10\fdg97$, and $47\fdg20$.

We focus on the \textit{N}-band data of HD~142527 as they allow us to probe the disc up to a few astronomical units. The correlated fluxes reveal contributions from spatially unresolved disc regions. At first order, longer baselines (i.e. higher spatial frequencies), probe disc regions closer to the centre and smaller in extent \citep{Buscher2015}. The lower spatial frequencies of the \textit{N} band, coupled with its sensitivity to cooler dust, make it a good indicator of the structure and composition of the extended inner disc regions (up to \qty{\sim10}{\astronomicalunit}). Moreover, HD~142527 shows strong silicate features in both total spectra and correlated fluxes (Fig.~\ref{fig:nbandData}). In particular, in the \textit{N} band, the silicate features disappear at several spatial frequencies (Figs.~\ref{fig:allDataSpatial} and \ref{fig:uv}). This could indicate a multiple-zone structure that cancels the silicate emission \citep[e.g.][]{Varga2024}. The prominence of the silicate feature aligns with previous observations by \citet{VanBoekel2005} and \citet{Juhasz2010}. The total spectra of the three epochs exhibit a very similar shape, with several crystalline silicate peaks, notably the \qty{11.3}{\micro\metre} forsterite peak. Moreover, we find a \qty{\sim10}{\percent} flux-level variability across the epochs. This variation remains within the flux calibration uncertainties and is consistent with archival \textit{N}-band data obtained with \ac{wise}, \ac{iso}/\ac{sws}, and \ac{spitzer}. The other \textit{N}-band observable is the closure phase $\Phi_{\nu,\text{cp}}$, which traces brightness asymmetries. HD~142527 shows large closure phases across epochs, suggesting a complex structure.

In contrast, the shorter-wavelength bands (\textit{H}, \textit{K}, \textit{L}, and \textit{M}) resolve compact structure close to the star (sub-\unit{\astronomicalunit} scales), due to the higher spatial frequencies from the \ac{at} observation. This wavelength range shows a less pronounced signal for the closure phase, which increases from the \textit{H} band to the \textit{M} band.

Comparing the \textit{N}-band observations to the shorter wavelengths reveals that HD~142527 is already quite resolved (i.e. the correlated fluxes are closer to zero), whereas the opposite holds for the shorter wavelengths. This results partly from the star contributing more at the shorter wavelengths, yielding higher correlated fluxes. Moreover, we observe a first zero-crossing of the complex correlated fluxes at spatial frequencies \qty{\sim15}{\mega\ensuremath{\lambda}}, with the next zero-crossing presumably occurring at spatial frequencies beyond our coverage (\qtyrange{\sim120}{150}{\mega\ensuremath{\lambda}}). The spatial frequency of zero-crossings is a particularly robust feature of our data, as it is unaffected by calibration uncertainties. An increase in the closure phase from the \textit{H} to the \textit{N} band is also apparent. This indicates a relatively symmetric structure on the smallest scales we resolve, whereas the larger scales of \qty{\sim1}{\astronomicalunit} show stronger asymmetry. This asymmetry suggests a two-component structure with an asymmetric, extended component at longer wavelengths and a more compact, relatively symmetric component at shorter wavelengths. We explore this further through geometric modelling in Sects.~\ref{sec:modelling:modelGeometries} and \ref{sec:discussion}.

\section{Modelling}
\label{sec:modelling}
We model our data using a geometrically thin disc model, first without and then with azimuthal asymmetries. We obtain the opacity of the disc material from a fit to the total spectrum in the \textit{N} band (Sect.~\ref{sec:modelling:dustOpacityModel}), which reveals a silicate emission feature with strong signs of high-temperature dust processing \citep[e.g.][]{Bouwman2001,VanBoekel2004,VanBoekel2005,Juhasz2010}. From this fit, we derive a `silicate' and `continuum' opacity component that we use in the interferometric modelling, allowing their relative ratio to vary across the different disc zones (Section~\ref{sec:modelling:modelComponents:asymTempGradDisc}).

As we lack sufficient sampling of spatial frequencies $(u,v)$ for image reconstruction, we applied a parametric modelling approach. We fitted our model $M$ to data $\vec{D}$ using Bayesian inference, implemented via the \texttt{dynesty} package \citep{Speagle2020}. The stopping criterion for this algorithm is the difference $\Delta\ln\mathcal{Z}_i$ between the current estimated $\mathcal{Z}_i$ and the remaining evidence $\Delta\mathcal{Z}_i$ at each iteration $i$. We chose $\Delta\ln\mathcal{Z}_i<0.01$, which ensures convergence when the sampler terminates. We employed flat priors $\pi(\vec{\Theta})\equiv P(\vec{\Theta},M)$ for all parameters. The number of live points $K_i$ (i.e. samples drawn from the prior volume $X$) was set to 1000. Parameter values $\vec{\Theta}$ are the sample median, while the lower and upper uncertainties are the samples at \qty{2.5}{\percent} and \qty{97.5}{\percent}, respectively. Appendix~\ref{app:bayesianFitting} provides further details on the fitting algorithm and Table~\ref{tab:modelParams} gives an overview of all model parameters.

We de-reddened the data, assuming foreground extinction (see Appendix~\ref{app:stellarSpectrum:methodDetails}) to enable the comparison of the model to the data. To ensure an equal weighting of the wavelength bands during the model fitting, we imposed a lower limit on the errors: a minimum of \qty{5}{\percent} (of the data) for the total spectrum and correlated fluxes, and \qty{5}{\degree} for the closure phases. For computational efficiency, we applied spectral binning with a window size of $\Delta\lambda=\qty{0.2}{\micro\metre}$ for the \textit{H} and \textit{K} bands, and $\Delta\lambda=\qty{0.1}{\micro\metre}$ for the remaining bands.

\subsection{Dust opacity model}
\label{sec:modelling:dustOpacityModel}

\begin{figure}[t!]
    \centering
    \includegraphics[width=0.5\textwidth]{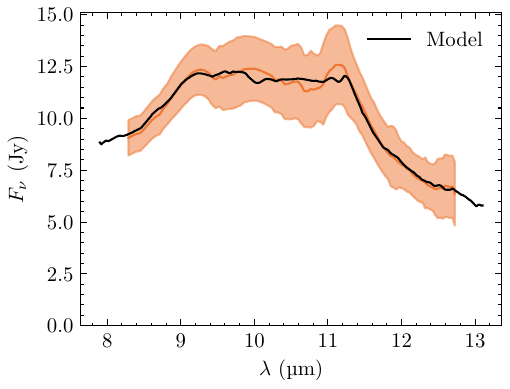}
    \caption{%
        Opacity fit. The model (black) overlays the averaged \textit{N}-band \ac{ut} data (orange). We exclude the \acs{pah} flux contribution from the model curve, as in the disc modelling in Sect.~\ref{sec:modelling:modelComponents}.
        }
    \label{fig:dustOpacityFit}
\end{figure}

We determined the opacity of disc material around HD~142527~A from a fit to the total \textit{N}-band spectrum.
This was not a quantitative dust-spectroscopy analysis; rather, we sought a well-fitting opacity across the disc model's wavelength range (Sect.~\ref{sec:modelling:modelComponents:asymTempGradDisc}). The dust opacity model \citep[see][]{VanBoekel2005} is described in Appendix~\ref{app:dustOpacityModel}. Table~\ref{tab:dustOpacityFit} and Fig.~\ref{fig:dustOpacityFit} present the resulting best-fit dust composition\footnote{Detailed analysis from \citet{VanBoekel2005} and/or \citet{Juhasz2010} (both using grains computed via the \acs{dhs} method) yield dust compositions significantly different from ours.}.

The HD~142527 system is unresolved in the total spectrum and thus does not provide insight into the disc's dust composition or its local distribution. In contrast, the correlated fluxes provide spatial information on the dust composition \citep[e.g.][]{Varga2024}. Nevertheless, for simplicity, we kept the silicate mineralogy spatially constant in our modelling, allowing only changes to the silicate-to-carbonaceous ratio (Eq.~\ref{eq:verticalOpticalDepth}). Consequently, our disc models do not reproduce the mineralogy gradient of the correlated fluxes (Fig.~\ref{fig:allDataSpatial}).

\subsection{Model components}
\label{sec:modelling:modelComponents}
Our disc models consist of a central star (i.e. a point source; Sect.~\ref{sec:modelling:modelComponents:star}), an asymmetric temperature-gradient disc (Sect.~\ref{sec:modelling:modelComponents:asymTempGradDisc}), and an off-centre Gaussian component (Sect.~\ref{sec:modelling:modelComponents:Gaussian}). We derive the complex correlated fluxes $\mathfrak{F}_{\nu}(q^\prime)$\footnote{For brevity, we shorten the de-projected spatial frequency $q^\prime=q^\prime(i,\theta)$ and omit it as a parameter of the complex correlated flux.} below. Appendix~\ref{app:computationOfObservables} shows the computation of the observables from the complex correlated flux, and an in-depth description of all applied models follows in Sect.~\ref{sec:modelling:modelGeometries}.

\subsubsection{Star}
\label{sec:modelling:modelComponents:star}
The primary star is the central component of the disc model. Its estimated angular diameter is only $\approx\qty{0.23}{\milliArcsecond}$ (see Appendix~\ref{app:stellarSpectrum}), so the star remains spatially unresolved in our observations. We treat it as a point source, whose complex correlated flux is given by
\begin{equation}
    \mathfrak{F}_{\nu,\star}=F_{\nu,\star}\exp(2\pi\mathrm{i}q^{\prime}(\alpha\cos(\psi)+\beta\sin(\psi))),
    \label{eq:complexVisStar}
\end{equation}
where $\alpha$ and $\beta$ are the angular coordinates (in radians) of the image plane in directions of \ac{ra}, and \ac{dec}, respectively, and $\psi$ is the angle corresponding to the de-projected spatial frequency $q^\prime$. For the fitting procedure, we used the polar representation of the angular coordinates, where $\rho$ is the separation from the centre and the position angle $\Phi$ (east of north) \citep{Berger2007}. Appendix~\ref{app:stellarSpectrum:methodDetails} details the derivation of the stellar flux $F_{\nu,\star}$.

\subsubsection{Asymmetric temperature-gradient disc}
\label{sec:modelling:modelComponents:asymTempGradDisc}
The disc model consists of one or more zones surrounding the primary star. Each zone (index $n$) is the result of integrating over infinitesimally thin rings \citep[see][]{Berger2007} with an emergent intensity $I_{\nu,n}$. These rings can include an additional azimuthal modulation \citep{Lazareff2017}. We now describe the emergent intensity using a power-law approach and show how it connects to the complex correlated flux.

We require the source function $B_\nu(T)$ to determine the zone intensity. The source function depends on a temperature distribution:
\begin{equation}
    T(r)=T_0\left(\frac{r}{R_0}\right)^q.
    \label{eq:tempPowerLaw}
\end{equation}
The radial behaviour of the temperature profile is determined by its power-law index $q$, where $T_0$ is the temperature at reference radius $R_0=\qty{1}{\astronomicalunit}$. Multiple zones share a single temperature profile. We also need the emissivity $\epsilon_{\nu,n}$ to describe the intensity. The emissivity depends on the optical depth $\tau_{\nu,n}$, itself determined by the zone surface density:
\begin{equation}
    \Sigma_n(r)=\Sigma_{0,n}\left(\frac{r}{R_0}\right)^{p_n}.
    \label{eq:opticalDepth}
\end{equation}
Here, $\Sigma_{0,n}$ is the surface density at $R_0$ and $p_n$ is the power-law index. Combined with the absorption-opacity curve of silicates $\kappa_{\nu,\text{abs,sil}}$, the carbonaceous grain continuum $\kappa_{\nu,\text{abs,cont}}$, and the mass fraction $w_{\text{cont},n}$, we obtain the vertical optical depth
\begin{equation}
    \tau_{\nu,n}(r)=\Sigma_n(r)[(1-w_{\text{cont},n})\kappa_{\nu,\text{abs,sil}}+w_{\text{cont},n}\kappa_{\nu,\text{abs,cont}}].
    \label{eq:verticalOpticalDepth}
\end{equation}
Accounting for inclination effects, the vertical optical depth is modified along the line of sight, and we obtain the emissivity
\begin{equation}
    \epsilon_{\nu,n}(r)=1-e^{-\tau_{\nu,n}\left(r\right)/\cos(i)}.
    \label{eq:emissivity}
\end{equation}
Multiplying this emissivity by the radially dependent source function gives the emergent intensity of the zone:
\begin{equation}
    I_{\nu,n}(r)=\epsilon_{\nu,n}(r)B_\nu\left(T(r)\right).
    \label{eq:emergentIntensityZone}
\end{equation}
From \citet{Lazareff2017} we obtain the complex correlated flux for an azimuthally modulated, infinitesimally thin ring:
\begin{equation}
    \mathfrak{F}_{\nu,\text{ring},n}=\sum^{\ell}_{m=0}(-\mathrm{i})^mA_{n,m}\cos(m(\psi-\phi_{n,m}))J_m\left(\frac{2\pi q^\prime\nu}{c}r\right).
\end{equation}
Here, $J_m$ is the Bessel function of the first kind of order $m$, $A_{n,m}$ is the modulation amplitude, and $\phi_{n,m}$ is the modulation angle. All orders up to $\ell$ contribute to the ring, which is symmetric for $\ell=0$ (i.e. $J_0$ and $A_{n,0}=1$), and azimuthally asymmetric for $\ell\geq1$. Using the intensity of Eq.~\eqref{eq:emergentIntensityZone} and integrating over the solid angle $\Omega$ yields the complex correlated flux for the azimuthally modulated disc zone,
\begin{equation}
    \mathfrak{F}_{\nu,n}=2\pi\cos(i)\int^{R_{\text{out},n}}_{R_{\text{in},n}}\mathfrak{F}_{\nu,\text{ring},n}I_{\nu,n}(r)r\odif{r}.
\end{equation}
The zone extends from an inner radius $R_{\text{in},n}$ to an outer radius $R_{\text{out},n}$.
For this component, we used the \ac{alma} observations of the outer disc from \citet{Bohn2022} in combination with the misalignment ($\Delta\vartheta_\text{in-out}=\qty{70}{\degree}$) from \citet{Marino2015} (see also Table~\ref{tab:stellarParameters}) to compute\footnote{For this computation, we rearranged and recursively solved Eq.~(7) from \citet{Bohn2022}: $i_\text{in}=\arccos[(\cos(\Delta\vartheta_\text{in-out})-\sin(i_\text{in})\sin(i_\text{out})\cos(\theta_\text{in}-\theta_\text{out}))/\cos(i_\text{out})]$} the inclination of the inner disc, and fix the inclination to $i_\text{in}=32\fdg05$ and the position angle to $\theta_\text{in}=\qty{352}{\degree}$. We followed this approach because \citet{Nowak2024} demonstrate that the angle convention of \citet{Bohn2022} is inconsistent with the observed shadows cast by the inner disc onto the outer disc of HD~142527.

\subsubsection{Gaussian}
\label{sec:modelling:modelComponents:Gaussian}
The correlated-flux behaviour (i.e. the zero-crossing; see Sect.~\ref{sec:results}) suggests a more spatially confined asymmetric component. We therefore chose an off-centre Gaussian whose complex correlated flux \citep[given by][]{Berger2007} we multiply by a source function $B_\nu(T(\rho))$, a scale factor $f$, and the emissivity $\epsilon_\nu$ as follows:
\begin{equation}
    \mathfrak{F}_{\nu,\text{Gauss}}=f\epsilon_\nu B_\nu(T(\rho))\exp\left(-\frac{(\pi aq^\prime)^2}{4\ln2}\right).
\end{equation}
Here, the temperature is computed via Eq.~\eqref{eq:tempPowerLaw}, using the power law of the accompanying disc component(s) $T(r)$ at Gaussian position $\rho$. The Gaussian size is given by the \ac{fwhm} $a$. The emissivity follows Eq.~\eqref{eq:emissivity}, except that $\Sigma$ is a fit parameter rather than a power-law derivation.

As we cannot resolve this component and assume that it is compact, we fixed the \ac{fwhm} to $a=\qty{2.57}{\milliArcsecond}$ ($\approx\qty{0.41}{\astronomicalunit}$ for $d=\qty{159.3}{\parsec}$). This choice provides sufficient flux to reproduce the signatures seen in the data while keeping the component unresolved in the \textit{N} band.

\subsection{Model geometries}
\label{sec:modelling:modelGeometries}
The following section describes the modelling approaches used in this work. Each model is centred on a point source (Sect.~\ref{sec:modelling:modelComponents:star}) representing HD~142527~A. We excluded the companion HD~142527~B from our model because it is \qty{4.5}{\magnitude} fainter than the primary in the \textit{H} band \citep[see][]{Lacour2016} and shows no definite signal in our data.

Previous studies indicate a gapped disc around HD~142527~A \citep[e.g.][]{Menu2015}. We therefore first compared two azimuthally symmetric disc models (Sect.~\ref{sec:modelling:modelComponents:asymTempGradDisc}): a one-zone (continuous) and a two-zone (gapped) disc model. Using this symmetric approach, the gapped model provided the best fit to the data.

To reproduce the large closure phases in the \textit{N} band (Sect.~\ref{sec:results}), we introduced an asymmetry. We selected the simplest asymmetric model; a first-order azimuthal modulation ($\ell=1$). Using this asymmetric two-zone disc model, we observe three effects. First, the large errors on the closure phases of the \textit{H}, \textit{K}, and \textit{L} bands prevented us from constraining the modulation of the inner zone, although the formal goodness of fit improved. Furthermore, the closure phases of these bands exhibit only a weak signal. This wavelength region is dominated by the innermost disc. To reduce the number of free parameters, we removed the modulation of the inner zone and retrained only the outer zone modulation. Second, a global asymmetry failed to adequately reproduce all visibility and phase signals, particularly the zero-point crossing, which is the most robust feature of our dataset (see Fig.~\ref{fig:allDataSpatial}) and is unaffected by systematics or calibration issues. We infer that fitting this feature requires a more localised, smaller asymmetric emission than the global $\ell=1$ modulation. Third, fitting all \textit{N}-band data simultaneously results in a worse $\chi_\text{r}^2$ than fitting the individual \textit{N}-band epochs with epoch-dependent parameters. This suggests temporal variability in the \textit{N}-band intensity distribution.

We therefore tested simple localised asymmetric emission, specifically an off-centre Gaussian. This phenomenological modelling approach yields significantly better fits than the first-order modulation. Epochs-specific fits again yield better results than fitting one model to all epochs simultaneously. This supports the assumed temporal variability, which we interpret in Sect.~\ref{sec:discussion:natureOfTheAsymmetry}.

Due to the limited $(u,v)$ coverage per epoch and the evidence for temporal variability, we did not attempt to fit more complex model geometries.

\section{Discussion}
\label{sec:discussion}

\subsection{The role of the stellar companion}
\label{sec:discussion:theStellarCompanion}

\begin{figure}[t!]
   \centering
   \includegraphics[width=0.5\textwidth]{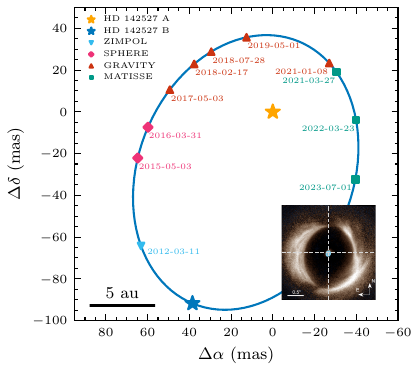}
   \caption{%
       System sketch. The host star HD~142527~A (orange star) is orbited by its companion HD~142527~B (blue star at apoapsis). We computed the orbit (blue) using parameters from \citet{Nowak2024} and used the same parameters to derive the positions of the \ac{matisse} epochs (teal squares). \textit{Bottom right}: \ac{sphere} observations of the outer disc \citep{Avenhaus2017} showing the orbit (light blue) of the companion.
   }
   \label{fig:systemSketch}
\end{figure}

We calculated the orbit of the companion HD~142527~B using the parameters of \citet{Nowak2024} listed in Table~\ref{tab:stellarParameters} and find that the three \textit{N}-band \ac{ut} observations occurred near periapsis passage (Fig.~\ref{fig:systemSketch}). We suggest that this close passage, combined with the short orbital period (\qty{\sim24}{\year}), strongly perturbs the disc, creating a complex and time-variable disc structure.

To infer the companion's influence on the primary's disc, we compute the Hill radius \citep{Eggleton1983,Hamilton1992}:
\begin{equation}
    R_\text{H}\approx a(1-e)\sqrt[3]{\frac{M_\text{B}}{3(M_\text{A}+M_\text{B})}}.
    \label{eq:hillRadius}
\end{equation}
Here, $a$ is the semi-major axis, $e$ is the eccentricity, and $M_\text{A}$ and $M_\text{B}$ are the masses of the primary and secondary, respectively. Applying the stellar parameters from Table~\ref{tab:stellarParameters}, Eq.~\eqref{eq:hillRadius} results in $R_\text{H}\approx\qty{2}{\astronomicalunit}$, roughly half the separation between the primary and secondary (\qty{\sim5}{\astronomicalunit}). This strongly influences the disc ($r_\text{out}\lesssim\qty{6}{\astronomicalunit}$) and likely generates complex structures. Moreover, perturbations caused by interactions with the secondary may deplete disc material, necessitating replenishment processes to explain the continued existence of the disc \citep[e.g. streamers, see][]{Casassus2015}.

The companion could also induce a time-variable asymmetry through alternative mechanisms, such as heating the disc mid-plane or via shock processes. This would also explain the high crystallinity observed in HD~142527 by providing temperatures sufficient to crystallise additional dust farther from the primary \citep{Harker2002,VanBoekel2004}. The \ac{sphere} observations by \citet{Avenhaus2017} support this scenario, revealing complex emissions just beyond \qty{4}{\astronomicalunit}.

\subsection{Nature of the asymmetry}
\label{sec:discussion:natureOfTheAsymmetry}

\begin{table}[ht!]
    \caption{Comparison of fit goodness (i.e. $\chi^2_\text{r,tot}$) per model.}
    \begin{center}
        \bgroup
        \def\arraystretch{1.2}
        \begin{tabular}{l | c c c }
             \toprule\toprule
             & M1 & M2 & M3 \\
             \midrule
             All epochs & 10.54 & 8.26 & 4.47 \\
             \midrule
             Epoch 1 & 10.54 & 5.50  & 2.44 \\
             Epoch 2 & 10.54 & 7.30 & 1.89 \\
             Epoch 3 & 10.54 & 3.07  & 1.64 \\
             \bottomrule
        \end{tabular}
        \egroup
    \end{center}
    \tablefoot{%
        For the calculation of these values, see Appendix~\ref{app:bayesianFitting}. \textit{First row}: time-invariant models fit to all data simultaneously. \textit{Second to last row}: the same models with additional free parameters per \textit{N}-band epoch. `M1': symmetric (time-invariant) two-zone disc model; `M2': M1 with an asymmetry in the outer zone; `M3': one-zone disc model with an off-centre Gaussian asymmetry.
    }
    \label{tab:fitGoodness}
\end{table}

\begin{table}[ht!]
    \caption{Parameters of the best-fit, time-dependent, one-zone disc model with a Gaussian asymmetry (corresponding to model M3 in Table~\ref{tab:fitGoodness}).}
    \begin{center}
        \bgroup
        \def\arraystretch{1.2}
        \begin{tabular}{l c c }
             \toprule\toprule
             Parameter & Unit & Value \\
             \midrule
             \textbf{\textsc{Free}} & \\
             \DTLforeach{bestFit}{\sI=Column1,\sII=Column2,\sIII=Column3}{\sI & \sII & \sIII \DTLiflastrow{}{\tabularnewline}}\\
             \textbf{\textsc{Fixed}} & \\
             $R_{\text{in},1}$ & (\unit{\astronomicalunit}) & 0.1 \\
             $q$ & & $-0.55$ \\
             $d$ & (\unit{\parsec}) & 159.3 \\
             $i_\text{in}$ & (\unit{\degree}) & 32.05 \\
             $\theta_\text{in}$ & (\unit{\degree}) & 352 \\
             $a$ & (\unit{\astronomicalunit}) & 0.41 \\
             \midrule
             $\chi^2_{r,F_\nu}$ & & 4.13 \\
             $\chi^2_{r,F_{\nu,\text{corr}}}$ & & 3.00 \\
             $\chi^2_{r,\Phi_{\nu,\text{cp}}}$ & & 3.86 \\
             $\chi^2_\text{r, tot}$ & & 3.36 \\
             \bottomrule
        \end{tabular}
        \egroup
    \end{center}
    \tablefoot{%
        For each parameter, the value is the median of the samples, with the uncertainties being the samples at \qty{2.5}{\percent} and \qty{97.5}{\percent}. The index $t$ indicates a time-variability (`t0': epoch 2021, `t1': epoch 2021, and `t2': epoch 2023). All angles are given in the east of north direction.
    }
    \label{tab:bestFitParams}
\end{table}

To investigate the prominent \textit{N}-band asymmetry, we build on the step-by-step evolution of our modelling approaches described in Sect.~\ref{sec:modelling:modelGeometries} and demonstrate that time-variable models, which account for the possibility of interactions between the disc and companion (Sect.~\ref{sec:discussion:theStellarCompanion}), yield the best results. Table~\ref{tab:fitGoodness} compares the goodness-of-fit of the different model geometries.

Fully symmetric models (e.g. M1) cannot reproduce the non-zero closure phases observed in the \textit{N} band. We first confirmed that a gapped disc (i.e. two-zone disc) model could be a good fit to the data. This is supported by the disappearance of the silicate feature in the correlated fluxes at some spatial frequencies (Figs.~\ref{fig:nbandData} and \ref{fig:allDataSpatial}) similar to HD~144432, where a secondary component counteracted the silicate emission from the primary component \citep{Varga2024}. In the asymmetric two-zone model, further evidence for a gap comes from the resolved component in the \textit{N}-band correlated fluxes and the decreasing closure phases at shorter wavelengths (Sect.~\ref{sec:results}), which indicates a less asymmetric structure closer to the star.

However, in Sect.~\ref{sec:modelling:modelGeometries}, we note that the zero-crossing, a robust feature of our data, is not reproduced by the time-variable two-zone disc model, while the first and third epochs (2021 and 2023) are reasonably well-fitted. This suggests that the true intensity distribution on-sky exhibits a more spatially confined, off-axis emission component than a disc with a global cosine-like azimuthal intensity modulation can provide.

\begin{figure*}[ht!]
    \centering
    \includegraphics[width=\textwidth]{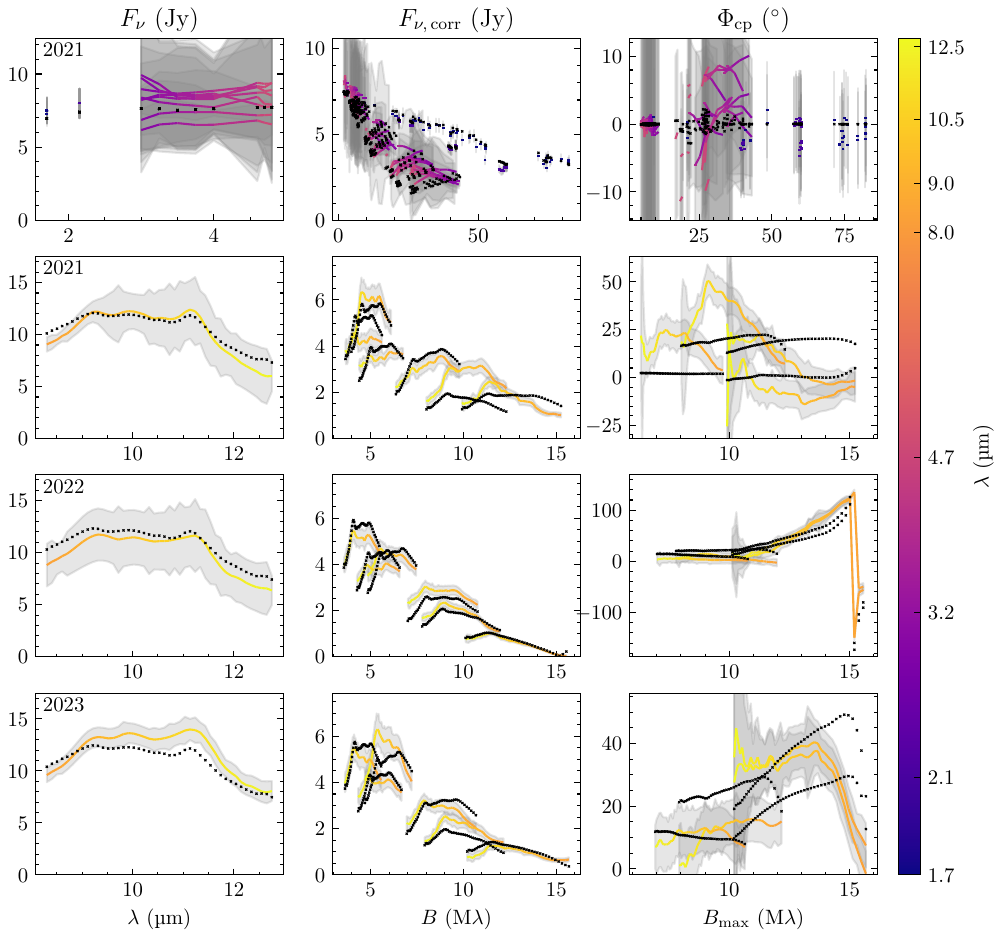}
    \caption{%
        Best-fit, one-zone disc model with off-centre Gaussian asymmetry. The data (coloured lines) are overlaid with the model (black crosses). Figure~\ref{fig:bestFitResiduals} shows the residuals of the plots. \textit{Left}: Total spectrum $F_\nu$. \textit{Middle}: Correlated fluxes $F_{\nu,\text{corr}}$. \textit{Right}: Closure phases $\Phi_{\nu,\text{cp}}$. \textit{Top}: \acs{pionier} (\textit{H} band), GRAVITY (\textit{K} band), and \acs{matisse} (\textit{L} and \textit{M} band) data, fitted with all \textit{N}-band datasets. \textit{Row 1}: Fit to the shorter wavelengths shown at the example of the first \textit{N}-band epoch. \textit{Rows 2-4}: First to third \textit{N}-band epochs (2021, 2022, and 2023).
    }
    \label{fig:bestFit}
\end{figure*}

\begin{figure*}[t!]
    \centering
    \includegraphics[width=\textwidth]{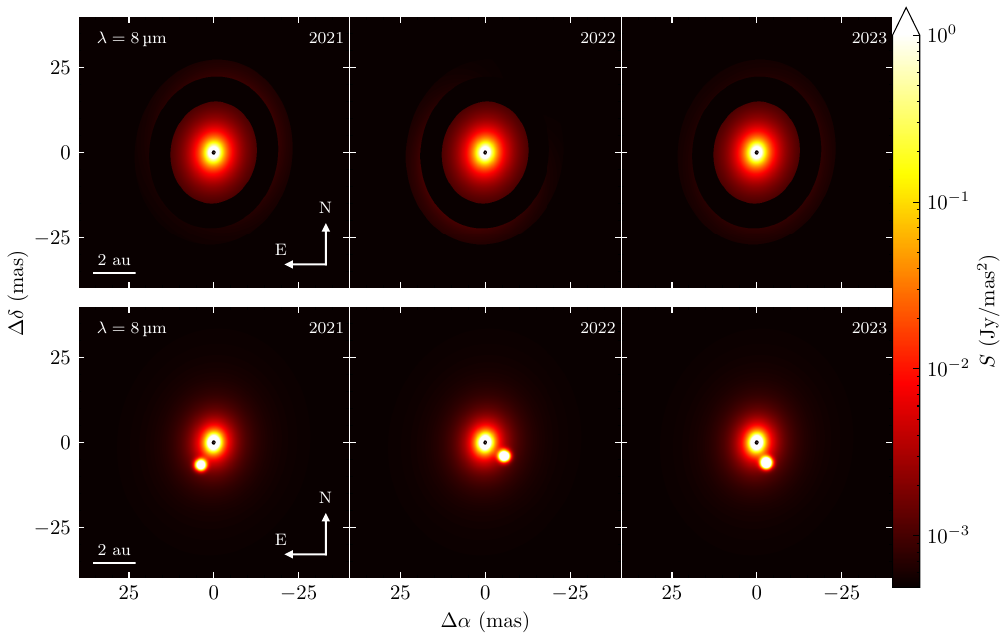}
    \caption{Model surface-brightness images. \textit{Left to right}: Epochs 2021, 2022, and 2023. \textit{Top}: Two-zone disc model. \textit{Bottom}: Best-fit, one-zone disc model with a Gaussian.}
    \label{fig:modelImages}
\end{figure*}

We therefore constructed a model with a central star, an azimuthally symmetric disc component, and a relatively compact off-axis component. We assume that the off-axis component is too compact to be significantly spatially resolved in the \textit{N} band and, for simplicity, model it as a symmetric Gaussian with a \ac{fwhm} of $a\approx\qty{2.6}{\milliArcsecond}$. The free parameters in this model are the Gaussian's position relative to the disc centre for each epoch and its relative intensity across all epochs. This approach successfully reproduces the zero-crossing in the 2022 epoch. Figs.~\ref{fig:bestFit} and \ref{fig:modelImages} illustrate this model, while Table~\ref{tab:bestFitParams} lists the best-fit parameters. 

We tested whether the angular position of the asymmetry correlates with the companion's position\footnote{$\theta_\text{B}=301\fdg882,\,264\fdg368,\,230\fdg611$ east of north, for epochs 2021, 2022, and 2023, respectively.}, but find no clear correlation between the companion and the best-fit position of the asymmetry per epoch. Neither of the explored models fully reproduces the observations, and in both cases the asymmetric element changes substantially from one epoch to the next, even between the second and third epochs, where the difference in the hour angle of the observations, and hence $(u,v)$ coverage, is small. This suggests that the true intensity distribution is more complex than the models explored here and varies on the $\sim$\qty{1}{\year} timescale that our observations probe.

Given the limited $(u,v)$ coverage of a single observation, fitting a model with significantly more complexity than the geometries explored here is not justified. At the same time, temporal variability complicates the combination of observations obtained over several years. Objects such as HD~142527 require dedicated imaging-like campaigns with the \acp{ut}, in which good $(u,v)$ coverage is collected within a time span $\ll\qty{1}{\year}$.

Nevertheless, our qualitative findings (i.e. a geometry deviating strongly from an azimuthally symmetric configuration with temporal variability on $\lesssim\qty{1}{\year}$ timescales) provide useful insight into the gravitational interaction between the circumprimary disc and the low-mass stellar companion on its highly eccentric orbit. \cite{Price2018} presents simulations of this system\footnote{Available at \url{https://users.monash.edu.au/~dprice/pubs/HD142527}.} in which the inner disc is strongly distorted during each periapsis passage of the companion and does not have time to settle between periastron passages. The disturbance proceeds from the outside-in, such that disc regions of roughly 1-few \unit{\astronomicalunit}, to which our \textit{N}-band observations are sensitive, are more strongly affected than the regions closer to the star probed by our \textit{L}-band observations. In the \textit{L} band, we see a more symmetric, less distorted geometry, as indicated by the smaller closure phases compared to the \textit{N} band. The orbital parameters of the HD~142527 system have been substantially revised since \citet{Price2018} with the orbit derived by \citet{Nowak2024}, having both a shorter period and a closer periapsis distance than any of the configurations explored in the earlier simulations. A strongly distorted and temporally variable geometry is therefore likely, and we conclude that our observations qualitatively support this scenario, but emphasise the need for a dedicated interferometric imaging-like observation strategy with the \acp{ut}.

\subsection{A very close-in rim}
\label{sec:discussion:aVeryCloseInRim}
Throughout the different model geometries, we consistently find the innder disc rim at $R_\text{rim}\approx\qty{0.1}{\astronomicalunit}$. Assuming a sublimation temperature of $T_\text{rim,literature}=\qty{1500}{\kelvin}$ and inserting this, together with the values from Table~\ref{tab:stellarParameters}, into the expression for the sublimation radius \citep{Dullemond2010},
\begin{equation}
    R_\text{rim}=\sqrt{\frac{L_\star}{4\pi\sigma T_\text{rim}^4}}=R_\star\left(\frac{T_\star}{T_\text{rim}}\right)^2    \label{eq:sublimationRadius},
\end{equation}
results in $R_\text{rim}=\qty{0.3}{\astronomicalunit}$. Here, $T_\text{rim}$ is the temperature at the inner rim of the disc, $\sigma$ is the Stefan-Boltzmann constant, and $L_\star$, $R_\star$, and $T_\star$ are the stellar luminosity, radius, and effective temperature, respectively. The sublimation radius computed from the theoretical assumption does not agree with the value derived from our data. However, such a close-in inner rim is supported by previous studies \citep[e.g.][]{Benisty2010,Benisty2011,Varga2024}. These studies reported material located beyond the sublimation radius and/or at temperatures higher than expected. In addition, earlier studies on this target by the \ac{pionier} \citep{Lazareff2017} survey determined $R_\text{rim}\approx\qty{0.05}{\astronomicalunit}$, and  $R_\text{rim}\approx\qty{0.13}{\astronomicalunit}$ with the GRAVITY \citep{GRAVITYcollaboration2019} survey. We compare the theoretically assumed temperature at the inner rim by computing the temperature at $R_\text{rim}=\qty{0.1}{\astronomicalunit}$ with the power law of our best-fit model ($T_0=\qty{599.74}{\kelvin}$ and $q=-0.55$; Table~\ref{tab:bestFitParams}). This yields a sublimation temperature $T_\text{rim}\approx\qty{2100}{\kelvin}$ that differs by $\sim$\qty{600}{\kelvin} from literature values. Several factors may contribute to this discrepancy between the calculation and the model fit. First, the assumed sublimation temperature $T_\text{rim,literature}=\qty{1500}{\kelvin}$, is an oversimplification, as it does not account for the diversity of dust grains \citep{Gail2004}. Various materials sublimate at higher temperatures. \citet{Varga2024} and \citet{Flock2025} present different materials that could explain a close-in inner rim. Examples include corundum ($T_\text{rim}\approx\qty{1850}{\kelvin}$) and tungsten ($T_\text{rim}\approx\qty{2000}{\kelvin}$). However, chemical equilibrium modelling by \cite{Varga2024} predicts that these species occur at abundances too low to produce observable spectral features, which are indeed absent. In our modelling, we therefore chose amorphous carbonaceous grains for the continuum as these are featureless across our wavelengths. Iron grains would perform better than carbonaceous grains (see Appendix~\ref{app:dustOpacityModel}). Their presence in the inner regions is supported by observations of atomic jets containing iron \citep{Assani2024,CarattioGaratti2024}, which require high speeds only available closer to the star. However, because iron is also featureless, we cannot distinguish it from carbonaceous grains in our modelling. Another potential explanation for the very close-in inner rim is emission from hot gases. \citet{Benisty2010a} showed that emission from hot gas in these regions, would require the presence of spectral lines in the \ac{sed}. We therefore inspected archival X-shooter spectra of HD~142527 \citep{Mendigutia2014,Fairlamb2015} for emission lines in the near infrared. These spectra show hydrogen lines (Br$_\gamma$, Pa$_{\beta,\gamma}$) and a few weak, isolated lines that we did not identify. This is inconsistent with dense gas being the dominant source of opacity in the inner disc region. We therefore conclude that dust likely survives farther inward and is detected by our observations \citep[e.g.][]{Klarmann2017}.

\section{Summary and conclusions}
\label{sec:summaryAndConclusions}
We present a chromatic and geometric disc model (Sect.~\ref{sec:modelling}) to probe the composition and structure of the circumstellar disc around HD~142527~A. We used photometric and interferometric data in the \acl{ir} spanning a wide wavelength range, from archival \ac{vlti}/\ac{pionier} observations (\textit{H} band) to \ac{vlti}/GRAVITY data (\textit{K} band), and new observations from our \ac{gto} \ac{vlti}/\ac{matisse} survey (\textit{L}, \textit{M}, and \textit{N} bands).

Our key findings can be summarised as follows. 
\begin{enumerate}
    \item Our \ac{matisse} data reveal large closure phases $\Phi_{\nu,\text{cp}}$ and strong silicate features in both total spectra $F_\nu$ and correlated fluxes $F_{\nu,\text{corr}}$ (Sect.~\ref{sec:results}), confirming earlier observations by \citet{VanBoekel2005} and \citet{Juhasz2010}. The closure phase signals increase in amplitude with wavelength, while the silicate features weaken towards longer baselines and vanish entirely at some spatial frequencies.
    \item Both models we applied, a two-zone disc (Sect.~\ref{sec:modelling:modelComponents:asymTempGradDisc}) with a first-order modulation in the outer zone and a one-zone disc with a localised off-centre Gaussian asymmetry (Sect.~\ref{sec:modelling:modelComponents:Gaussian}), fail reproduce the \textit{N} band of all epochs simultaneously. Only the best-fit model incorporating an off-centre Gaussian reproduces the robust zero-crossing measurement observed in the 2022 epoch (Fig.~\ref{fig:bestFit}).
    \item We obtain the best fit to our data by implementing a time-dependent asymmetry for each \textit{N}-band epoch and incorporating information from the shorter wavelength bands. To achieve this, the parameters defining the asymmetry have an iteration for each epoch, while the remaining parameters are shared across epochs (Sect.~\ref{sec:discussion:theStellarCompanion}).
    \item By combining the near-\ac{ir} with the new \textit{L}-band information, we place the inner rim at $R_\text{rim}\approx\qty{0.1}{\astronomicalunit}$ (Sect.~\ref{sec:discussion:aVeryCloseInRim}). This is consistent with earlier \textit{H}- and \textit{K}-band surveys \citep{Lazareff2017,GRAVITYcollaboration2019} and implies rim temperatures of $T_\text{rim}\approx$~\qtyrange{2100}{2200}{\kelvin} for both the one-zone-plus-Gaussian and the two-zone disc models. In contrast, assuming a wall-like inner rim and a sublimation temperature of $T_\text{rim,literature}=\qty{1500}{\kelvin}$ \citep{Dullemond2010} places the inner rim farther out at $R_\text{rim,literature}=\qty{0.3}{\astronomicalunit}$.
\end{enumerate}
From this, we conclude that: 
\begin{enumerate}
    \item The circumstellar disc around HD~142527~A exhibits a complex, strongly non-point symmetric, and temporally variable geometry. Our best-fit model shows no direct connection between the separation and/or position of the azimuthal asymmetry and the stellar companion HD~142527~B. Nevertheless, the very complex state of the innermost disc region is, likely, caused by the close pass (Fig.~\ref{fig:systemSketch}) and short orbit ($P\approx\qty{24}{\year}$) of the companion, which together with its large Hill radius of $R_\text{H}\approx\qty{2}{\astronomicalunit}$, strongly perturbs the disc ($r_\text{out}\lesssim\qty{6}{\astronomicalunit}$) around the primary (Sect.~\ref{sec:discussion:theStellarCompanion}).
    This is supported by hydrodynamical simulations from \citet{Price2018}.
    \item We cannot fully constrain the movement of the asymmetry or explain its position given the interval between epochs (\qtyrange{\sim1}{1.3}{\year}). This suggests that the asymmetry is intrinsically more complex (potentially involving multiple clumps or spirals), as observed in other planet-forming discs \citep[see][]{Setterholm2025}. Further investigation would require more sophisticated models (e.g. higher orders of modulation $\ell>1$ for the two-zone model) or image reconstruction. However, we lack sufficient data coverage to justify these approaches.
    \item The discrepancy between the spatial extent of the inner edge from the literature and our findings likely reflects one of several factors. These may include an oversimplified view of the sublimation radius \citep[i.e. not accounting for different dust grain species, see][]{Gail2004}, hot and dense gas \citep{Benisty2010a}, or material present farther inwards \citep{Klarmann2017}. Analysis of X-shooter spectra \citep{Mendigutia2014,Fairlamb2015} excludes dense, hot gas as a potential source for the rim located farther inwards. We therefore conclude that the inner rim of $R_\text{rim}\approx\qty{0.1}{\astronomicalunit}$ most likely reflects a different dust composition or material farther inward.
\end{enumerate}
Due to the complexity of the object, an in-depth study of the position and movement of the asymmetry likely requires more data. This could be achieved with a dedicated \ac{matisse} imaging-style campaign using the \acp{ut} obtained during a single epoch (ideally the same night, or at least the same period) with denser sampling in $(u,v)$ space (roughly five hour angles spaced by an hour). Additionally, observations obtained farther from the periapsis passage (e.g. around 2027, near apoapsis) could provide additional insight into the companion's effect on the disc. Finally, radial velocity measurements (e.g. with the \acs{vlt}/\ac{espresso} instrument) could constrain the orbiting speed and direction of the asymmetry.

\section*{Data availability}
\label{sec:dataAvailability}
The code used for data analysis and plotting (incl. csv. tables), which employs the \texttt{oiplot} package (\url{https://zenodo.org/records/16727743}), is available at \url{https://zenodo.org/records/18922120}. The model-fitting code from the \texttt{ppdmod} package is available at \url{https://zenodo.org/records/16728341}.

\begin{acknowledgements}
    We thank the anonymous referee for their constructive and detailed comments, which significantly contributed to the improvement of this work. The \ac{matisse} consortium is composed of multiple institutes: The Côte d’Azur observatory with the J-L Lagrange laboratory, the \ac{insu} at the \ac{cnrs}, the University of Nice Sophia-Antipolis, the \ac{mpia}, the \ac{mpifr}, the University of Kiel, the University of Leiden and the \ac{nova}, the University of Vienna, the University of Cologne, and the Konkoly observatory. GRAVITY has been developed in a collaboration by the Max-Planck-Institute for Extraterrestrial Physics, LESIA of Paris Observatory-PSL/\ac{cnrs}/Sorbonne université/université Paris Cité and IPAG of Université Grenoble Alpes/\ac{cnrs}, the \ac{mpia}, the university of Cologne, the Centro Multidisciplinar de Astrofisica Lisbon and Porto, and the \ac{eso}.\\\indent
    We also acknowledge the support of M. Fousneau and the \href{https://www.mpia.de/institute/research-and-development/data-science}{data science group} of the \ac{mpia} for their assistance with Bayesian inference.\\\indent
    This work was supported by \ac{cnrs}/INSU, by the ”Programme National de Physique Stellaire” (PNPS) of \ac{cnrs}/INSU co-funded by CEA and CNES, and by Action Spécifique ASHRA of CNRS/INSU co-funded by CNES. This work has been supported by the French National Research Agency (ANR) in the framework of the “Investissements d’Avenir” program (ANR-15-IDEX-02) and in the framework of the “ANR-23-EDIR0001-01” project. J. Varga is funded from the Hungarian NKFIH OTKA project no. K-132406, and K-147380. This work was also supported by the NKFIH NKKP grant ADVANCED 149943. Project no. 149943 has been implemented with the support provided by the Ministry of Culture and Innovation of Hungary from the National Research, Development and Innovation Fund, financed under the NKKP ADVANCED funding scheme. J. Varga acknowledges support from the Fizeau exchange visitors programme. The research leading to these results has received funding from the European Union’s Horizon 2020 research and innovation programme under Grant Agreement 101004719 (ORP). F. Lykou acknowledges support from the NKFIH OTKA project no. K-147380.\\\indent
    Furthermore, this publication made use of the \href{https://ui.adsabs.harvard.edu/}{\ac{nasa} \ac{ads}}, the \href{https://simbad.cds.unistra.fr/simbad/}{\ac{simbad}} database, operated at \ac{cds}, and the \href{https://oidb.jmmc.fr}{\ac{oidb}} of the \ac{jmmc}.\\\indent
    This work is based on observations collected at the \acl{eso} under \acs{eso} programmes 190.C-0963(B), 190.C-0963(D), 190.C-0963(E), 190.C-0963(F) for \acs{vlti}/\acs{pionier};  098.D-0488(A) for \acs{vlti}/GRAVITY; and 106.21Q8.007, 108.225V.006, 108.225V.009, 108.225V.011, 111.254P.001, and 111.254P.002 for \acs{vlti}/\acs{matisse}. In addition, we made use of data from the following surveys or libraries: the \ac{2mass} \citep{Skrutskie2006}, a joint project of the university of Massachusetts and the \ac{ipac}/\ac{caltech}, funded by \ac{nasa} and the \ac{nsf}; the \href{http://xsl.u-strasbg.fr/}{X-shooter spectral library} \citep{Chen2014}; the \ac{esa} mission \href{https://www.cosmos.esa.int/gaia}{GAIA}, processed by the GAIA \href{https://www.cosmos.esa.int/web/gaia/dpac/consortium}{\ac{dpac}}. Funding for the \ac{dpac} has been provided by national institutions, in particular the institutions participating in the GAIA multilateral agreement; and from \ac{wise} \citep{Mainzer2011}, a joint project of the University of California, Los Angeles, and the \ac{jpl}/\ac{caltech}, and \acs{neowise} \citep{Wright2010}, a project of the \ac{jpl}/\ac{caltech}. \ac{wise} and \acs{neowise} are funded by \ac{nasa}.\\\indent
    This paper utilised the following software: the \acp{llm} \href{https://openai.com/chatgpt}{\texttt{ChatGPT 4}} \citep{OpenAI2023}, \href{https://www.deepseek.com}{\texttt{DeepSeek-R1}} \citep{Guo2025}, and \href{https://proton.me/blog/lumo-1-1}{\texttt{lumo 1.1}} for code assistance/completion; \href{http://www.jmmc.fr/aspro}{\texttt{Aspro}} of the \ac{jmmc}. Furthermore, the following Python packages: \href{http://www.astropy.org}{\texttt{astropy}} \citep{AstropyCollaboration2013,AstropyCollaboration2018,AstropyCollaboration2022}; \href{https://astroquery.readthedocs.io/en/latest/index.html}{\texttt{astroquery}} \citep{Ginsburg2019,Ginsburg2024}; \href{https://corner.readthedocs.io/en/latest/index.html}{\texttt{corner}} \citep{ForemanMackey2016,ForemanMackey2024}; \href{https://dynesty.readthedocs.io/en/stable/}{\texttt{dynesty}} \citep{Speagle2020}; \href{https://matplotlib.org/}{\texttt{matplotlib}} \citep{Hunter2007}; \href{https://numpy.org/}{\texttt{numpy}} \citep{Harris2020}; \href{https://codeberg.org/MBSck/oiplot}{\texttt{oiplot}} \citep{Scheuck2025a}; \href{https://github.com/cdominik/optool}{\texttt{optool}} \citep{Dominik2021}; \href{https://pandas.pydata.org/}{\texttt{pandas}} \citep{McKinney2010,Team2025} with \href{https://openpyxl.readthedocs.io/en}{\texttt{openpyxl}}; \href{https://codeberg.org/MBSck/ppdmod}{\texttt{ppdmod}} \citep{Scheuck2025}; \href{https://scipy.org/}{\texttt{scipy}} \citep{Virtanen2020,Gommers2025} from which the following functions were used: $j0$ (wrapper of \href{https://www.netlib.org/cephes}{\texttt{Cephes}} library), $jv$ \citep[wrapper of \href{https://netlib.org/amos}{\texttt{AMOS}} $zbesj$ routine;][]{Amos1995}, and \texttt{gaussian\_kde} \citep{Gray1969,Bashtannyk2001,Scott2015,Silverman2018}; \href{https://github.com/garrettj403/SciencePlots}{\texttt{SciencePlots}} \citep{Garrett2023}; \href{https://tqdm.github.io}{\texttt{tqdm}} \citep{daCostaLuis2024}. Some of the acknowledgements were compiled using the \href{https://astrofrog.github.io/acknowledgment-generator}{\texttt{astronomy acknowledgement generator}} with some of the software citation information being aggregated using \texttt{\href{https://www.tomwagg.com/software-citation-station/}{the software citation station}} \citep{Wagg2024,Wagg2024a}.
\end{acknowledgements}
\bibliographystyle{aa}
\bibliography{references}

@Software{Ginsburg2024,
  adsurl    = {https://ui.adsabs.harvard.edu/abs/2024zndo..10799414G},
  author    = {{Ginsburg}, Adam and {Sip{\H{o}}cz}, Brigitta and {Brasseur}, C.~E. and {Parikh}, Madhura and {jcsegovia} and {Groener}, Austen and {Norman}, Henrik and {derdon} and {Kelley}, Michael and {Robitaille}, Thomas and {Lim}, P.~L. and {Vaher}, Eero and {Deil}, Christoph and {Mommert}, Michael and {Medina}, Jenny V and {Tollerud}, Erik and {Nilsson}, Ricky and {Baumann}, Matthieu and {Craig}, Matt and {de Val-Borro}, Miguel and {Weaver}, Benjamin Alan and {jespinosaar} and {Davies}, James and {Adeleke}, Tinuade and {Space Cowboy}, Nick and {Persson}, Magnus Vilhelm and {Dempsey}, James and {syed-gilani} and {Mesh}, Christian and {Mirocha}, Jordan},
  doi       = {10.5281/zenodo.10799414},
  eid       = {10.5281/zenodo.10799414},
  month     = mar,
  publisher = {Zenodo},
  title     = {{astropy/astroquery: v0.4.7}},
  version   = {v0.4.7},
  year      = {2024},
}

@Article{ALMAPartnership2015,
  author        = {{ALMA Partnership} and {Fomalont}, E.~B. and {Vlahakis}, C. and {Corder}, S. and {Remijan}, A. and {Barkats}, D. and {Lucas}, R. and {Hunter}, T.~R. and {Brogan}, C.~L. and {Asaki}, Y. and {Matsushita}, S. and {Dent}, W.~R.~F. and {Hills}, R.~E. and {Phillips}, N. and {Richards}, A.~M.~S. and {Cox}, P. and {Amestica}, R. and {Broguiere}, D. and {Cotton}, W. and {Hales}, A.~S. and {Hiriart}, R. and {Hirota}, A. and {Hodge}, J.~A. and {Impellizzeri}, C.~M.~V. and {Kern}, J. and {Kneissl}, R. and {Liuzzo}, E. and {Marcelino}, N. and {Marson}, R. and {Mignano}, A. and {Nakanishi}, K. and {Nikolic}, B. and {Perez}, J.~E. and {P{\'e}rez}, L.~M. and {Toledo}, I. and {Aladro}, R. and {Butler}, B. and {Cortes}, J. and {Cortes}, P. and {Dhawan}, V. and {Di Francesco}, J. and {Espada}, D. and {Galarza}, F. and {Garcia-Appadoo}, D. and {Guzman-Ramirez}, L. and {Humphreys}, E.~M. and {Jung}, T. and {Kameno}, S. and {Laing}, R.~A. and {Leon}, S. and {Mangum}, J. and {Marconi}, G. and {Nagai}, H. and {Nyman}, L.-A. and {Radiszcz}, M. and {Rod{\'o}n}, J.~A. and {Sawada}, T. and {Takahashi}, S. and {Tilanus}, R.~P.~J. and {van Kempen}, T. and {Vila Vilaro}, B. and {Watson}, L.~C. and {Wiklind}, T. and {Gueth}, F. and {Tatematsu}, K. and {Wootten}, A. and {Castro-Carrizo}, A. and {Chapillon}, E. and {Dumas}, G. and {de Gregorio-Monsalvo}, I. and {Francke}, H. and {Gallardo}, J. and {Garcia}, J. and {Gonzalez}, S. and {Hibbard}, J.~E. and {Hill}, T. and {Kaminski}, T. and {Karim}, A. and {Krips}, M. and {Kurono}, Y. and {Lopez}, C. and {Martin}, S. and {Maud}, L. and {Morales}, F. and {Pietu}, V. and {Plarre}, K. and {Schieven}, G. and {Testi}, L. and {Videla}, L. and {Villard}, E. and {Whyborn}, N. and {Zwaan}, M.~A. and {Alves}, F. and {Andreani}, P. and {Avison}, A. and {Barta}, M. and {Bedosti}, F. and {Bendo}, G.~J. and {Bertoldi}, F. and {Bethermin}, M. and {Biggs}, A. and {Boissier}, J. and {Brand}, J. and {Burkutean}, S. and {Casasola}, V. and {Conway}, J. and {Cortese}, L. and {Dabrowski}, B. and {Davis}, T.~A. and {Diaz Trigo}, M. and {Fontani}, F. and {Franco-Hernandez}, R. and {Fuller}, G. and {Galvan Madrid}, R. and {Giannetti}, A. and {Ginsburg}, A. and {Graves}, S.~F. and {Hatziminaoglou}, E. and {Hogerheijde}, M. and {Jachym}, P. and {Jimenez Serra}, I. and {Karlicky}, M. and {Klaasen}, P. and {Kraus}, M. and {Kunneriath}, D. and {Lagos}, C. and {Longmore}, S. and {Leurini}, S. and {Maercker}, M. and {Magnelli}, B. and {Marti Vidal}, I. and {Massardi}, M. and {Maury}, A. and {Muehle}, S. and {Muller}, S. and {Muxlow}, T. and {O'Gorman}, E. and {Paladino}, R. and {Petry}, D. and {Pineda}, J.~E. and {Randall}, S. and {Richer}, J.~S. and {Rossetti}, A. and {Rushton}, A. and {Rygl}, K. and {Sanchez Monge}, A. and {Schaaf}, R. and {Schilke}, P. and {Stanke}, T. and {Schmalzl}, M. and {Stoehr}, F. and {Urban}, S. and {van Kampen}, E. and {Vlemmings}, W. and {Wang}, K. and {Wild}, W. and {Yang}, Y. and {Iguchi}, S. and {Hasegawa}, T. and {Saito}, M. and {Inatani}, J. and {Mizuno}, N. and {Asayama}, S. and {Kosugi}, G. and {Morita}, K.-I. and {Chiba}, K. and {Kawashima}, S. and {Okumura}, S.~K. and {Ohashi}, N. and {Ogasawara}, R. and {Sakamoto}, S. and {Noguchi}, T. and {Huang}, Y.-D. and {Liu}, S.-Y. and {Kemper}, F. and {Koch}, P.~M. and {Chen}, M.-T. and {Chikada}, Y. and {Hiramatsu}, M. and {Iono}, D. and {Shimojo}, M. and {Komugi}, S. and {Kim}, J. and {Lyo}, A.-R. and {Muller}, E. and {Herrera}, C. and {Miura}, R.~E. and {Ueda}, J. and {Chibueze}, J. and {Su}, Y.-N. and {Trejo-Cruz}, A. and {Wang}, K.-S. and {Kiuchi}, H. and {Ukita}, N. and {Sugimoto}, M. and {Kawabe}, R. and {Hayashi}, M. and {Miyama}, S. and {Ho}, P.~T.~P. and {Kaifu}, N.},
  journal       = {\apjl},
  title         = {{The 2014 ALMA Long Baseline Campaign: An Overview}},
  year          = {2015},
  month         = jul,
  number        = {1},
  pages         = {L1},
  volume        = {808},
  adsurl        = {https://ui.adsabs.harvard.edu/abs/2015ApJ...808L...1A},
  archiveprefix = {arXiv},
  doi           = {10.1088/2041-8205/808/1/L1},
  eid           = {L1},
  eprint        = {1504.04877},
  primaryclass  = {astro-ph.IM},
}

@Article{Amos1995,
  author     = {Amos, D. E.},
  journal    = {ACM Trans. Math. Softw.},
  title      = {A remark on Algorithm 644: “A portable package for Bessel functions of a complex argument and nonnegative order”},
  year       = {1995},
  issn       = {0098-3500},
  month      = dec,
  number     = {4},
  pages      = {388–393},
  volume     = {21},
  address    = {New York, NY, USA},
  doi        = {10.1145/212066.212078},
  issue_date = {Dec. 1995},
  numpages   = {6},
  publisher  = {Association for Computing Machinery},
}

@Article{Andrae2010,
  author        = {{Andrae}, Rene and {Schulze-Hartung}, Tim and {Melchior}, Peter},
  journal       = {arXiv e-prints},
  title         = {{Dos and don'ts of reduced chi-squared}},
  year          = {2010},
  month         = dec,
  pages         = {arXiv:1012.3754},
  adsurl        = {https://ui.adsabs.harvard.edu/abs/2010arXiv1012.3754A},
  archiveprefix = {arXiv},
  doi           = {10.48550/arXiv.1012.3754},
  eid           = {arXiv:1012.3754},
  eprint        = {1012.3754},
  primaryclass  = {astro-ph.IM},
}

@Article{Andrews2018,
  author        = {{Andrews}, Sean M. and {Huang}, Jane and {P{\'e}rez}, Laura M. and {Isella}, Andrea and {Dullemond}, Cornelis P. and {Kurtovic}, Nicol{\'a}s T. and {Guzm{\'a}n}, Viviana V. and {Carpenter}, John M. and {Wilner}, David J. and {Zhang}, Shangjia and {Zhu}, Zhaohuan and {Birnstiel}, Tilman and {Bai}, Xue-Ning and {Benisty}, Myriam and {Hughes}, A. Meredith and {{\"O}berg}, Karin I. and {Ricci}, Luca},
  journal       = {\apjl},
  title         = {{The Disk Substructures at High Angular Resolution Project (DSHARP). I. Motivation, Sample, Calibration, and Overview}},
  year          = {2018},
  month         = dec,
  number        = {2},
  pages         = {L41},
  volume        = {869},
  adsurl        = {https://ui.adsabs.harvard.edu/abs/2018ApJ...869L..41A},
  archiveprefix = {arXiv},
  doi           = {10.3847/2041-8213/aaf741},
  eid           = {L41},
  eprint        = {1812.04040},
  primaryclass  = {astro-ph.SR},
}

@Article{Assani2024,
  author        = {{Assani}, K.~D. and {Harsono}, D. and {Ramsey}, J.~P. and {Li}, Z.-Y. and {Bjerkeli}, P. and {Pontoppidan}, K.~M. and {Tychoniec}, {\L}. and {Calcutt}, H. and {Kristensen}, L.~E. and {J{\o}rgensen}, J.~K. and {Plunkett}, A. and {van Gelder}, M.~L. and {Francis}, L.},
  journal       = {\aap},
  title         = {{The asymmetric bipolar [Fe II] jet and H$_{2}$ outflow of TMC1A resolved with the JWST NIRSpec Integral Field Unit}},
  year          = {2024},
  month         = aug,
  pages         = {A26},
  volume        = {688},
  adsurl        = {https://ui.adsabs.harvard.edu/abs/2024A&A...688A..26A},
  archiveprefix = {arXiv},
  doi           = {10.1051/0004-6361/202449745},
  eid           = {A26},
  eprint        = {2404.18334},
  primaryclass  = {astro-ph.SR},
}

@Article{Avenhaus2017,
  author        = {{Avenhaus}, H. and {Quanz}, S.~P. and {Schmid}, H.~M. and {Dominik}, C. and {Stolker}, T. and {Ginski}, C. and {de Boer}, J. and {Szul{\'a}gyi}, J. and {Garufi}, A. and {Zurlo}, A. and {Hagelberg}, J. and {Benisty}, M. and {Henning}, T. and {M{\'e}nard}, F. and {Meyer}, M.~R. and {Baruffolo}, A. and {Bazzon}, A. and {Beuzit}, J.~L. and {Costille}, A. and {Dohlen}, K. and {Girard}, J.~H. and {Gisler}, D. and {Kasper}, M. and {Mouillet}, D. and {Pragt}, J. and {Roelfsema}, R. and {Salasnich}, B. and {Sauvage}, J.-F.},
  journal       = {\aj},
  title         = {{Exploring Dust around HD 142527 down to 0.″025 (4 au) Using SPHERE/ZIMPOL}},
  year          = {2017},
  month         = jul,
  number        = {1},
  pages         = {33},
  volume        = {154},
  adsurl        = {https://ui.adsabs.harvard.edu/abs/2017AJ....154...33A},
  archiveprefix = {arXiv},
  doi           = {10.3847/1538-3881/aa7560},
  eid           = {33},
  eprint        = {1705.09680},
  primaryclass  = {astro-ph.EP},
}

@Article{Bashtannyk2001,
  author   = {David M. Bashtannyk and Rob J. Hyndman},
  journal  = {Comput. Stat. Data An.},
  title    = {Bandwidth selection for kernel conditional density estimation},
  year     = {2001},
  issn     = {0167-9473},
  number   = {3},
  pages    = {279-298},
  volume   = {36},
  doi      = {10.1016/S0167-9473(00)00046-3},
  fjournal = {Computational Statistics & Data Analysis},
  url      = {https://www.sciencedirect.com/science/article/pii/S0167947300000463},
}

@Article{Benisty2010,
  author        = {{Benisty}, M. and {Tatulli}, E. and {M{\'e}nard}, F. and {Swain}, M.~R.},
  journal       = {\aap},
  title         = {{The complex structure of the disk around HD 100546. The inner few astronomical units}},
  year          = {2010},
  month         = feb,
  pages         = {A75},
  volume        = {511},
  adsurl        = {https://ui.adsabs.harvard.edu/abs/2010A&A...511A..75B},
  archiveprefix = {arXiv},
  doi           = {10.1051/0004-6361/200913590},
  eid           = {A75},
  eprint        = {1001.2491},
  primaryclass  = {astro-ph.SR},
}

@Article{Benisty2011,
  author        = {{Benisty}, M. and {Renard}, S. and {Natta}, A. and {Berger}, J.~P. and {Massi}, F. and {Malbet}, F. and {Garcia}, P.~J.~V. and {Isella}, A. and {M{\'e}rand}, A. and {Monin}, J.~L. and {Testi}, L. and {Thi{\'e}baut}, E. and {Vannier}, M. and {Weigelt}, G.},
  journal       = {\aap},
  title         = {{A low optical depth region in the inner disk of the Herbig Ae star HR 5999}},
  year          = {2011},
  month         = jul,
  pages         = {A84},
  volume        = {531},
  adsurl        = {https://ui.adsabs.harvard.edu/abs/2011A&A...531A..84B},
  archiveprefix = {arXiv},
  doi           = {10.1051/0004-6361/201016091},
  eid           = {A84},
  eprint        = {1106.4150},
  primaryclass  = {astro-ph.SR},
}

@InProceedings{Benisty2023,
  author        = {{Benisty}, M. and {Dominik}, C. and {Follette}, K. and {Garufi}, A. and {Ginski}, C. and {Hashimoto}, J. and {Keppler}, M. and {Kley}, W. and {Monnier}, J.},
  booktitle     = {Protostars and Planets VII},
  title         = {{Optical and Near-infrared View of Planet-forming Disks and Protoplanets}},
  year          = {2023},
  editor        = {{Inutsuka}, S. and {Aikawa}, Y. and {Muto}, T. and {Tomida}, K. and {Tamura}, M.},
  month         = jul,
  pages         = {605},
  series        = {Astronomical Society of the Pacific Conference Series},
  volume        = {534},
  adsurl        = {https://ui.adsabs.harvard.edu/abs/2023ASPC..534..605B},
  archiveprefix = {arXiv},
  doi           = {10.48550/arXiv.2203.09991},
  eprint        = {2203.09991},
  primaryclass  = {astro-ph.EP},
}

@Article{Benisty2010a,
  author        = {{Benisty}, M. and {Natta}, A. and {Isella}, A. and {Berger}, J.-P. and {Massi}, F. and {Le Bouquin}, J.-B. and {M{\'e}rand}, A. and {Duvert}, G. and {Kraus}, S. and {Malbet}, F. and {Olofsson}, J. and {Robbe-Dubois}, S. and {Testi}, L. and {Vannier}, M. and {Weigelt}, G.},
  journal       = {\aap},
  title         = {{Strong near-infrared emission in the sub-AU disk of the Herbig Ae star HD 163296: evidence of refractory dust?}},
  year          = {2010},
  month         = feb,
  pages         = {A74},
  volume        = {511},
  adsurl        = {https://ui.adsabs.harvard.edu/abs/2010A&A...511A..74B},
  archiveprefix = {arXiv},
  doi           = {10.1051/0004-6361/200912898},
  eid           = {A74},
  eprint        = {0911.4363},
  primaryclass  = {astro-ph.SR},
}

@Article{Berger2007,
  author  = {{Berger}, Jean Philippe and {Segransan}, Damien},
  journal = {\nar},
  title   = {{An introduction to visibility modeling}},
  year    = {2007},
  month   = oct,
  number  = {8-9},
  pages   = {576-582},
  volume  = {51},
  adsurl  = {https://ui.adsabs.harvard.edu/abs/2007NewAR..51..576B},
  doi     = {10.1016/j.newar.2007.06.003},
}

@Article{Biller2012,
  author        = {{Biller}, Beth and {Lacour}, Sylvestre and {Juh{\'a}sz}, Attila and {Benisty}, Myriam and {Chauvin}, Gael and {Olofsson}, Johan and {Pott}, J{\"o}rg-Uwe and {M{\"u}ller}, Andr{\'e} and {Sicilia-Aguilar}, Aurora and {Bonnefoy}, Micka{\"e}l and {Tuthill}, Peter and {Thebault}, Philippe and {Henning}, Thomas and {Crida}, Aurelien},
  journal       = {\apjl},
  title         = {{A Likely Close-in Low-mass Stellar Companion to the Transitional Disk Star HD 142527}},
  year          = {2012},
  month         = jul,
  number        = {2},
  pages         = {L38},
  volume        = {753},
  adsurl        = {https://ui.adsabs.harvard.edu/abs/2012ApJ...753L..38B},
  archiveprefix = {arXiv},
  doi           = {10.1088/2041-8205/753/2/L38},
  eid           = {L38},
  eprint        = {1206.2654},
  primaryclass  = {astro-ph.SR},
}

@Article{Birnstiel2024,
  author        = {{Birnstiel}, Tilman},
  journal       = {\araa},
  title         = {{Dust Growth and Evolution in Protoplanetary Disks}},
  year          = {2024},
  month         = sep,
  number        = {1},
  pages         = {157-202},
  volume        = {62},
  adsurl        = {https://ui.adsabs.harvard.edu/abs/2024ARA&A..62..157B},
  archiveprefix = {arXiv},
  doi           = {10.1146/annurev-astro-071221-052705},
  eprint        = {2312.13287},
  primaryclass  = {astro-ph.EP},
}

@Article{Boccaletti2020,
  author        = {{Boccaletti}, A. and {Di Folco}, E. and {Pantin}, E. and {Dutrey}, A. and {Guilloteau}, S. and {Tang}, Y.~W. and {Pi{\'e}tu}, V. and {Habart}, E. and {Milli}, J. and {Beck}, T.~L. and {Maire}, A.-L.},
  journal       = {\aap},
  title         = {{Possible evidence of ongoing planet formation in AB Aurigae. A showcase of the SPHERE/ALMA synergy}},
  year          = {2020},
  month         = may,
  pages         = {L5},
  volume        = {637},
  adsurl        = {https://ui.adsabs.harvard.edu/abs/2020A&A...637L...5B},
  archiveprefix = {arXiv},
  doi           = {10.1051/0004-6361/202038008},
  eid           = {L5},
  eprint        = {2005.09064},
  primaryclass  = {astro-ph.EP},
}

@Article{Bohn2022,
  author        = {{Bohn}, A.~J. and {Benisty}, M. and {Perraut}, K. and {van der Marel}, N. and {W{\"o}lfer}, L. and {van Dishoeck}, E.~F. and {Facchini}, S. and {Manara}, C.~F. and {Teague}, R. and {Francis}, L. and {Berger}, J.-P. and {Garcia-Lopez}, R. and {Ginski}, C. and {Henning}, T. and {Kenworthy}, M. and {Kraus}, S. and {M{\'e}nard}, F. and {M{\'e}rand}, A. and {P{\'e}rez}, L.~M.},
  journal       = {\aap},
  title         = {{Probing inner and outer disk misalignments in transition disks. Constraints from VLTI/GRAVITY and ALMA observations}},
  year          = {2022},
  month         = feb,
  pages         = {A183},
  volume        = {658},
  adsurl        = {https://ui.adsabs.harvard.edu/abs/2022A&A...658A.183B},
  archiveprefix = {arXiv},
  doi           = {10.1051/0004-6361/202142070},
  eid           = {A183},
  eprint        = {2112.00123},
  primaryclass  = {astro-ph.EP},
}

@Article{Bouwman2001,
  author   = {{Bouwman}, J. and {Meeus}, G. and {de Koter}, A. and {Hony}, S. and {Dominik}, C. and {Waters}, L.~B.~F.~M.},
  journal  = {\aap},
  title    = {{Processing of silicate dust grains in Herbig Ae/Be systems}},
  year     = {2001},
  month    = sep,
  pages    = {950-962},
  volume   = {375},
  adsurl   = {https://ui.adsabs.harvard.edu/abs/2001A&A...375..950B},
  doi      = {10.1051/0004-6361:20010878},
}

@Book{Buscher2015,
  author     = {Buscher, David F.},
  publisher  = {Cambridge University Press},
  title      = {Practical Optical Interferometry: Imaging at Visible and Infrared Wavelengths},
  year       = {2015},
  series     = {Cambridge Observing Handbooks for Research Astronomers},
  collection = {Cambridge Observing Handbooks for Research Astronomers},
  place      = {Cambridge},
}

@Article{CarattioGaratti2024,
  author        = {{Caratti o Garatti}, A. and {Ray}, T.~P. and {Kavanagh}, P.~J. and {McCaughrean}, M.~J. and {Gieser}, C. and {Giannini}, T. and {van Dishoeck}, E.~F. and {Justtanont}, K. and {van Gelder}, M.~L. and {Francis}, L. and {Beuther}, H. and {Tychoniec}, {\L}. and {Nisini}, B. and {Navarro}, M.~G. and {Devaraj}, R. and {Reyes}, S. and {Nazari}, P. and {Klaassen}, P. and {G{\"u}del}, M. and {Henning}, Th. and {Lagage}, P.~O. and {{\"O}stlin}, G. and {Vandenbussche}, B. and {Waelkens}, C. and {Wright}, G.},
  journal       = {\aap},
  title         = {{JWST Observations of Young protoStars (JOYS): HH211: Textbook case of a protostellar jet and outflow}},
  year          = {2024},
  month         = nov,
  pages         = {A134},
  volume        = {691},
  adsurl        = {https://ui.adsabs.harvard.edu/abs/2024A&A...691A.134C},
  archiveprefix = {arXiv},
  doi           = {10.1051/0004-6361/202451350},
  eid           = {A134},
  eprint        = {2409.16061},
  primaryclass  = {astro-ph.SR},
}

@Article{Casassus2015,
  author        = {{Casassus}, S. and {Marino}, S. and {P{\'e}rez}, S. and {Roman}, P. and {Dunhill}, A. and {Armitage}, P.~J. and {Cuadra}, J. and {Wootten}, A. and {van der Plas}, G. and {Cieza}, L. and {Moral}, Victor and {Christiaens}, V. and {Montesinos}, Mat{\'\i}as},
  journal       = {\apj},
  title         = {{Accretion Kinematics through the Warped Transition Disk in HD142527 from Resolved CO(6-5) Observations}},
  year          = {2015},
  month         = oct,
  number        = {2},
  pages         = {92},
  volume        = {811},
  adsurl        = {https://ui.adsabs.harvard.edu/abs/2015ApJ...811...92C},
  archiveprefix = {arXiv},
  doi           = {10.1088/0004-637X/811/2/92},
  eid           = {92},
  eprint        = {1505.07732},
  primaryclass  = {astro-ph.SR},
}

@Article{Casassus2013,
  author        = {{Casassus}, Simon and {van der Plas}, Gerrit M. and {Perez}, Sebastian and {Dent}, William R.~F. and {Fomalont}, Ed and {Hagelberg}, Janis and {Hales}, Antonio and {Jord{\'a}n}, Andr{\'e}s and {Mawet}, Dimitri and {M{\'e}nard}, Francois and {Wootten}, Al and {Wilner}, David and {Hughes}, A. Meredith and {Schreiber}, Matthias R. and {Girard}, Julien H. and {Ercolano}, Barbara and {Canovas}, Hector and {Rom{\'a}n}, Pablo E. and {Salinas}, Vachail},
  journal       = {\nat},
  title         = {{Flows of gas through a protoplanetary gap}},
  year          = {2013},
  month         = jan,
  number        = {7431},
  pages         = {191-194},
  volume        = {493},
  adsurl        = {https://ui.adsabs.harvard.edu/abs/2013Natur.493..191C},
  archiveprefix = {arXiv},
  doi           = {10.1038/nature11769},
  eprint        = {1305.6062},
  primaryclass  = {astro-ph.GA},
}

@Software{daCostaLuis2024,
  adsurl    = {https://ui.adsabs.harvard.edu/abs/2024zndo..14231923D},
  author    = {{da Costa-Luis}, Casper and {Larroque}, Stephen Karl and {Altendorf}, Kyle and {Mary}, Hadrien and {richardsheridan} and {Korobov}, Mikhail and {Yorav-Raphael}, Noam and {Ivanov}, Ivan and {Bargull}, Marcel and {Rodrigues}, Nishant and {Shawn} and {Dektyarev}, Mikhail and {G{\'o}rny}, Micha{\l} and {mjstevens777} and {Pagel}, Matthew D. and {Zugnoni}, Martin and {JC} and {CrazyPython} and {Newey}, Charles and {Lee}, Antony and {pgajdos} and {Todd} and {Malmgren}, Staffan and {redbug312} and {Desh}, Orivej and {Nechaev}, Nikolay and {Boyle}, Mike and {Nordlund}, Max and {MapleCCC} and {McCracken}, Jack},
  doi       = {10.5281/zenodo.14231923},
  eid       = {10.5281/zenodo.14231923},
  month     = nov,
  publisher = {Zenodo},
  title     = {{tqdm: A fast, Extensible Progress Bar for Python and CLI}},
  version   = {v4.67.1},
  year      = {2024},
}

@Article{Chen2014,
  author  = {{Chen}, Y.-P. and {Trager}, S.~C. and {Peletier}, R.~F. and {Lan{\c{c}}on}, A. and {Vazdekis}, A. and {Prugniel}, P. and {Silva}, D. and {Gonneau}, A. and {Lyubenova}, M. and {Koleva}, M. and {Barroso}, J.~F. and {Bl{\'a}zquez}, P.~S. and {Walcher}, C.~J. and {Choudhury}, O.~S. and {Meneses-Goytia}, S.},
  journal = {The Messenger},
  title   = {{The X-shooter Spectral Library (XSL) and its First Data Release}},
  year    = {2014},
  month   = dec,
  pages   = {30-34},
  volume  = {158},
  adsurl  = {https://ui.adsabs.harvard.edu/abs/2014Msngr.158...30C},
}

@Article{Choi2016,
  author        = {{Choi}, Jieun and {Dotter}, Aaron and {Conroy}, Charlie and {Cantiello}, Matteo and {Paxton}, Bill and {Johnson}, Benjamin D.},
  journal       = {\apj},
  title         = {{Mesa Isochrones and Stellar Tracks (MIST). I. Solar-scaled Models}},
  year          = {2016},
  month         = jun,
  number        = {2},
  pages         = {102},
  volume        = {823},
  adsurl        = {https://ui.adsabs.harvard.edu/abs/2016ApJ...823..102C},
  archiveprefix = {arXiv},
  doi           = {10.3847/0004-637X/823/2/102},
  eid           = {102},
  eprint        = {1604.08592},
  primaryclass  = {astro-ph.SR},
}

@Article{Christiaens2018,
  author        = {{Christiaens}, V. and {Casassus}, S. and {Absil}, O. and {Kimeswenger}, S. and {Gomez Gonzalez}, C.~A. and {Girard}, J. and {Ram{\'\i}rez}, R. and {Wertz}, O. and {Zurlo}, A. and {Wahhaj}, Z. and {Flores}, C. and {Salinas}, V. and {Jord{\'a}n}, A. and {Mawet}, D.},
  journal       = {\aap},
  title         = {{Characterization of low-mass companion HD 142527 B}},
  year          = {2018},
  month         = sep,
  pages         = {A37},
  volume        = {617},
  adsurl        = {https://ui.adsabs.harvard.edu/abs/2018A&A...617A..37C},
  archiveprefix = {arXiv},
  doi           = {10.1051/0004-6361/201629454},
  eid           = {A37},
  eprint        = {1806.04792},
  primaryclass  = {astro-ph.EP},
}

@Article{Claudi2019,
  author        = {{Claudi}, R. and {Maire}, A.-L. and {Mesa}, D. and {Cheetham}, A. and {Fontanive}, C. and {Gratton}, R. and {Zurlo}, A. and {Avenhaus}, H. and {Bhowmik}, T. and {Biller}, B. and {Boccaletti}, A. and {Bonavita}, M. and {Bonnefoy}, M. and {Cascone}, E. and {Chauvin}, G. and {Delboulb{\'e}}, A. and {Desidera}, S. and {D'Orazi}, V. and {Feautrier}, P. and {Feldt}, M. and {Flammini Dotti}, F. and {Girard}, J.~H. and {Giro}, E. and {Janson}, M. and {Hagelberg}, J. and {Keppler}, M. and {Kopytova}, T. and {Lacour}, S. and {Lagrange}, A.-M. and {Langlois}, M. and {Lannier}, J. and {Le Coroller}, H. and {Menard}, F. and {Messina}, S. and {Meyer}, M. and {Millward}, M. and {Olofsson}, J. and {Pavlov}, A. and {Peretti}, S. and {Perrot}, C. and {Pinte}, C. and {Pragt}, J. and {Ramos}, J. and {Rochat}, S. and {Rodet}, L. and {Roelfsema}, R. and {Rouan}, D. and {Salter}, G. and {Schmidt}, T. and {Sissa}, E. and {Thebault}, P. and {Udry}, S. and {Vigan}, A.},
  journal       = {\aap},
  title         = {{SPHERE dynamical and spectroscopic characterization of HD 142527B}},
  year          = {2019},
  month         = feb,
  pages         = {A96},
  volume        = {622},
  adsurl        = {https://ui.adsabs.harvard.edu/abs/2019A&A...622A..96C},
  archiveprefix = {arXiv},
  doi           = {10.1051/0004-6361/201833990},
  eid           = {A96},
  eprint        = {1812.07814},
  primaryclass  = {astro-ph.SR},
}

@Article{Cohen2003,
  author        = {{Cohen}, Martin and {Wheaton}, Wm. A. and {Megeath}, S.~T.},
  journal       = {\aj},
  title         = {{Spectral Irradiance Calibration in the Infrared. XIV. The Absolute Calibration of 2MASS}},
  year          = {2003},
  month         = aug,
  number        = {2},
  pages         = {1090-1096},
  volume        = {126},
  adsurl        = {https://ui.adsabs.harvard.edu/abs/2003AJ....126.1090C},
  archiveprefix = {arXiv},
  doi           = {10.1086/376474},
  eprint        = {astro-ph/0304350},
  primaryclass  = {astro-ph},
}

@Book{Cutri2003,
  author  = {{Cutri}, R.~M. and {Skrutskie}, M.~F. and {van Dyk}, S. and {Beichman}, C.~A. and {Carpenter}, J.~M. and {Chester}, T. and {Cambresy}, L. and {Evans}, T. and {Fowler}, J. and {Gizis}, J. and {Howard}, E. and {Huchra}, J. and {Jarrett}, T. and {Kopan}, E.~L. and {Kirkpatrick}, J.~D. and {Light}, R.~M. and {Marsh}, K.~A. and {McCallon}, H. and {Schneider}, S. and {Stiening}, R. and {Sykes}, M. and {Weinberg}, M. and {Wheaton}, W.~A. and {Wheelock}, S. and {Zacarias}, N.},
  title   = {{2MASS All Sky Catalog of point sources.}},
  year    = {2003},
  adsurl  = {https://ui.adsabs.harvard.edu/abs/2003tmc..book.....C},
}

@Software{ForemanMackey2024,
  adsurl    = {https://ui.adsabs.harvard.edu/abs/2024zndo..14209694F},
  author    = {{Foreman-Mackey}, Dan and {Price-Whelan}, Adrian and {Vousden}, Will and {Ryan}, Geoffrey and {Pitkin}, Matt and {Zabalza}, V{\'\i}ctor and {jsheyl} and {Rice}, Emily and {Smith}, Michael and {Singer}, Leo and {Ashton}, Gregory and {Smith}, Arfon and {Cruz}, Kelle and {Weiner}, Zach and {Kerzendorf}, Wolfgang and {Vandal}, Thomas and {Caswell}, Thomas A and {Li}, Zhaozhou and {Hoyer}, Stephan and {Castillo Hair}, Sebastian M. and {Prechelt}, Remy and {Marlon} and {Barbary}, Kyle and {Buchner}, Johannes and {Tocknell}, James and {Matthews}, James and {NeutralKaon} and {Czekala}, Ian},
  doi       = {10.5281/zenodo.14209694},
  eid       = {10.5281/zenodo.14209694},
  month     = nov,
  publisher = {Zenodo},
  title     = {{dfm/corner.py: v2.2.3}},
  version   = {v2.2.3},
  year      = {2024},
}

@Article{Guo2025,
  author        = {{Guo}, Daya and {Yang}, Dejian and {Zhang}, Haowei and {Song}, Junxiao and {Wang}, Peiyi and {Zhu}, Qihao and {Xu}, Runxin and {Zhang}, Ruoyu and {Ma}, Shirong and {Bi}, Xiao and {Zhang}, Xiaokang and {Yu}, Xingkai and {Wu}, Yu and {Wu}, Z.~F. and {Gou}, Zhibin and {Shao}, Zhihong and {Li}, Zhuoshu and {Gao}, Ziyi and {Liu}, Aixin and {Xue}, Bing and {Wang}, Bingxuan and {Wu}, Bochao and {Feng}, Bei and {Lu}, Chengda and {Zhao}, Chenggang and {Deng}, Chengqi and {Ruan}, Chong and {Dai}, Damai and {Chen}, Deli and {Ji}, Dongjie and {Li}, Erhang and {Lin}, Fangyun and {Dai}, Fucong and {Luo}, Fuli and {Hao}, Guangbo and {Chen}, Guanting and {Li}, Guowei and {Zhang}, H. and {Xu}, Hanwei and {Ding}, Honghui and {Gao}, Huazuo and {Qu}, Hui and {Li}, Hui and {Guo}, Jianzhong and {Li}, Jiashi and {Chen}, Jingchang and {Yuan}, Jingyang and {Tu}, Jinhao and {Qiu}, Junjie and {Li}, Junlong and {Cai}, J.~L. and {Ni}, Jiaqi and {Liang}, Jian and {Chen}, Jin and {Dong}, Kai and {Hu}, Kai and {You}, Kaichao and {Gao}, Kaige and {Guan}, Kang and {Huang}, Kexin and {Yu}, Kuai and {Wang}, Lean and {Zhang}, Lecong and {Zhao}, Liang and {Wang}, Litong and {Zhang}, Liyue and {Xu}, Lei and {Xia}, Leyi and {Zhang}, Mingchuan and {Zhang}, Minghua and {Tang}, Minghui and {Zhou}, Mingxu and {Li}, Meng and {Wang}, Miaojun and {Li}, Mingming and {Tian}, Ning and {Huang}, Panpan and {Zhang}, Peng and {Wang}, Qiancheng and {Chen}, Qinyu and {Du}, Qiushi and {Ge}, Ruiqi and {Zhang}, Ruisong and {Pan}, Ruizhe and {Wang}, Runji and {Chen}, R.~J. and {Jin}, R.~L. and {Chen}, Ruyi and {Lu}, Shanghao and {Zhou}, Shangyan and {Chen}, Shanhuang and {Ye}, Shengfeng and {Wang}, Shiyu and {Yu}, Shuiping and {Zhou}, Shunfeng and {Pan}, Shuting and {Li}, S.~S. and {Zhou}, Shuang and {Wu}, Shaoqing and {Yun}, Tao and {Pei}, Tian and {Sun}, Tianyu and {Wang}, T. and {Zeng}, Wangding and {Liu}, Wen and {Liang}, Wenfeng and {Gao}, Wenjun and {Yu}, Wenqin and {Zhang}, Wentao and {Xiao}, W.~L. and {An}, Wei and {Liu}, Xiaodong and {Wang}, Xiaohan and {Chen}, Xiaokang and {Nie}, Xiaotao and {Cheng}, Xin and {Liu}, Xin and {Xie}, Xin and {Liu}, Xingchao and {Yang}, Xinyu and {Li}, Xinyuan and {Su}, Xuecheng and {Lin}, Xuheng and {Li}, X.~Q. and {Jin}, Xiangyue and {Shen}, Xiaojin and {Chen}, Xiaosha and {Sun}, Xiaowen and {Wang}, Xiaoxiang and {Song}, Xinnan and {Zhou}, Xinyi and {Wang}, Xianzu and {Shan}, Xinxia and {Li}, Y.~K. and {Wang}, Y.~Q. and {Wei}, Y.~X. and {Zhang}, Yang and {Xu}, Yanhong and {Li}, Yao and {Zhao}, Yao and {Sun}, Yaofeng and {Wang}, Yaohui and {Yu}, Yi and {Zhang}, Yichao and {Shi}, Yifan and {Xiong}, Yiliang and {He}, Ying and {Piao}, Yishi and {Wang}, Yisong and {Tan}, Yixuan and {Ma}, Yiyang and {Liu}, Yiyuan and {Guo}, Yongqiang and {Ou}, Yuan and {Wang}, Yuduan and {Gong}, Yue and {Zou}, Yuheng and {He}, Yujia and {Xiong}, Yunfan and {Luo}, Yuxiang and {You}, Yuxiang and {Liu}, Yuxuan and {Zhou}, Yuyang and {Zhu}, Y.~X. and {Huang}, Yanping and {Li}, Yaohui and {Zheng}, Yi and {Zhu}, Yuchen and {Ma}, Yunxian and {Tang}, Ying and {Zha}, Yukun and {Yan}, Yuting and {Ren}, Z.~Z. and {Ren}, Zehui and {Sha}, Zhangli and {Fu}, Zhe and {Xu}, Zhean and {Xie}, Zhenda and {Zhang}, Zhengyan and {Hao}, Zhewen and {Ma}, Zhicheng and {Yan}, Zhigang and {Wu}, Zhiyu and {Gu}, Zihui and {Zhu}, Zijia and {Liu}, Zijun and {Li}, Zilin and {Xie}, Ziwei and {Song}, Ziyang and {Pan}, Zizheng and {Huang}, Zhen and {Xu}, Zhipeng and {Zhang}, Zhongyu and {Zhang}, Zhen},
  journal       = {\nat},
  title         = {{DeepSeek-R1 incentivizes reasoning in LLMs through reinforcement learning}},
  year          = {2025},
  month         = sep,
  number        = {8081},
  pages         = {633-638},
  volume        = {645},
  adsurl        = {https://ui.adsabs.harvard.edu/abs/2025Natur.645..633G},
  archiveprefix = {arXiv},
  doi           = {10.1038/s41586-025-09422-z},
  eprint        = {2501.12948},
  primaryclass  = {cs.CL},
}

@Software{Dominik2021,
  adsurl        = {https://ui.adsabs.harvard.edu/abs/2021ascl.soft04010D},
  archiveprefix = {ascl},
  author        = {{Dominik}, Carsten and {Min}, Michiel and {Tazaki}, Ryo},
  eid           = {ascl:2104.010},
  eprint        = {2104.010},
  howpublished  = {Astrophysics Source Code Library, record ascl:2104.010},
  month         = apr,
  title         = {{OpTool: Command-line driven tool for creating complex dust opacities}},
  year          = {2021},
}

@Article{Dorschner1995,
  author   = {{Dorschner}, J. and {Begemann}, B. and {Henning}, T. and {Jaeger}, C. and {Mutschke}, H.},
  journal  = {\aap},
  title    = {{Steps toward interstellar silicate mineralogy. II. Study of Mg-Fe-silicate glasses of variable composition.}},
  year     = {1995},
  month    = aug,
  pages    = {503},
  volume   = {300},
  adsurl   = {https://ui.adsabs.harvard.edu/abs/1995A&A...300..503D},
}

@Article{Dullemond2010,
  author        = {{Dullemond}, C.~P. and {Monnier}, J.~D.},
  journal       = {\araa},
  title         = {{The Inner Regions of Protoplanetary Disks}},
  year          = {2010},
  month         = sep,
  pages         = {205-239},
  volume        = {48},
  adsurl        = {https://ui.adsabs.harvard.edu/abs/2010ARA&A..48..205D},
  archiveprefix = {arXiv},
  doi           = {10.1146/annurev-astro-081309-130932},
  eprint        = {1006.3485},
  primaryclass  = {astro-ph.SR},
}

@Article{Eggleton1983,
  author   = {{Eggleton}, P.~P.},
  journal  = {\apj},
  title    = {{Aproximations to the radii of Roche lobes.}},
  year     = {1983},
  month    = may,
  pages    = {368-369},
  volume   = {268},
  adsurl   = {https://ui.adsabs.harvard.edu/abs/1983ApJ...268..368E},
  doi      = {10.1086/160960},
}

@Article{Fairlamb2015,
  author        = {{Fairlamb}, J.~R. and {Oudmaijer}, R.~D. and {Mendigut{\'\i}a}, I. and {Ilee}, J.~D. and {van den Ancker}, M.~E.},
  journal       = {\mnras},
  title         = {{A spectroscopic survey of Herbig Ae/Be stars with X-shooter - I. Stellar parameters and accretion rates}},
  year          = {2015},
  month         = oct,
  number        = {1},
  pages         = {976-1001},
  volume        = {453},
  adsurl        = {https://ui.adsabs.harvard.edu/abs/2015MNRAS.453..976F},
  archiveprefix = {arXiv},
  doi           = {10.1093/mnras/stv1576},
  eprint        = {1507.05967},
  primaryclass  = {astro-ph.SR},
}

@Article{Fedele2017,
  author        = {{Fedele}, D. and {Carney}, M. and {Hogerheijde}, M.~R. and {Walsh}, C. and {Miotello}, A. and {Klaassen}, P. and {Bruderer}, S. and {Henning}, Th. and {van Dishoeck}, E.~F.},
  journal       = {\aap},
  title         = {{ALMA unveils rings and gaps in the protoplanetary system <ASTROBJ>HD 169142</ASTROBJ>: signatures of two giant protoplanets}},
  year          = {2017},
  month         = apr,
  pages         = {A72},
  volume        = {600},
  adsurl        = {https://ui.adsabs.harvard.edu/abs/2017A&A...600A..72F},
  archiveprefix = {arXiv},
  doi           = {10.1051/0004-6361/201629860},
  eid           = {A72},
  eprint        = {1702.02844},
  primaryclass  = {astro-ph.SR},
}

@Article{Fitzpatrick2009,
  author        = {{Fitzpatrick}, E.~L. and {Massa}, D.},
  journal       = {\apj},
  title         = {{An Analysis of the Shapes of Interstellar Extinction Curves. VI. The Near-IR Extinction Law}},
  year          = {2009},
  month         = jul,
  number        = {2},
  pages         = {1209-1222},
  volume        = {699},
  adsurl        = {https://ui.adsabs.harvard.edu/abs/2009ApJ...699.1209F},
  archiveprefix = {arXiv},
  doi           = {10.1088/0004-637X/699/2/1209},
  eprint        = {0905.0133},
  primaryclass  = {astro-ph.GA},
}

@Article{Flock2025,
  author        = {{Flock}, Mario and {Chrenko}, Ond{\v{r}}ej and {Ueda}, Takahiro and {Benisty}, Myriam and {Varga}, Jozsef and {van Boekel}, Roy},
  journal       = {\aap},
  title         = {{Effect of multi-dust species on the inner rim of magnetized protoplanetary disks}},
  year          = {2025},
  month         = sep,
  pages         = {A259},
  volume        = {701},
  adsurl        = {https://ui.adsabs.harvard.edu/abs/2025A&A...701A.259F},
  archiveprefix = {arXiv},
  doi           = {10.1051/0004-6361/202453124},
  eid           = {A259},
  eprint        = {2508.04254},
  primaryclass  = {astro-ph.EP},
}

@Article{ForemanMackey2016,
  author  = {{Foreman-Mackey}, Daniel},
  journal = {JOSS},
  title   = {{corner.py: Scatterplot matrices in Python}},
  year    = {2016},
  month   = jun,
  pages   = {24},
  volume  = {1},
  adsurl  = {https://ui.adsabs.harvard.edu/abs/2016JOSS....1...24F},
  doi     = {10.21105/joss.00024},
}

@Article{Fukagawa2006,
  author   = {{Fukagawa}, Misato and {Tamura}, Motohide and {Itoh}, Yoichi and {Kudo}, Tomoyuki and {Imaeda}, Yusuke and {Oasa}, Yumiko and {Hayashi}, Saeko S. and {Hayashi}, Masahiko},
  journal  = {\apjl},
  title    = {{Near-Infrared Images of Protoplanetary Disk Surrounding HD 142527}},
  year     = {2006},
  month    = jan,
  number   = {2},
  pages    = {L153-L156},
  volume   = {636},
  adsurl   = {https://ui.adsabs.harvard.edu/abs/2006ApJ...636L.153F},
  doi      = {10.1086/500128},
}

@Article{GaiaCollaboration2018,
  author        = {{Gaia Collaboration} and {Brown}, A.~G.~A. and {Vallenari}, A. and {Prusti}, T. and {de Bruijne}, J.~H.~J. and {Babusiaux}, C. and {Bailer-Jones}, C.~A.~L. and {Biermann}, M. and {Evans}, D.~W. and {Eyer}, L. and {Jansen}, F. and {Jordi}, C. and {Klioner}, S.~A. and {Lammers}, U. and {Lindegren}, L. and {Luri}, X. and {Mignard}, F. and {Panem}, C. and {Pourbaix}, D. and {Randich}, S. and {Sartoretti}, P. and {Siddiqui}, H.~I. and {Soubiran}, C. and {van Leeuwen}, F. and {Walton}, N.~A. and {Arenou}, F. and {Bastian}, U. and {Cropper}, M. and {Drimmel}, R. and {Katz}, D. and {Lattanzi}, M.~G. and {Bakker}, J. and {Cacciari}, C. and {Casta{\~n}eda}, J. and {Chaoul}, L. and {Cheek}, N. and {De Angeli}, F. and {Fabricius}, C. and {Guerra}, R. and {Holl}, B. and {Masana}, E. and {Messineo}, R. and {Mowlavi}, N. and {Nienartowicz}, K. and {Panuzzo}, P. and {Portell}, J. and {Riello}, M. and {Seabroke}, G.~M. and {Tanga}, P. and {Th{\'e}venin}, F. and {Gracia-Abril}, G. and {Comoretto}, G. and {Garcia-Reinaldos}, M. and {Teyssier}, D. and {Altmann}, M. and {Andrae}, R. and {Audard}, M. and {Bellas-Velidis}, I. and {Benson}, K. and {Berthier}, J. and {Blomme}, R. and {Burgess}, P. and {Busso}, G. and {Carry}, B. and {Cellino}, A. and {Clementini}, G. and {Clotet}, M. and {Creevey}, O. and {Davidson}, M. and {De Ridder}, J. and {Delchambre}, L. and {Dell'Oro}, A. and {Ducourant}, C. and {Fern{\'a}ndez-Hern{\'a}ndez}, J. and {Fouesneau}, M. and {Fr{\'e}mat}, Y. and {Galluccio}, L. and {Garc{\'\i}a-Torres}, M. and {Gonz{\'a}lez-N{\'u}{\~n}ez}, J. and {Gonz{\'a}lez-Vidal}, J.~J. and {Gosset}, E. and {Guy}, L.~P. and {Halbwachs}, J.-L. and {Hambly}, N.~C. and {Harrison}, D.~L. and {Hern{\'a}ndez}, J. and {Hestroffer}, D. and {Hodgkin}, S.~T. and {Hutton}, A. and {Jasniewicz}, G. and {Jean-Antoine-Piccolo}, A. and {Jordan}, S. and {Korn}, A.~J. and {Krone-Martins}, A. and {Lanzafame}, A.~C. and {Lebzelter}, T. and {L{\"o}ffler}, W. and {Manteiga}, M. and {Marrese}, P.~M. and {Mart{\'\i}n-Fleitas}, J.~M. and {Moitinho}, A. and {Mora}, A. and {Muinonen}, K. and {Osinde}, J. and {Pancino}, E. and {Pauwels}, T. and {Petit}, J.-M. and {Recio-Blanco}, A. and {Richards}, P.~J. and {Rimoldini}, L. and {Robin}, A.~C. and {Sarro}, L.~M. and {Siopis}, C. and {Smith}, M. and {Sozzetti}, A. and {S{\"u}veges}, M. and {Torra}, J. and {van Reeven}, W. and {Abbas}, U. and {Abreu Aramburu}, A. and {Accart}, S. and {Aerts}, C. and {Altavilla}, G. and {{\'A}lvarez}, M.~A. and {Alvarez}, R. and {Alves}, J. and {Anderson}, R.~I. and {Andrei}, A.~H. and {Anglada Varela}, E. and {Antiche}, E. and {Antoja}, T. and {Arcay}, B. and {Astraatmadja}, T.~L. and {Bach}, N. and {Baker}, S.~G. and {Balaguer-N{\'u}{\~n}ez}, L. and {Balm}, P. and {Barache}, C. and {Barata}, C. and {Barbato}, D. and {Barblan}, F. and {Barklem}, P.~S. and {Barrado}, D. and {Barros}, M. and {Barstow}, M.~A. and {Bartholom{\'e} Mu{\~n}oz}, S. and {Bassilana}, J.-L. and {Becciani}, U. and {Bellazzini}, M. and {Berihuete}, A. and {Bertone}, S. and {Bianchi}, L. and {Bienaym{\'e}}, O. and {Blanco-Cuaresma}, S. and {Boch}, T. and {Boeche}, C. and {Bombrun}, A. and {Borrachero}, R. and {Bossini}, D. and {Bouquillon}, S. and {Bourda}, G. and {Bragaglia}, A. and {Bramante}, L. and {Breddels}, M.~A. and {Bressan}, A. and {Brouillet}, N. and {Br{\"u}semeister}, T. and {Brugaletta}, E. and {Bucciarelli}, B. and {Burlacu}, A. and {Busonero}, D. and {Butkevich}, A.~G. and {Buzzi}, R. and {Caffau}, E. and {Cancelliere}, R. and {Cannizzaro}, G. and {Cantat-Gaudin}, T. and {Carballo}, R. and {Carlucci}, T. and {Carrasco}, J.~M. and {Casamiquela}, L. and {Castellani}, M. and {Castro-Ginard}, A. and {Charlot}, P. and {Chemin}, L. and {Chiavassa}, A. and {Cocozza}, G. and {Costigan}, G. and {Cowell}, S. and {Crifo}, F. and {Crosta}, M. and {Crowley}, C. and {Cuypers}, J. and {Dafonte}, C. and {Damerdji}, Y. and {Dapergolas}, A. and {David}, P. and {David}, M. and {de Laverny}, P. and {De Luise}, F.},
  journal       = {\aap},
  title         = {{Gaia Data Release 2. Summary of the contents and survey properties}},
  year          = {2018},
  month         = aug,
  pages         = {A1},
  volume        = {616},
  adsurl        = {https://ui.adsabs.harvard.edu/abs/2018A&A...616A...1G},
  archiveprefix = {arXiv},
  doi           = {10.1051/0004-6361/201833051},
  eid           = {A1},
  eprint        = {1804.09365},
  primaryclass  = {astro-ph.GA},
}

@Article{GaiaCollaboration2023,
  author        = {{Gaia Collaboration} and {Vallenari}, A. and {Brown}, A.~G.~A. and {Prusti}, T. and {de Bruijne}, J.~H.~J. and {Arenou}, F. and {Babusiaux}, C. and {Biermann}, M. and {Creevey}, O.~L. and {Ducourant}, C. and {Evans}, D.~W. and {Eyer}, L. and {Guerra}, R. and {Hutton}, A. and {Jordi}, C. and {Klioner}, S.~A. and {Lammers}, U.~L. and {Lindegren}, L. and {Luri}, X. and {Mignard}, F. and {Panem}, C. and {Pourbaix}, D. and {Randich}, S. and {Sartoretti}, P. and {Soubiran}, C. and {Tanga}, P. and {Walton}, N.~A. and {Bailer-Jones}, C.~A.~L. and {Bastian}, U. and {Drimmel}, R. and {Jansen}, F. and {Katz}, D. and {Lattanzi}, M.~G. and {van Leeuwen}, F. and {Bakker}, J. and {Cacciari}, C. and {Casta{\~n}eda}, J. and {De Angeli}, F. and {Fabricius}, C. and {Fouesneau}, M. and {Fr{\'e}mat}, Y. and {Galluccio}, L. and {Guerrier}, A. and {Heiter}, U. and {Masana}, E. and {Messineo}, R. and {Mowlavi}, N. and {Nicolas}, C. and {Nienartowicz}, K. and {Pailler}, F. and {Panuzzo}, P. and {Riclet}, F. and {Roux}, W. and {Seabroke}, G.~M. and {Sordo}, R. and {Th{\'e}venin}, F. and {Gracia-Abril}, G. and {Portell}, J. and {Teyssier}, D. and {Altmann}, M. and {Andrae}, R. and {Audard}, M. and {Bellas-Velidis}, I. and {Benson}, K. and {Berthier}, J. and {Blomme}, R. and {Burgess}, P.~W. and {Busonero}, D. and {Busso}, G. and {C{\'a}novas}, H. and {Carry}, B. and {Cellino}, A. and {Cheek}, N. and {Clementini}, G. and {Damerdji}, Y. and {Davidson}, M. and {de Teodoro}, P. and {Nu{\~n}ez Campos}, M. and {Delchambre}, L. and {Dell'Oro}, A. and {Esquej}, P. and {Fern{\'a}ndez-Hern{\'a}ndez}, J. and {Fraile}, E. and {Garabato}, D. and {Garc{\'\i}a-Lario}, P. and {Gosset}, E. and {Haigron}, R. and {Halbwachs}, J.-L. and {Hambly}, N.~C. and {Harrison}, D.~L. and {Hern{\'a}ndez}, J. and {Hestroffer}, D. and {Hodgkin}, S.~T. and {Holl}, B. and {Jan{\ss}en}, K. and {Jevardat de Fombelle}, G. and {Jordan}, S. and {Krone-Martins}, A. and {Lanzafame}, A.~C. and {L{\"o}ffler}, W. and {Marchal}, O. and {Marrese}, P.~M. and {Moitinho}, A. and {Muinonen}, K. and {Osborne}, P. and {Pancino}, E. and {Pauwels}, T. and {Recio-Blanco}, A. and {Reyl{\'e}}, C. and {Riello}, M. and {Rimoldini}, L. and {Roegiers}, T. and {Rybizki}, J. and {Sarro}, L.~M. and {Siopis}, C. and {Smith}, M. and {Sozzetti}, A. and {Utrilla}, E. and {van Leeuwen}, M. and {Abbas}, U. and {{\'A}brah{\'a}m}, P. and {Abreu Aramburu}, A. and {Aerts}, C. and {Aguado}, J.~J. and {Ajaj}, M. and {Aldea-Montero}, F. and {Altavilla}, G. and {{\'A}lvarez}, M.~A. and {Alves}, J. and {Anders}, F. and {Anderson}, R.~I. and {Anglada Varela}, E. and {Antoja}, T. and {Baines}, D. and {Baker}, S.~G. and {Balaguer-N{\'u}{\~n}ez}, L. and {Balbinot}, E. and {Balog}, Z. and {Barache}, C. and {Barbato}, D. and {Barros}, M. and {Barstow}, M.~A. and {Bartolom{\'e}}, S. and {Bassilana}, J.-L. and {Bauchet}, N. and {Becciani}, U. and {Bellazzini}, M. and {Berihuete}, A. and {Bernet}, M. and {Bertone}, S. and {Bianchi}, L. and {Binnenfeld}, A. and {Blanco-Cuaresma}, S. and {Blazere}, A. and {Boch}, T. and {Bombrun}, A. and {Bossini}, D. and {Bouquillon}, S. and {Bragaglia}, A. and {Bramante}, L. and {Breedt}, E. and {Bressan}, A. and {Brouillet}, N. and {Brugaletta}, E. and {Bucciarelli}, B. and {Burlacu}, A. and {Butkevich}, A.~G. and {Buzzi}, R. and {Caffau}, E. and {Cancelliere}, R. and {Cantat-Gaudin}, T. and {Carballo}, R. and {Carlucci}, T. and {Carnerero}, M.~I. and {Carrasco}, J.~M. and {Casamiquela}, L. and {Castellani}, M. and {Castro-Ginard}, A. and {Chaoul}, L. and {Charlot}, P. and {Chemin}, L. and {Chiaramida}, V. and {Chiavassa}, A. and {Chornay}, N. and {Comoretto}, G. and {Contursi}, G. and {Cooper}, W.~J. and {Cornez}, T. and {Cowell}, S. and {Crifo}, F. and {Cropper}, M. and {Crosta}, M. and {Crowley}, C. and {Dafonte}, C. and {Dapergolas}, A. and {David}, M. and {David}, P. and {de Laverny}, P. and {De Luise}, F. and {De March}, R.},
  journal       = {\aap},
  title         = {{Gaia Data Release 3. Summary of the content and survey properties}},
  year          = {2023},
  month         = jun,
  pages         = {A1},
  volume        = {674},
  adsurl        = {https://ui.adsabs.harvard.edu/abs/2023A&A...674A...1G},
  archiveprefix = {arXiv},
  doi           = {10.1051/0004-6361/202243940},
  eid           = {A1},
  eprint        = {2208.00211},
  primaryclass  = {astro-ph.GA},
}

@Article{GaiaCollaboration2021,
  author        = {{Gaia Collaboration} and {Brown}, A.~G.~A. and {Vallenari}, A. and {Prusti}, T. and {de Bruijne}, J.~H.~J. and {Babusiaux}, C. and {Biermann}, M. and {Creevey}, O.~L. and {Evans}, D.~W. and {Eyer}, L. and {Hutton}, A. and {Jansen}, F. and {Jordi}, C. and {Klioner}, S.~A. and {Lammers}, U. and {Lindegren}, L. and {Luri}, X. and {Mignard}, F. and {Panem}, C. and {Pourbaix}, D. and {Randich}, S. and {Sartoretti}, P. and {Soubiran}, C. and {Walton}, N.~A. and {Arenou}, F. and {Bailer-Jones}, C.~A.~L. and {Bastian}, U. and {Cropper}, M. and {Drimmel}, R. and {Katz}, D. and {Lattanzi}, M.~G. and {van Leeuwen}, F. and {Bakker}, J. and {Cacciari}, C. and {Casta{\~n}eda}, J. and {De Angeli}, F. and {Ducourant}, C. and {Fabricius}, C. and {Fouesneau}, M. and {Fr{\'e}mat}, Y. and {Guerra}, R. and {Guerrier}, A. and {Guiraud}, J. and {Jean-Antoine Piccolo}, A. and {Masana}, E. and {Messineo}, R. and {Mowlavi}, N. and {Nicolas}, C. and {Nienartowicz}, K. and {Pailler}, F. and {Panuzzo}, P. and {Riclet}, F. and {Roux}, W. and {Seabroke}, G.~M. and {Sordo}, R. and {Tanga}, P. and {Th{\'e}venin}, F. and {Gracia-Abril}, G. and {Portell}, J. and {Teyssier}, D. and {Altmann}, M. and {Andrae}, R. and {Bellas-Velidis}, I. and {Benson}, K. and {Berthier}, J. and {Blomme}, R. and {Brugaletta}, E. and {Burgess}, P.~W. and {Busso}, G. and {Carry}, B. and {Cellino}, A. and {Cheek}, N. and {Clementini}, G. and {Damerdji}, Y. and {Davidson}, M. and {Delchambre}, L. and {Dell'Oro}, A. and {Fern{\'a}ndez-Hern{\'a}ndez}, J. and {Galluccio}, L. and {Garc{\'\i}a-Lario}, P. and {Garcia-Reinaldos}, M. and {Gonz{\'a}lez-N{\'u}{\~n}ez}, J. and {Gosset}, E. and {Haigron}, R. and {Halbwachs}, J.-L. and {Hambly}, N.~C. and {Harrison}, D.~L. and {Hatzidimitriou}, D. and {Heiter}, U. and {Hern{\'a}ndez}, J. and {Hestroffer}, D. and {Hodgkin}, S.~T. and {Holl}, B. and {Jan{\ss}en}, K. and {Jevardat de Fombelle}, G. and {Jordan}, S. and {Krone-Martins}, A. and {Lanzafame}, A.~C. and {L{\"o}ffler}, W. and {Lorca}, A. and {Manteiga}, M. and {Marchal}, O. and {Marrese}, P.~M. and {Moitinho}, A. and {Mora}, A. and {Muinonen}, K. and {Osborne}, P. and {Pancino}, E. and {Pauwels}, T. and {Petit}, J.-M. and {Recio-Blanco}, A. and {Richards}, P.~J. and {Riello}, M. and {Rimoldini}, L. and {Robin}, A.~C. and {Roegiers}, T. and {Rybizki}, J. and {Sarro}, L.~M. and {Siopis}, C. and {Smith}, M. and {Sozzetti}, A. and {Ulla}, A. and {Utrilla}, E. and {van Leeuwen}, M. and {van Reeven}, W. and {Abbas}, U. and {Abreu Aramburu}, A. and {Accart}, S. and {Aerts}, C. and {Aguado}, J.~J. and {Ajaj}, M. and {Altavilla}, G. and {{\'A}lvarez}, M.~A. and {{\'A}lvarez Cid-Fuentes}, J. and {Alves}, J. and {Anderson}, R.~I. and {Anglada Varela}, E. and {Antoja}, T. and {Audard}, M. and {Baines}, D. and {Baker}, S.~G. and {Balaguer-N{\'u}{\~n}ez}, L. and {Balbinot}, E. and {Balog}, Z. and {Barache}, C. and {Barbato}, D. and {Barros}, M. and {Barstow}, M.~A. and {Bartolom{\'e}}, S. and {Bassilana}, J.-L. and {Bauchet}, N. and {Baudesson-Stella}, A. and {Becciani}, U. and {Bellazzini}, M. and {Bernet}, M. and {Bertone}, S. and {Bianchi}, L. and {Blanco-Cuaresma}, S. and {Boch}, T. and {Bombrun}, A. and {Bossini}, D. and {Bouquillon}, S. and {Bragaglia}, A. and {Bramante}, L. and {Breedt}, E. and {Bressan}, A. and {Brouillet}, N. and {Bucciarelli}, B. and {Burlacu}, A. and {Busonero}, D. and {Butkevich}, A.~G. and {Buzzi}, R. and {Caffau}, E. and {Cancelliere}, R. and {C{\'a}novas}, H. and {Cantat-Gaudin}, T. and {Carballo}, R. and {Carlucci}, T. and {Carnerero}, M.~I. and {Carrasco}, J.~M. and {Casamiquela}, L. and {Castellani}, M. and {Castro-Ginard}, A. and {Castro Sampol}, P. and {Chaoul}, L. and {Charlot}, P. and {Chemin}, L. and {Chiavassa}, A. and {Cioni}, M.-R.~L. and {Comoretto}, G. and {Cooper}, W.~J. and {Cornez}, T. and {Cowell}, S. and {Crifo}, F. and {Crosta}, M. and {Crowley}, C. and {Dafonte}, C. and {Dapergolas}, A. and {David}, M. and {David}, P.},
  journal       = {\aap},
  title         = {{Gaia Early Data Release 3. Summary of the contents and survey properties}},
  year          = {2021},
  month         = may,
  pages         = {A1},
  volume        = {649},
  adsurl        = {https://ui.adsabs.harvard.edu/abs/2021A&A...649A...1G},
  archiveprefix = {arXiv},
  doi           = {10.1051/0004-6361/202039657},
  eid           = {A1},
  eprint        = {2012.01533},
  primaryclass  = {astro-ph.GA},
}

@Article{GaiaCollaboration2016a,
  author        = {{Gaia Collaboration} and {Prusti}, T. and {de Bruijne}, J.~H.~J. and {Brown}, A.~G.~A. and {Vallenari}, A. and {Babusiaux}, C. and {Bailer-Jones}, C.~A.~L. and {Bastian}, U. and {Biermann}, M. and {Evans}, D.~W. and {Eyer}, L. and {Jansen}, F. and {Jordi}, C. and {Klioner}, S.~A. and {Lammers}, U. and {Lindegren}, L. and {Luri}, X. and {Mignard}, F. and {Milligan}, D.~J. and {Panem}, C. and {Poinsignon}, V. and {Pourbaix}, D. and {Randich}, S. and {Sarri}, G. and {Sartoretti}, P. and {Siddiqui}, H.~I. and {Soubiran}, C. and {Valette}, V. and {van Leeuwen}, F. and {Walton}, N.~A. and {Aerts}, C. and {Arenou}, F. and {Cropper}, M. and {Drimmel}, R. and {H{\o}g}, E. and {Katz}, D. and {Lattanzi}, M.~G. and {O'Mullane}, W. and {Grebel}, E.~K. and {Holland}, A.~D. and {Huc}, C. and {Passot}, X. and {Bramante}, L. and {Cacciari}, C. and {Casta{\~n}eda}, J. and {Chaoul}, L. and {Cheek}, N. and {De Angeli}, F. and {Fabricius}, C. and {Guerra}, R. and {Hern{\'a}ndez}, J. and {Jean-Antoine-Piccolo}, A. and {Masana}, E. and {Messineo}, R. and {Mowlavi}, N. and {Nienartowicz}, K. and {Ord{\'o}{\~n}ez-Blanco}, D. and {Panuzzo}, P. and {Portell}, J. and {Richards}, P.~J. and {Riello}, M. and {Seabroke}, G.~M. and {Tanga}, P. and {Th{\'e}venin}, F. and {Torra}, J. and {Els}, S.~G. and {Gracia-Abril}, G. and {Comoretto}, G. and {Garcia-Reinaldos}, M. and {Lock}, T. and {Mercier}, E. and {Altmann}, M. and {Andrae}, R. and {Astraatmadja}, T.~L. and {Bellas-Velidis}, I. and {Benson}, K. and {Berthier}, J. and {Blomme}, R. and {Busso}, G. and {Carry}, B. and {Cellino}, A. and {Clementini}, G. and {Cowell}, S. and {Creevey}, O. and {Cuypers}, J. and {Davidson}, M. and {De Ridder}, J. and {de Torres}, A. and {Delchambre}, L. and {Dell'Oro}, A. and {Ducourant}, C. and {Fr{\'e}mat}, Y. and {Garc{\'\i}a-Torres}, M. and {Gosset}, E. and {Halbwachs}, J.-L. and {Hambly}, N.~C. and {Harrison}, D.~L. and {Hauser}, M. and {Hestroffer}, D. and {Hodgkin}, S.~T. and {Huckle}, H.~E. and {Hutton}, A. and {Jasniewicz}, G. and {Jordan}, S. and {Kontizas}, M. and {Korn}, A.~J. and {Lanzafame}, A.~C. and {Manteiga}, M. and {Moitinho}, A. and {Muinonen}, K. and {Osinde}, J. and {Pancino}, E. and {Pauwels}, T. and {Petit}, J.-M. and {Recio-Blanco}, A. and {Robin}, A.~C. and {Sarro}, L.~M. and {Siopis}, C. and {Smith}, M. and {Smith}, K.~W. and {Sozzetti}, A. and {Thuillot}, W. and {van Reeven}, W. and {Viala}, Y. and {Abbas}, U. and {Abreu Aramburu}, A. and {Accart}, S. and {Aguado}, J.~J. and {Allan}, P.~M. and {Allasia}, W. and {Altavilla}, G. and {{\'A}lvarez}, M.~A. and {Alves}, J. and {Anderson}, R.~I. and {Andrei}, A.~H. and {Anglada Varela}, E. and {Antiche}, E. and {Antoja}, T. and {Ant{\'o}n}, S. and {Arcay}, B. and {Atzei}, A. and {Ayache}, L. and {Bach}, N. and {Baker}, S.~G. and {Balaguer-N{\'u}{\~n}ez}, L. and {Barache}, C. and {Barata}, C. and {Barbier}, A. and {Barblan}, F. and {Baroni}, M. and {Barrado y Navascu{\'e}s}, D. and {Barros}, M. and {Barstow}, M.~A. and {Becciani}, U. and {Bellazzini}, M. and {Bellei}, G. and {Bello Garc{\'\i}a}, A. and {Belokurov}, V. and {Bendjoya}, P. and {Berihuete}, A. and {Bianchi}, L. and {Bienaym{\'e}}, O. and {Billebaud}, F. and {Blagorodnova}, N. and {Blanco-Cuaresma}, S. and {Boch}, T. and {Bombrun}, A. and {Borrachero}, R. and {Bouquillon}, S. and {Bourda}, G. and {Bouy}, H. and {Bragaglia}, A. and {Breddels}, M.~A. and {Brouillet}, N. and {Br{\"u}semeister}, T. and {Bucciarelli}, B. and {Budnik}, F. and {Burgess}, P. and {Burgon}, R. and {Burlacu}, A. and {Busonero}, D. and {Buzzi}, R. and {Caffau}, E. and {Cambras}, J. and {Campbell}, H. and {Cancelliere}, R. and {Cantat-Gaudin}, T. and {Carlucci}, T. and {Carrasco}, J.~M. and {Castellani}, M. and {Charlot}, P. and {Charnas}, J. and {Charvet}, P. and {Chassat}, F. and {Chiavassa}, A. and {Clotet}, M. and {Cocozza}, G. and {Collins}, R.~S. and {Collins}, P. and {Costigan}, G.},
  journal       = {\aap},
  title         = {{The Gaia mission}},
  year          = {2016},
  month         = nov,
  pages         = {A1},
  volume        = {595},
  adsurl        = {https://ui.adsabs.harvard.edu/abs/2016A&A...595A...1G},
  archiveprefix = {arXiv},
  doi           = {10.1051/0004-6361/201629272},
  eid           = {A1},
  eprint        = {1609.04153},
  primaryclass  = {astro-ph.IM},
}

@Article{Gail2004,
  author   = {{Gail, H.-P.}},
  journal  = {\aap},
  title    = {Radial mixing in protoplanetary accretion disks - IV. Metamorphosis of the silicate dust complex},
  year     = {2004},
  number   = {2},
  pages    = {571-591},
  volume   = {413},
  doi      = {10.1051/0004-6361:20031554},
  fjournal = {Astronomy and Astrophysics},
}

@Software{Garrett2023,
  adsurl    = {https://ui.adsabs.harvard.edu/abs/2021zndo...4106649G},
  author    = {{Garrett}, John and {Luis}, Echedey and {Peng}, H.-H. and {Cera}, Tim and {gobinathj} and {Borrow}, Josh and {Ke{\c{c}}eci}, Mehmet and {Splines} and {Iyer}, Suraj and {Liu}, Yuming and {cjw} and {Gasanov}, Mikhail},
  doi       = {10.5281/zenodo.4106649},
  eid       = {10.5281/zenodo.4106649},
  month     = nov,
  publisher = {Zenodo},
  title     = {{garrettj403/SciencePlots: 2.1.1}},
  version   = {2.1.1},
  year      = {2023},
}

@Article{Garufi2017,
  author        = {{Garufi}, A. and {Meeus}, G. and {Benisty}, M. and {Quanz}, S.~P. and {Banzatti}, A. and {Kama}, M. and {Canovas}, H. and {Eiroa}, C. and {Schmid}, H.~M. and {Stolker}, T. and {Pohl}, A. and {Rigliaco}, E. and {M{\'e}nard}, F. and {Meyer}, M.~R. and {van Boekel}, R. and {Dominik}, C.},
  journal       = {\aap},
  title         = {{Evolution of protoplanetary disks from their taxonomy in scattered light: Group I vs. Group II}},
  year          = {2017},
  month         = jul,
  pages         = {A21},
  volume        = {603},
  adsurl        = {https://ui.adsabs.harvard.edu/abs/2017A&A...603A..21G},
  archiveprefix = {arXiv},
  doi           = {10.1051/0004-6361/201630320},
  eid           = {A21},
  eprint        = {1703.01512},
  primaryclass  = {astro-ph.EP},
}

@Article{Garufi2018,
  author        = {{Garufi}, A. and {Benisty}, M. and {Pinilla}, P. and {Tazzari}, M. and {Dominik}, C. and {Ginski}, C. and {Henning}, Th. and {Kral}, Q. and {Langlois}, M. and {M{\'e}nard}, F. and {Stolker}, T. and {Szulagyi}, J. and {Villenave}, M. and {van der Plas}, G.},
  journal       = {\aap},
  title         = {{Evolution of protoplanetary disks from their taxonomy in scattered light: spirals, rings, cavities, and shadows}},
  year          = {2018},
  month         = dec,
  pages         = {A94},
  volume        = {620},
  adsurl        = {https://ui.adsabs.harvard.edu/abs/2018A&A...620A..94G},
  archiveprefix = {arXiv},
  doi           = {10.1051/0004-6361/201833872},
  eid           = {A94},
  eprint        = {1810.04564},
  primaryclass  = {astro-ph.SR},
}

@Article{Ginsburg2019,
  author        = {{Ginsburg}, Adam and {Sip{\H{o}}cz}, Brigitta M. and {Brasseur}, C.~E. and {Cowperthwaite}, Philip S. and {Craig}, Matthew W. and {Deil}, Christoph and {Guillochon}, James and {Guzman}, Giannina and {Liedtke}, Simon and {Lian Lim}, Pey and {Lockhart}, Kelly E. and {Mommert}, Michael and {Morris}, Brett M. and {Norman}, Henrik and {Parikh}, Madhura and {Persson}, Magnus V. and {Robitaille}, Thomas P. and {Segovia}, Juan-Carlos and {Singer}, Leo P. and {Tollerud}, Erik J. and {de Val-Borro}, Miguel and {Valtchanov}, Ivan and {Woillez}, Julien and {Astroquery Collaboration} and {a subset of astropy Collaboration}},
  journal       = {\aj},
  title         = {{astroquery: An Astronomical Web-querying Package in Python}},
  year          = {2019},
  month         = mar,
  number        = {3},
  pages         = {98},
  volume        = {157},
  adsurl        = {https://ui.adsabs.harvard.edu/abs/2019AJ....157...98G},
  archiveprefix = {arXiv},
  doi           = {10.3847/1538-3881/aafc33},
  eid           = {98},
  eprint        = {1901.04520},
  primaryclass  = {astro-ph.IM},
}

@Article{GRAVITYCollaboration2017,
  author        = {{GRAVITY Collaboration} and {Abuter}, R. and {Accardo}, M. and {Amorim}, A. and {Anugu}, N. and {{\'A}vila}, G. and {Azouaoui}, N. and {Benisty}, M. and {Berger}, J.~P. and {Blind}, N. and {Bonnet}, H. and {Bourget}, P. and {Brandner}, W. and {Brast}, R. and {Buron}, A. and {Burtscher}, L. and {Cassaing}, F. and {Chapron}, F. and {Choquet}, {\'E}. and {Cl{\'e}net}, Y. and {Collin}, C. and {Coud{\'e} Du Foresto}, V. and {de Wit}, W. and {de Zeeuw}, P.~T. and {Deen}, C. and {Delplancke-Str{\"o}bele}, F. and {Dembet}, R. and {Derie}, F. and {Dexter}, J. and {Duvert}, G. and {Ebert}, M. and {Eckart}, A. and {Eisenhauer}, F. and {Esselborn}, M. and {F{\'e}dou}, P. and {Finger}, G. and {Garcia}, P. and {Garcia Dabo}, C.~E. and {Garcia Lopez}, R. and {Gendron}, E. and {Genzel}, R. and {Gillessen}, S. and {Gonte}, F. and {Gordo}, P. and {Grould}, M. and {Gr{\"o}zinger}, U. and {Guieu}, S. and {Haguenauer}, P. and {Hans}, O. and {Haubois}, X. and {Haug}, M. and {Haussmann}, F. and {Henning}, Th. and {Hippler}, S. and {Horrobin}, M. and {Huber}, A. and {Hubert}, Z. and {Hubin}, N. and {Hummel}, C.~A. and {Jakob}, G. and {Janssen}, A. and {Jochum}, L. and {Jocou}, L. and {Kaufer}, A. and {Kellner}, S. and {Kendrew}, S. and {Kern}, L. and {Kervella}, P. and {Kiekebusch}, M. and {Klein}, R. and {Kok}, Y. and {Kolb}, J. and {Kulas}, M. and {Lacour}, S. and {Lapeyr{\`e}re}, V. and {Lazareff}, B. and {Le Bouquin}, J.-B. and {L{\`e}na}, P. and {Lenzen}, R. and {L{\'e}v{\^e}que}, S. and {Lippa}, M. and {Magnard}, Y. and {Mehrgan}, L. and {Mellein}, M. and {M{\'e}rand}, A. and {Moreno-Ventas}, J. and {Moulin}, T. and {M{\"u}ller}, E. and {M{\"u}ller}, F. and {Neumann}, U. and {Oberti}, S. and {Ott}, T. and {Pallanca}, L. and {Panduro}, J. and {Pasquini}, L. and {Paumard}, T. and {Percheron}, I. and {Perraut}, K. and {Perrin}, G. and {Pfl{\"u}ger}, A. and {Pfuhl}, O. and {Phan Duc}, T. and {Plewa}, P.~M. and {Popovic}, D. and {Rabien}, S. and {Ram{\'\i}rez}, A. and {Ramos}, J. and {Rau}, C. and {Riquelme}, M. and {Rohloff}, R.-R. and {Rousset}, G. and {Sanchez-Bermudez}, J. and {Scheithauer}, S. and {Sch{\"o}ller}, M. and {Schuhler}, N. and {Spyromilio}, J. and {Straubmeier}, C. and {Sturm}, E. and {Suarez}, M. and {Tristram}, K.~R.~W. and {Ventura}, N. and {Vincent}, F. and {Waisberg}, I. and {Wank}, I. and {Weber}, J. and {Wieprecht}, E. and {Wiest}, M. and {Wiezorrek}, E. and {Wittkowski}, M. and {Woillez}, J. and {Wolff}, B. and {Yazici}, S. and {Ziegler}, D. and {Zins}, G.},
  journal       = {\aap},
  title         = {{First light for GRAVITY: Phase referencing optical interferometry for the Very Large Telescope Interferometer}},
  year          = {2017},
  month         = jun,
  pages         = {A94},
  volume        = {602},
  adsurl        = {https://ui.adsabs.harvard.edu/abs/2017A&A...602A..94G},
  archiveprefix = {arXiv},
  doi           = {10.1051/0004-6361/201730838},
  eid           = {A94},
  eprint        = {1705.02345},
  primaryclass  = {astro-ph.IM},
}

@Article{GRAVITYCollaboration2019,
  author        = {{GRAVITY Collaboration} and {Perraut}, K. and {Labadie}, L. and {Lazareff}, B. and {Klarmann}, L. and {Segura-Cox}, D. and {Benisty}, M. and {Bouvier}, J. and {Brandner}, W. and {Caratti O Garatti}, A. and {Caselli}, P. and {Dougados}, C. and {Garcia}, P. and {Garcia-Lopez}, R. and {Kendrew}, S. and {Koutoulaki}, M. and {Kervella}, P. and {Lin}, C.-C. and {Pineda}, J. and {Sanchez-Bermudez}, J. and {van Dishoeck}, E. and {Abuter}, R. and {Amorim}, A. and {Berger}, J.-P. and {Bonnet}, H. and {Buron}, A. and {Cantalloube}, F. and {Cl{\'e}net}, Y. and {Coud{\'e} Du Foresto}, V. and {Dexter}, J. and {de Zeeuw}, P.~T. and {Duvert}, G. and {Eckart}, A. and {Eisenhauer}, F. and {Eupen}, F. and {Gao}, F. and {Gendron}, E. and {Genzel}, R. and {Gillessen}, S. and {Gordo}, P. and {Grellmann}, R. and {Haubois}, X. and {Haussmann}, F. and {Henning}, T. and {Hippler}, S. and {Horrobin}, M. and {Hubert}, Z. and {Jocou}, L. and {Lacour}, S. and {Le Bouquin}, J.-B. and {L{\'e}na}, P. and {M{\'e}rand}, A. and {Ott}, T. and {Paumard}, T. and {Perrin}, G. and {Pfuhl}, O. and {Rabien}, S. and {Ray}, T. and {Rau}, C. and {Rousset}, G. and {Scheithauer}, S. and {Straub}, O. and {Straubmeier}, C. and {Sturm}, E. and {Vincent}, F. and {Waisberg}, I. and {Wank}, I. and {Widmann}, F. and {Wieprecht}, E. and {Wiest}, M. and {Wiezorrek}, E. and {Woillez}, J. and {Yazici}, S.},
  journal       = {\aap},
  title         = {{The GRAVITY Young Stellar Object survey. I. Probing the disks of Herbig Ae/Be stars in terrestrial orbits}},
  year          = {2019},
  month         = dec,
  pages         = {A53},
  volume        = {632},
  adsurl        = {https://ui.adsabs.harvard.edu/abs/2019A&A...632A..53G},
  archiveprefix = {arXiv},
  doi           = {10.1051/0004-6361/201936403},
  eid           = {A53},
  eprint        = {1911.00611},
  primaryclass  = {astro-ph.SR},
}

@Article{GRAVITYCollaboration2021,
  author        = {{GRAVITY Collaboration} and {Sanchez-Bermudez}, J. and {Caratti O Garatti}, A. and {Garcia Lopez}, R. and {Perraut}, K. and {Labadie}, L. and {Benisty}, M. and {Brandner}, W. and {Dougados}, C. and {Garcia}, P.~J.~V. and {Henning}, Th. and {Klarmann}, L. and {Amorim}, A. and {Baub{\"o}ck}, M. and {Berger}, J.~P. and {Le Bouquin}, J.~B. and {Caselli}, P. and {Cl{\'e}net}, Y. and {Coud{\'e} Du Foresto}, V. and {de Zeeuw}, P.~T. and {Drescher}, A. and {Duvert}, G. and {Eckart}, A. and {Eisenhauer}, F. and {Filho}, M. and {Gao}, F. and {Gendron}, E. and {Genzel}, R. and {Gillessen}, S. and {Grellmann}, R. and {Heissel}, G. and {Horrobin}, M. and {Hubert}, Z. and {Jim{\'e}nez-Rosales}, A. and {Jocou}, L. and {Kervella}, P. and {Lacour}, S. and {Lapeyr{\`e}re}, V. and {L{\'e}na}, P. and {Ott}, T. and {Paumard}, T. and {Perrin}, G. and {Pineda}, J.~E. and {Rodr{\'\i}guez-Coira}, G. and {Rousset}, G. and {Segura-Cox}, D.~M. and {Shangguan}, J. and {Shimizu}, T. and {Stadler}, J. and {Straub}, O. and {Straubmeier}, C. and {Sturm}, E. and {van Dishoeck}, E. and {Vincent}, F. and {von Fellenberg}, S.~D. and {Widmann}, F. and {Woillez}, J.},
  journal       = {\aap},
  title         = {{The GRAVITY young stellar object survey. VI. Mapping the variable inner disk of HD 163296 at sub-au scales}},
  year          = {2021},
  month         = oct,
  pages         = {A97},
  volume        = {654},
  adsurl        = {https://ui.adsabs.harvard.edu/abs/2021A&A...654A..97G},
  archiveprefix = {arXiv},
  doi           = {10.1051/0004-6361/202039600},
  eid           = {A97},
  eprint        = {2107.02391},
  primaryclass  = {astro-ph.SR},
}

@Article{GRAVITYCollaboration2024,
  author        = {{GRAVITY Collaboration} and {Ganci}, V. and {Labadie}, L. and {Perraut}, K. and {Wojtczak}, A. and {Kaufhold}, J. and {Benisty}, M. and {Alecian}, E. and {Bourdarot}, G. and {Brandner}, W. and {Caratti O Garatti}, A. and {Dougados}, C. and {Garcia Lopez}, R. and {Sanchez-Bermudez}, J. and {Soulain}, A. and {Amorim}, A. and {Berger}, J.-P. and {Caselli}, P. and {Cl{\'e}net}, Y. and {Drescher}, A. and {Eckart}, A. and {Eisenhauer}, F. and {Fabricius}, M. and {Feuchtgruber}, H. and {Garcia}, P. and {Gendron}, E. and {Genzel}, R. and {Gillessen}, S. and {Grant}, S. and {Hei{\ss}el}, G. and {Henning}, T. and {Horrobin}, M. and {Jocou}, L. and {Kervella}, P. and {Lacour}, S. and {Lapeyr{\`e}re}, V. and {Le Bouquin}, J.-B. and {L{\'e}na}, P. and {Lutz}, D. and {Mang}, F. and {Moruj{\~a}o}, N. and {Ott}, T. and {Paumard}, T. and {Perrin}, G. and {Ribeiro}, D. and {Sadun Bordoni}, M. and {Scheithauer}, S. and {Shangguan}, J. and {Shimizu}, T. and {Straubmeier}, C. and {Sturm}, E. and {Tacconi}, L. and {van Dishoeck}, E. and {Vincent}, F. and {Woillez}, J.},
  journal       = {\aap},
  title         = {{The GRAVITY young stellar object survey. XIII. Tracing the time-variable asymmetric disk structure in the inner AU of the Herbig star HD 98922}},
  year          = {2024},
  month         = apr,
  pages         = {A200},
  volume        = {684},
  adsurl        = {https://ui.adsabs.harvard.edu/abs/2024A&A...684A.200G},
  archiveprefix = {arXiv},
  doi           = {10.1051/0004-6361/202346926},
  eid           = {A200},
  eprint        = {2401.17764},
  primaryclass  = {astro-ph.SR},
}

@Article{Gray1969,
  author          = {P. G. Gray},
  journal         = {J. R. Stat. Soc. Series A (General)},
  title           = {Survey Sampling},
  year            = {1969},
  issn            = {0035-9238},
  number          = {2},
  pages           = {272--274},
  volume          = {132},
  publisher       = {[Royal Statistical Society, Wiley]},
  reviewed-author = {Leslie Kish},
  url             = {http://www.jstor.org/stable/2343791},
  urldate         = {2026-03-02},
}

@Article{GuzmanDiaz2021,
  author        = {{Guzm{\'a}n-D{\'\i}az}, J. and {Mendigut{\'\i}a}, I. and {Montesinos}, B. and {Oudmaijer}, R.~D. and {Vioque}, M. and {Rodrigo}, C. and {Solano}, E. and {Meeus}, G. and {Marcos-Arenal}, P.},
  journal       = {\aap},
  title         = {{Homogeneous study of Herbig Ae/Be stars from spectral energy distributions and Gaia EDR3}},
  year          = {2021},
  month         = jun,
  pages         = {A182},
  volume        = {650},
  adsurl        = {https://ui.adsabs.harvard.edu/abs/2021A&A...650A.182G},
  archiveprefix = {arXiv},
  doi           = {10.1051/0004-6361/202039519},
  eid           = {A182},
  eprint        = {2104.11759},
  primaryclass  = {astro-ph.SR},
}

@Article{Haffert2019,
  author        = {{Haffert}, S.~Y. and {Bohn}, A.~J. and {de Boer}, J. and {Snellen}, I.~A.~G. and {Brinchmann}, J. and {Girard}, J.~H. and {Keller}, C.~U. and {Bacon}, R.},
  journal       = {Nat. Astron.},
  title         = {{Two accreting protoplanets around the young star PDS 70}},
  year          = {2019},
  month         = jun,
  pages         = {749-754},
  volume        = {3},
  adsurl        = {https://ui.adsabs.harvard.edu/abs/2019NatAs...3..749H},
  archiveprefix = {arXiv},
  doi           = {10.1038/s41550-019-0780-5},
  eprint        = {1906.01486},
  primaryclass  = {astro-ph.EP},
}

@Article{Hamilton1992,
  author   = {Douglas P. Hamilton and Joseph A. Burns},
  journal  = {Icarus},
  title    = {Orbital stability zones about asteroids: II. The destabilizing effects of eccentric orbits and of solar radiation},
  year     = {1992},
  issn     = {0019-1035},
  number   = {1},
  pages    = {43-64},
  volume   = {96},
  doi      = {10.1016/0019-1035(92)90005-R},
  url      = {https://www.sciencedirect.com/science/article/pii/001910359290005R},
}

@Article{Harker2002,
  author        = {{Harker}, David E. and {Desch}, Steven J.},
  journal       = {\apjl},
  title         = {{Annealing of Silicate Dust by Nebular Shocks at 10 AU}},
  year          = {2002},
  month         = feb,
  number        = {2},
  pages         = {L109-L112},
  volume        = {565},
  adsurl        = {https://ui.adsabs.harvard.edu/abs/2002ApJ...565L.109H},
  archiveprefix = {arXiv},
  doi           = {10.1086/339363},
  eprint        = {astro-ph/0112494},
  primaryclass  = {astro-ph},
}

@Article{Harris2020,
  author    = {Harris, Charles R. and Millman, K. Jarrod and van der Walt, Stéfan J. and Gommers, Ralf and Virtanen, Pauli and Cournapeau, David and Wieser, Eric and Taylor, Julian and Berg, Sebastian and Smith, Nathaniel J. and Kern, Robert and Picus, Matti and Hoyer, Stephan and van Kerkwijk, Marten H. and Brett, Matthew and Haldane, Allan and del Río, Jaime Fernández and Wiebe, Mark and Peterson, Pearu and Gérard-Marchant, Pierre and Sheppard, Kevin and Reddy, Tyler and Weckesser, Warren and Abbasi, Hameer and Gohlke, Christoph and Oliphant, Travis E.},
  journal   = {\nat},
  title     = {Array programming with NumPy},
  year      = {2020},
  issn      = {1476-4687},
  month     = sep,
  number    = {7825},
  pages     = {357--362},
  volume    = {585},
  doi       = {10.1038/s41586-020-2649-2},
  fjournal  = {Nature},
  publisher = {Springer Science and Business Media LLC},
}

@Article{Hauschildt2025,
  author        = {{Hauschildt}, P.~H. and {Barman}, T. and {Baron}, E. and {Aufdenberg}, J.~P. and {Schweitzer}, A.},
  journal       = {\aap},
  title         = {{The NewEra model grid}},
  year          = {2025},
  month         = jun,
  pages         = {A47},
  volume        = {698},
  adsurl        = {https://ui.adsabs.harvard.edu/abs/2025A&A...698A..47H},
  archiveprefix = {arXiv},
  doi           = {10.1051/0004-6361/202554171},
  eid           = {A47},
  eprint        = {2504.17597},
  primaryclass  = {astro-ph.SR},
}

@Article{Henning1997,
  author   = {{Henning}, T. and {Mutschke}, H.},
  journal  = {\aap},
  title    = {{Low-temperature infrared properties of cosmic dust analogues.}},
  year     = {1997},
  month    = nov,
  pages    = {743-754},
  volume   = {327},
  adsurl   = {https://ui.adsabs.harvard.edu/abs/1997A&A...327..743H},
}

@Article{Hunter2007,
  author   = {{Hunter}, John D.},
  journal  = {CSE},
  title    = {{Matplotlib: A 2D Graphics Environment}},
  year     = {2007},
  month    = jan,
  number   = {3},
  pages    = {90-95},
  volume   = {9},
  adsurl   = {https://ui.adsabs.harvard.edu/abs/2007CSE.....9...90H},
  doi      = {10.1109/MCSE.2007.55},
}

@Article{Hunziker2021,
  author        = {{Hunziker}, S. and {Schmid}, H.~M. and {Ma}, J. and {Menard}, F. and {Avenhaus}, H. and {Boccaletti}, A. and {Beuzit}, J.~L. and {Chauvin}, G. and {Dohlen}, K. and {Dominik}, C. and {Engler}, N. and {Ginski}, C. and {Gratton}, R. and {Henning}, T. and {Langlois}, M. and {Milli}, J. and {Mouillet}, D. and {Tschudi}, C. and {van Holstein}, R.~G. and {Vigan}, A.},
  journal       = {\aap},
  title         = {{HD 142527: quantitative disk polarimetry with SPHERE}},
  year          = {2021},
  month         = apr,
  pages         = {A110},
  volume        = {648},
  adsurl        = {https://ui.adsabs.harvard.edu/abs/2021A&A...648A.110H},
  archiveprefix = {arXiv},
  doi           = {10.1051/0004-6361/202040166},
  eid           = {A110},
  eprint        = {2103.08462},
  primaryclass  = {astro-ph.EP},
}

@Article{Indebetouw2005,
  author        = {{Indebetouw}, R. and {Mathis}, J.~S. and {Babler}, B.~L. and {Meade}, M.~R. and {Watson}, C. and {Whitney}, B.~A. and {Wolff}, M.~J. and {Wolfire}, M.~G. and {Cohen}, M. and {Bania}, T.~M. and {Benjamin}, R.~A. and {Clemens}, D.~P. and {Dickey}, J.~M. and {Jackson}, J.~M. and {Kobulnicky}, H.~A. and {Marston}, A.~P. and {Mercer}, E.~P. and {Stauffer}, J.~R. and {Stolovy}, S.~R. and {Churchwell}, E.},
  journal       = {\apj},
  title         = {{The Wavelength Dependence of Interstellar Extinction from 1.25 to 8.0 {\ensuremath{\mu}}m Using GLIMPSE Data}},
  year          = {2005},
  month         = feb,
  number        = {2},
  pages         = {931-938},
  volume        = {619},
  adsurl        = {https://ui.adsabs.harvard.edu/abs/2005ApJ...619..931I},
  archiveprefix = {arXiv},
  doi           = {10.1086/426679},
  eprint        = {astro-ph/0406403},
  primaryclass  = {astro-ph},
}

@Article{Jaeger1998,
  author   = {{Jaeger}, C. and {Molster}, F.~J. and {Dorschner}, J. and {Henning}, Th. and {Mutschke}, H. and {Waters}, L.~B.~F.~M.},
  journal  = {\aap},
  title    = {{Steps toward interstellar silicate mineralogy. IV. The crystalline revolution}},
  year     = {1998},
  month    = nov,
  pages    = {904-916},
  volume   = {339},
  adsurl   = {https://ui.adsabs.harvard.edu/abs/1998A&A...339..904J},
}

@Article{Juhasz2010,
  author        = {{Juh{\'a}sz}, A. and {Bouwman}, J. and {Henning}, Th. and {Acke}, B. and {van den Ancker}, M.~E. and {Meeus}, G. and {Dominik}, C. and {Min}, M. and {Tielens}, A.~G.~G.~M. and {Waters}, L.~B.~F.~M.},
  journal       = {\apj},
  title         = {{Dust Evolution in Protoplanetary Disks Around Herbig Ae/Be Stars{\textemdash}the Spitzer View}},
  year          = {2010},
  month         = sep,
  number        = {1},
  pages         = {431-455},
  volume        = {721},
  adsurl        = {https://ui.adsabs.harvard.edu/abs/2010ApJ...721..431J},
  archiveprefix = {arXiv},
  doi           = {10.1088/0004-637X/721/1/431},
  eprint        = {1008.0083},
  primaryclass  = {astro-ph.SR},
}

@Article{Keppler2018,
  author        = {{Keppler}, M. and {Benisty}, M. and {M{\"u}ller}, A. and {Henning}, Th. and {van Boekel}, R. and {Cantalloube}, F. and {Ginski}, C. and {van Holstein}, R.~G. and {Maire}, A.-L. and {Pohl}, A. and {Samland}, M. and {Avenhaus}, H. and {Baudino}, J.-L. and {Boccaletti}, A. and {de Boer}, J. and {Bonnefoy}, M. and {Chauvin}, G. and {Desidera}, S. and {Langlois}, M. and {Lazzoni}, C. and {Marleau}, G.-D. and {Mordasini}, C. and {Pawellek}, N. and {Stolker}, T. and {Vigan}, A. and {Zurlo}, A. and {Birnstiel}, T. and {Brandner}, W. and {Feldt}, M. and {Flock}, M. and {Girard}, J. and {Gratton}, R. and {Hagelberg}, J. and {Isella}, A. and {Janson}, M. and {Juhasz}, A. and {Kemmer}, J. and {Kral}, Q. and {Lagrange}, A.-M. and {Launhardt}, R. and {Matter}, A. and {M{\'e}nard}, F. and {Milli}, J. and {Molli{\`e}re}, P. and {Olofsson}, J. and {P{\'e}rez}, L. and {Pinilla}, P. and {Pinte}, C. and {Quanz}, S.~P. and {Schmidt}, T. and {Udry}, S. and {Wahhaj}, Z. and {Williams}, J.~P. and {Buenzli}, E. and {Cudel}, M. and {Dominik}, C. and {Galicher}, R. and {Kasper}, M. and {Lannier}, J. and {Mesa}, D. and {Mouillet}, D. and {Peretti}, S. and {Perrot}, C. and {Salter}, G. and {Sissa}, E. and {Wildi}, F. and {Abe}, L. and {Antichi}, J. and {Augereau}, J.-C. and {Baruffolo}, A. and {Baudoz}, P. and {Bazzon}, A. and {Beuzit}, J.-L. and {Blanchard}, P. and {Brems}, S.~S. and {Buey}, T. and {De Caprio}, V. and {Carbillet}, M. and {Carle}, M. and {Cascone}, E. and {Cheetham}, A. and {Claudi}, R. and {Costille}, A. and {Delboulb{\'e}}, A. and {Dohlen}, K. and {Fantinel}, D. and {Feautrier}, P. and {Fusco}, T. and {Giro}, E. and {Gluck}, L. and {Gry}, C. and {Hubin}, N. and {Hugot}, E. and {Jaquet}, M. and {Le Mignant}, D. and {Llored}, M. and {Madec}, F. and {Magnard}, Y. and {Martinez}, P. and {Maurel}, D. and {Meyer}, M. and {M{\"o}ller-Nilsson}, O. and {Moulin}, T. and {Mugnier}, L. and {Orign{\'e}}, A. and {Pavlov}, A. and {Perret}, D. and {Petit}, C. and {Pragt}, J. and {Puget}, P. and {Rabou}, P. and {Ramos}, J. and {Rigal}, F. and {Rochat}, S. and {Roelfsema}, R. and {Rousset}, G. and {Roux}, A. and {Salasnich}, B. and {Sauvage}, J.-F. and {Sevin}, A. and {Soenke}, C. and {Stadler}, E. and {Suarez}, M. and {Turatto}, M. and {Weber}, L.},
  journal       = {\aap},
  title         = {{Discovery of a planetary-mass companion within the gap of the transition disk around PDS 70}},
  year          = {2018},
  month         = sep,
  pages         = {A44},
  volume        = {617},
  adsurl        = {https://ui.adsabs.harvard.edu/abs/2018A&A...617A..44K},
  archiveprefix = {arXiv},
  doi           = {10.1051/0004-6361/201832957},
  eid           = {A44},
  eprint        = {1806.11568},
  primaryclass  = {astro-ph.EP},
}

@Article{Klarmann2017,
  author        = {{Klarmann}, L. and {Benisty}, M. and {Min}, M. and {Dominik}, C. and {Berger}, J.-P. and {Waters}, L.~B.~F.~M. and {Kluska}, J. and {Lazareff}, B. and {Le Bouquin}, J.-B.},
  journal       = {\aap},
  title         = {{Interferometric evidence for quantum heated particles in the inner region of protoplanetary disks around Herbig stars}},
  year          = {2017},
  month         = mar,
  pages         = {A80},
  volume        = {599},
  adsurl        = {https://ui.adsabs.harvard.edu/abs/2017A&A...599A..80K},
  archiveprefix = {arXiv},
  doi           = {10.1051/0004-6361/201628820},
  eid           = {A80},
  eprint        = {1612.06311},
  primaryclass  = {astro-ph.EP},
}

@Article{Kluska2022,
  author        = {{Kluska}, J. and {Van Winckel}, H. and {Copp{\'e}e}, Q. and {Oomen}, G.-M. and {Dsilva}, K. and {Kamath}, D. and {Bujarrabal}, V. and {Min}, M.},
  journal       = {\aap},
  title         = {{A population of transition disks around evolved stars: Fingerprints of planets. Catalog of disks surrounding Galactic post-AGB binaries}},
  year          = {2022},
  month         = feb,
  pages         = {A36},
  volume        = {658},
  adsurl        = {https://ui.adsabs.harvard.edu/abs/2022A&A...658A..36K},
  archiveprefix = {arXiv},
  doi           = {10.1051/0004-6361/202141690},
  eid           = {A36},
  eprint        = {2201.13155},
  primaryclass  = {astro-ph.EP},
}

@Article{Lacour2016,
  author        = {{Lacour}, S. and {Biller}, B. and {Cheetham}, A. and {Greenbaum}, A. and {Pearce}, T. and {Marino}, S. and {Tuthill}, P. and {Pueyo}, L. and {Mamajek}, E.~E. and {Girard}, J.~H. and {Sivaramakrishnan}, A. and {Bonnefoy}, M. and {Baraffe}, I. and {Chauvin}, G. and {Olofsson}, J. and {Juhasz}, A. and {Benisty}, M. and {Pott}, J.-U. and {Sicilia-Aguilar}, A. and {Henning}, T. and {Cardwell}, A. and {Goodsell}, S. and {Graham}, J.~R. and {Hibon}, P. and {Ingraham}, P. and {Konopacky}, Q. and {Macintosh}, B. and {Oppenheimer}, R. and {Perrin}, M. and {Rantakyr{\"o}}, F. and {Sadakuni}, N. and {Thomas}, S.},
  journal       = {\aap},
  title         = {{An M-dwarf star in the transition disk of Herbig HD 142527. Physical parameters and orbital elements}},
  year          = {2016},
  month         = may,
  pages         = {A90},
  volume        = {590},
  adsurl        = {https://ui.adsabs.harvard.edu/abs/2016A&A...590A..90L},
  archiveprefix = {arXiv},
  doi           = {10.1051/0004-6361/201527863},
  eid           = {A90},
  eprint        = {1511.09390},
  primaryclass  = {astro-ph.SR},
}

@Article{Lazareff2017,
  author        = {{Lazareff}, B. and {Berger}, J.-P. and {Kluska}, J. and {Le Bouquin}, J.-B. and {Benisty}, M. and {Malbet}, F. and {Koen}, C. and {Pinte}, C. and {Thi}, W.-F. and {Absil}, O. and {Baron}, F. and {Delboulb{\'e}}, A. and {Duvert}, G. and {Isella}, A. and {Jocou}, L. and {Juhasz}, A. and {Kraus}, S. and {Lachaume}, R. and {M{\'e}nard}, F. and {Millan-Gabet}, R. and {Monnier}, J.~D. and {Moulin}, T. and {Perraut}, K. and {Rochat}, S. and {Soulez}, F. and {Tallon}, M. and {Thi{\'e}baut}, E. and {Traub}, W. and {Zins}, G.},
  journal       = {\aap},
  title         = {{Structure of Herbig AeBe disks at the milliarcsecond scale . A statistical survey in the H band using PIONIER-VLTI}},
  year          = {2017},
  month         = mar,
  pages         = {A85},
  volume        = {599},
  adsurl        = {https://ui.adsabs.harvard.edu/abs/2017A&A...599A..85L},
  archiveprefix = {arXiv},
  doi           = {10.1051/0004-6361/201629305},
  eid           = {A85},
  eprint        = {1611.08428},
  primaryclass  = {astro-ph.SR},
}

@Article{Lopez2022,
  author        = {{Lopez}, B. and {Lagarde}, S. and {Petrov}, R.~G. and {Jaffe}, W. and {Antonelli}, P. and {Allouche}, F. and {Berio}, P. and {Matter}, A. and {Meilland}, A. and {Millour}, F. and {Robbe-Dubois}, S. and {Henning}, Th. and {Weigelt}, G. and {Glindemann}, A. and {Agocs}, T. and {Bailet}, Ch. and {Beckmann}, U. and {Bettonvil}, F. and {van Boekel}, R. and {Bourget}, P. and {Bresson}, Y. and {Bristow}, P. and {Cruzal{\`e}bes}, P. and {Eldswijk}, E. and {Fante{\"\i} Caujolle}, Y. and {Gonz{\'a}lez Herrera}, J.~C. and {Graser}, U. and {Guajardo}, P. and {Heininger}, M. and {Hofmann}, K.-H. and {Kroes}, G. and {Laun}, W. and {Lehmitz}, M. and {Leinert}, C. and {Meisenheimer}, K. and {Morel}, S. and {Neumann}, U. and {Paladini}, C. and {Percheron}, I. and {Riquelme}, M. and {Schoeller}, M. and {Stee}, Ph. and {Venema}, L. and {Woillez}, J. and {Zins}, G. and {{\'A}brah{\'a}m}, P. and {Abadie}, S. and {Abuter}, R. and {Accardo}, M. and {Adler}, T. and {Alonso}, J. and {Augereau}, J.-C. and {B{\"o}hm}, A. and {Bazin}, G. and {Beltran}, J. and {Bensberg}, A. and {Boland}, W. and {Brast}, R. and {Burtscher}, L. and {Castillo}, R. and {Chelli}, A. and {Cid}, C. and {Clausse}, J.-M. and {Connot}, C. and {Conzelmann}, R.~D. and {Danchi}, W.-C. and {Delbo}, M. and {Drevon}, J. and {Dominik}, C. and {van Duin}, A. and {Ebert}, M. and {Eisenhauer}, F. and {Flament}, S. and {Frahm}, R. and {G{\'a}mez Rosas}, V. and {Gabasch}, A. and {Gallenne}, A. and {Garces}, E. and {Girard}, P. and {Glazenborg}, A. and {Gont{\'e}}, F.~Y.~J. and {Guitton}, F. and {de Haan}, M. and {Hanenburg}, H. and {Haubois}, X. and {Hocd{\'e}}, V. and {Hogerheijde}, M. and {ter Horst}, R. and {Hron}, J. and {Hummel}, C.~A. and {Hubin}, N. and {Huerta}, R. and {Idserda}, J. and {Isbell}, J.~W. and {Ives}, D. and {Jakob}, G. and {Jask{\'o}}, A. and {Jochum}, L. and {Klarmann}, L. and {Klein}, R. and {Kragt}, J. and {Kuindersma}, S. and {Kokoulina}, E. and {Labadie}, L. and {Lacour}, S. and {Leftley}, J. and {Le Poole}, R. and {Lizon}, J.-L. and {Lopez}, M. and {Lykou}, F. and {M{\'e}rand}, A. and {Marcotto}, A. and {Mauclert}, N. and {Maurer}, T. and {Mehrgan}, L.~H. and {Meisner}, J. and {Meixner}, K. and {Mellein}, M. and {Menut}, J.~L. and {Mohr}, L. and {Mosoni}, L. and {Navarro}, R. and {Nu{\ss}baum}, E. and {Pallanca}, L. and {Pantin}, E. and {Pasquini}, L. and {Phan Duc}, T. and {Pott}, J.-U. and {Pozna}, E. and {Richichi}, A. and {Ridinger}, A. and {Rigal}, F. and {Rivinius}, Th. and {Roelfsema}, R. and {Rohloff}, R.-R. and {Rousseau}, S. and {Salabert}, D. and {Schertl}, D. and {Schuhler}, N. and {Schuil}, M. and {Shabun}, K. and {Soulain}, A. and {Stephan}, C. and {Toledo}, P. and {Tristram}, K. and {Tromp}, N. and {Vakili}, F. and {Varga}, J. and {Vinther}, J. and {Waters}, L.~B.~F.~M. and {Wittkowski}, M. and {Wolf}, S. and {Wrhel}, F. and {Yoffe}, G.},
  journal       = {\aap},
  title         = {{MATISSE, the VLTI mid-infrared imaging spectro-interferometer}},
  year          = {2022},
  month         = mar,
  pages         = {A192},
  volume        = {659},
  adsurl        = {https://ui.adsabs.harvard.edu/abs/2022A&A...659A.192L},
  archiveprefix = {arXiv},
  doi           = {10.1051/0004-6361/202141785},
  eid           = {A192},
  eprint        = {2110.15556},
  primaryclass  = {astro-ph.IM},
}

@Article{Mainzer2011,
  author        = {{Mainzer}, A. and {Bauer}, J. and {Grav}, T. and {Masiero}, J. and {Cutri}, R.~M. and {Dailey}, J. and {Eisenhardt}, P. and {McMillan}, R.~S. and {Wright}, E. and {Walker}, R. and {Jedicke}, R. and {Spahr}, T. and {Tholen}, D. and {Alles}, R. and {Beck}, R. and {Brandenburg}, H. and {Conrow}, T. and {Evans}, T. and {Fowler}, J. and {Jarrett}, T. and {Marsh}, K. and {Masci}, F. and {McCallon}, H. and {Wheelock}, S. and {Wittman}, M. and {Wyatt}, P. and {DeBaun}, E. and {Elliott}, G. and {Elsbury}, D. and {Gautier}, IV, T. and {Gomillion}, S. and {Leisawitz}, D. and {Maleszewski}, C. and {Micheli}, M. and {Wilkins}, A.},
  journal       = {\apj},
  title         = {{Preliminary Results from NEOWISE: An Enhancement to the Wide-field Infrared Survey Explorer for Solar System Science}},
  year          = {2011},
  month         = apr,
  number        = {1},
  pages         = {53},
  volume        = {731},
  adsurl        = {https://ui.adsabs.harvard.edu/abs/2011ApJ...731...53M},
  archiveprefix = {arXiv},
  doi           = {10.1088/0004-637X/731/1/53},
  eid           = {53},
  eprint        = {1102.1996},
  primaryclass  = {astro-ph.EP},
}

@Article{Marino2015,
  author        = {{Marino}, S. and {Perez}, S. and {Casassus}, S.},
  journal       = {\apjl},
  title         = {{Shadows Cast by a Warp in the HD 142527 Protoplanetary Disk}},
  year          = {2015},
  month         = jan,
  number        = {2},
  pages         = {L44},
  volume        = {798},
  adsurl        = {https://ui.adsabs.harvard.edu/abs/2015ApJ...798L..44M},
  archiveprefix = {arXiv},
  doi           = {10.1088/2041-8205/798/2/L44},
  eid           = {L44},
  eprint        = {1412.4632},
  primaryclass  = {astro-ph.EP},
}

@Article{Matter2014,
  author        = {{Matter}, A. and {Labadie}, L. and {Kreplin}, A. and {Lopez}, B. and {Wolf}, S. and {Weigelt}, G. and {Ertel}, S. and {Pott}, J.-U. and {Danchi}, W.~C.},
  journal       = {\aap},
  title         = {{Evidence of a discontinuous disk structure around the Herbig Ae star HD 139614}},
  year          = {2014},
  month         = jan,
  pages         = {A26},
  volume        = {561},
  adsurl        = {https://ui.adsabs.harvard.edu/abs/2014A&A...561A..26M},
  archiveprefix = {arXiv},
  doi           = {10.1051/0004-6361/201322042},
  eid           = {A26},
  eprint        = {1311.5131},
  primaryclass  = {astro-ph.SR},
}

@Article{McKinney2010,
  author  = {McKinney, Wes},
  journal = {SciPy 2010},
  title   = {Data Structures for Statistical Computing in Python},
  year    = {2010},
  doi     = {10.25080/Majora-92bf1922-00a},
}

@Article{Mendigutia2014,
  author        = {{Mendigut{\'\i}a}, I. and {Fairlamb}, J. and {Montesinos}, B. and {Oudmaijer}, R.~D. and {Najita}, J.~R. and {Brittain}, S.~D. and {van den Ancker}, M.~E.},
  journal       = {\apj},
  title         = {{Stellar Parameters and Accretion Rate of the Transition Disk Star HD 142527 from X-Shooter}},
  year          = {2014},
  month         = jul,
  number        = {1},
  pages         = {21},
  volume        = {790},
  adsurl        = {https://ui.adsabs.harvard.edu/abs/2014ApJ...790...21M},
  archiveprefix = {arXiv},
  doi           = {10.1088/0004-637X/790/1/21},
  eid           = {21},
  eprint        = {1405.7378},
  primaryclass  = {astro-ph.SR},
}

@Article{Menu2015,
  author        = {{Menu}, J. and {van Boekel}, R. and {Henning}, Th. and {Leinert}, Ch. and {Waelkens}, C. and {Waters}, L.~B.~F.~M.},
  journal       = {\aap},
  title         = {{The structure of disks around intermediate-mass young stars from mid-infrared interferometry. Evidence for a population of group II disks with gaps}},
  year          = {2015},
  month         = sep,
  pages         = {A107},
  volume        = {581},
  adsurl        = {https://ui.adsabs.harvard.edu/abs/2015A&A...581A.107M},
  archiveprefix = {arXiv},
  doi           = {10.1051/0004-6361/201525654},
  eid           = {A107},
  eprint        = {1506.03274},
  primaryclass  = {astro-ph.SR},
}

@InProceedings{Millour2016,
  author        = {{Millour}, F. and {Berio}, P. and {Heininger}, M. and {Hofmann}, K.-H. and {Schertl}, D. and {Weigelt}, G. and {Guitton}, F. and {Jaffe}, W. and {Beckmann}, U. and {Petrov}, R. and {Allouche}, F. and {Robbe-Dubois}, S. and {Lagarde}, S. and {Soulain}, A. and {Meilland}, A. and {Matter}, A. and {Cruzal{\`e}bes}, P. and {Lopez}, B.},
  booktitle     = {Optical and Infrared Interferometry and Imaging V},
  title         = {{Data reduction for the MATISSE instrument}},
  year          = {2016},
  editor        = {{Malbet}, Fabien and {Creech-Eakman}, Michelle J. and {Tuthill}, Peter G.},
  month         = aug,
  pages         = {990723},
  series        = {Society of Photo-Optical Instrumentation Engineers (SPIE) Conference Series},
  volume        = {9907},
  adsurl        = {https://ui.adsabs.harvard.edu/abs/2016SPIE.9907E..23M},
  archiveprefix = {arXiv},
  doi           = {10.1117/12.2232283},
  eid           = {990723},
  eprint        = {1608.01913},
  primaryclass  = {astro-ph.IM},
}

@Article{Min2005,
  author        = {{Min}, M. and {Hovenier}, J.~W. and {de Koter}, A.},
  journal       = {\aap},
  title         = {{Modeling optical properties of cosmic dust grains using a distribution of hollow spheres}},
  year          = {2005},
  month         = mar,
  number        = {3},
  pages         = {909-920},
  volume        = {432},
  adsurl        = {https://ui.adsabs.harvard.edu/abs/2005A&A...432..909M},
  archiveprefix = {arXiv},
  doi           = {10.1051/0004-6361:20041920},
  eprint        = {astro-ph/0503068},
  primaryclass  = {astro-ph},
}

@Article{Min2007,
  author        = {{Min}, M. and {Waters}, L.~B.~F.~M. and {de Koter}, A. and {Hovenier}, J.~W. and {Keller}, L.~P. and {Markwick-Kemper}, F.},
  journal       = {\aap},
  title         = {{The shape and composition of interstellar silicate grains}},
  year          = {2007},
  month         = feb,
  number        = {2},
  pages         = {667-676},
  volume        = {462},
  adsurl        = {https://ui.adsabs.harvard.edu/abs/2007A&A...462..667M},
  archiveprefix = {arXiv},
  doi           = {10.1051/0004-6361:20065436},
  eprint        = {astro-ph/0611329},
  primaryclass  = {astro-ph},
}

@Article{Mueller2018,
  author        = {{M{\"u}ller}, A. and {Keppler}, M. and {Henning}, Th. and {Samland}, M. and {Chauvin}, G. and {Beust}, H. and {Maire}, A.-L. and {Molaverdikhani}, K. and {van Boekel}, R. and {Benisty}, M. and {Boccaletti}, A. and {Bonnefoy}, M. and {Cantalloube}, F. and {Charnay}, B. and {Baudino}, J.-L. and {Gennaro}, M. and {Long}, Z.~C. and {Cheetham}, A. and {Desidera}, S. and {Feldt}, M. and {Fusco}, T. and {Girard}, J. and {Gratton}, R. and {Hagelberg}, J. and {Janson}, M. and {Lagrange}, A.-M. and {Langlois}, M. and {Lazzoni}, C. and {Ligi}, R. and {M{\'e}nard}, F. and {Mesa}, D. and {Meyer}, M. and {Molli{\`e}re}, P. and {Mordasini}, C. and {Moulin}, T. and {Pavlov}, A. and {Pawellek}, N. and {Quanz}, S.~P. and {Ramos}, J. and {Rouan}, D. and {Sissa}, E. and {Stadler}, E. and {Vigan}, A. and {Wahhaj}, Z. and {Weber}, L. and {Zurlo}, A.},
  journal       = {\aap},
  title         = {{Orbital and atmospheric characterization of the planet within the gap of the PDS 70 transition disk}},
  year          = {2018},
  month         = sep,
  pages         = {L2},
  volume        = {617},
  adsurl        = {https://ui.adsabs.harvard.edu/abs/2018A&A...617L...2M},
  archiveprefix = {arXiv},
  doi           = {10.1051/0004-6361/201833584},
  eid           = {L2},
  eprint        = {1806.11567},
  primaryclass  = {astro-ph.EP},
}

@Article{Nowak2024,
  author        = {{Nowak}, M. and {Rowther}, S. and {Lacour}, S. and {Meru}, F. and {Nealon}, R. and {Price}, D.~J.},
  journal       = {\aap},
  title         = {{The orbit of HD 142527 B is too compact to explain many of the disc features}},
  year          = {2024},
  month         = mar,
  pages         = {A6},
  volume        = {683},
  adsurl        = {https://ui.adsabs.harvard.edu/abs/2024A&A...683A...6N},
  archiveprefix = {arXiv},
  doi           = {10.1051/0004-6361/202347748},
  eid           = {A6},
  eprint        = {2402.03595},
  primaryclass  = {astro-ph.SR},
}

@Article{OpenAI2023,
  author        = {{OpenAI} and {Achiam}, Josh and {Adler}, Steven and {Agarwal}, Sandhini and {Ahmad}, Lama and {Akkaya}, Ilge and {Leoni Aleman}, Florencia and {Almeida}, Diogo and {Altenschmidt}, Janko and {Altman}, Sam and {Anadkat}, Shyamal and {Avila}, Red and {Babuschkin}, Igor and {Balaji}, Suchir and {Balcom}, Valerie and {Baltescu}, Paul and {Bao}, Haiming and {Bavarian}, Mohammad and {Belgum}, Jeff and {Bello}, Irwan and {Berdine}, Jake and {Bernadett-Shapiro}, Gabriel and {Berner}, Christopher and {Bogdonoff}, Lenny and {Boiko}, Oleg and {Boyd}, Madelaine and {Brakman}, Anna-Luisa and {Brockman}, Greg and {Brooks}, Tim and {Brundage}, Miles and {Button}, Kevin and {Cai}, Trevor and {Campbell}, Rosie and {Cann}, Andrew and {Carey}, Brittany and {Carlson}, Chelsea and {Carmichael}, Rory and {Chan}, Brooke and {Chang}, Che and {Chantzis}, Fotis and {Chen}, Derek and {Chen}, Sully and {Chen}, Ruby and {Chen}, Jason and {Chen}, Mark and {Chess}, Ben and {Cho}, Chester and {Chu}, Casey and {Chung}, Hyung Won and {Cummings}, Dave and {Currier}, Jeremiah and {Dai}, Yunxing and {Decareaux}, Cory and {Degry}, Thomas and {Deutsch}, Noah and {Deville}, Damien and {Dhar}, Arka and {Dohan}, David and {Dowling}, Steve and {Dunning}, Sheila and {Ecoffet}, Adrien and {Eleti}, Atty and {Eloundou}, Tyna and {Farhi}, David and {Fedus}, Liam and {Felix}, Niko and {Posada Fishman}, Sim{\'o}n and {Forte}, Juston and {Fulford}, Isabella and {Gao}, Leo and {Georges}, Elie and {Gibson}, Christian and {Goel}, Vik and {Gogineni}, Tarun and {Goh}, Gabriel and {Gontijo-Lopes}, Rapha and {Gordon}, Jonathan and {Grafstein}, Morgan and {Gray}, Scott and {Greene}, Ryan and {Gross}, Joshua and {Gu}, Shixiang Shane and {Guo}, Yufei and {Hallacy}, Chris and {Han}, Jesse and {Harris}, Jeff and {He}, Yuchen and {Heaton}, Mike and {Heidecke}, Johannes and {Hesse}, Chris and {Hickey}, Alan and {Hickey}, Wade and {Hoeschele}, Peter and {Houghton}, Brandon and {Hsu}, Kenny and {Hu}, Shengli and {Hu}, Xin and {Huizinga}, Joost and {Jain}, Shantanu and {Jain}, Shawn and {Jang}, Joanne and {Jiang}, Angela and {Jiang}, Roger and {Jin}, Haozhun and {Jin}, Denny and {Jomoto}, Shino and {Jonn}, Billie and {Jun}, Heewoo and {Kaftan}, Tomer and {Kaiser}, {\L}ukasz and {Kamali}, Ali and {Kanitscheider}, Ingmar and {Shirish Keskar}, Nitish and {Khan}, Tabarak and {Kilpatrick}, Logan and {Kim}, Jong Wook and {Kim}, Christina and {Kim}, Yongjik and {Hendrik Kirchner}, Jan and {Kiros}, Jamie and {Knight}, Matt and {Kokotajlo}, Daniel and {Kondraciuk}, {\L}ukasz and {Kondrich}, Andrew and {Konstantinidis}, Aris and {Kosic}, Kyle and {Krueger}, Gretchen and {Kuo}, Vishal and {Lampe}, Michael and {Lan}, Ikai and {Lee}, Teddy and {Leike}, Jan and {Leung}, Jade and {Levy}, Daniel and {Li}, Chak Ming and {Lim}, Rachel and {Lin}, Molly and {Lin}, Stephanie and {Litwin}, Mateusz and {Lopez}, Theresa and {Lowe}, Ryan and {Lue}, Patricia and {Makanju}, Anna and {Malfacini}, Kim and {Manning}, Sam and {Markov}, Todor and {Markovski}, Yaniv and {Martin}, Bianca and {Mayer}, Katie and {Mayne}, Andrew and {McGrew}, Bob and {McKinney}, Scott Mayer and {McLeavey}, Christine and {McMillan}, Paul and {McNeil}, Jake and {Medina}, David and {Mehta}, Aalok and {Menick}, Jacob and {Metz}, Luke and {Mishchenko}, Andrey and {Mishkin}, Pamela and {Monaco}, Vinnie and {Morikawa}, Evan and {Mossing}, Daniel and {Mu}, Tong and {Murati}, Mira and {Murk}, Oleg and {M{\'e}ly}, David and {Nair}, Ashvin and {Nakano}, Reiichiro and {Nayak}, Rajeev and {Neelakantan}, Arvind and {Ngo}, Richard and {Noh}, Hyeonwoo and {Ouyang}, Long and {O'Keefe}, Cullen and {Pachocki}, Jakub and {Paino}, Alex and {Palermo}, Joe and {Pantuliano}, Ashley and {Parascandolo}, Giambattista and {Parish}, Joel and {Parparita}, Emy and {Passos}, Alex and {Pavlov}, Mikhail and {Peng}, Andrew and {Perelman}, Adam and {de Avila Belbute Peres}, Filipe and {Petrov}, Michael and {Ponde de Oliveira Pinto}, Henrique and {Michael} and {Pokorny} and {Pokrass}, Michelle and {Pong}, Vitchyr H. and {Powell}, Tolly and {Power}, Alethea and {Power}, Boris and {Proehl}, Elizabeth and {Puri}, Raul and {Radford}, Alec},
  journal       = {arXiv e-prints},
  title         = {{GPT-4 Technical Report}},
  year          = {2023},
  month         = mar,
  pages         = {arXiv:2303.08774},
  adsurl        = {https://ui.adsabs.harvard.edu/abs/2023arXiv230308774O},
  archiveprefix = {arXiv},
  doi           = {10.48550/arXiv.2303.08774},
  eid           = {arXiv:2303.08774},
  eprint        = {2303.08774},
  primaryclass  = {cs.CL},
}

@Article{Paunzen2022,
  author        = {{Paunzen}, E.},
  journal       = {\aap},
  title         = {{Catalogue of stars measured in the Geneva seven-colour photometric system}},
  year          = {2022},
  month         = may,
  pages         = {A89},
  volume        = {661},
  adsurl        = {https://ui.adsabs.harvard.edu/abs/2022A&A...661A..89P},
  archiveprefix = {arXiv},
  doi           = {10.1051/0004-6361/202142355},
  eid           = {A89},
  eprint        = {2111.04810},
  primaryclass  = {astro-ph.SR},
}

@InProceedings{Petrov2020,
  author    = {{Petrov}, Romain G. and {Allouche}, Fatm{\'e} and {Matter}, Alexis and {Meilland}, Anthony and {Lagarde}, St{\'e}phane and {Berio}, Philippe and {Cruzal{\`e}bes}, Pierre and {Millour}, Florentin and {Robbe-Dubois}, Sylvie and {Jaffe}, Walter and {Hofmann}, Karl-Heinz and {Varga}, Jozsef and {Schertl}, Dieter and {Burtscher}, Leonard and {Meisenheimer}, Klaus and {Chelli}, Alain and {Zins}, Gerard and {Woillez}, Julien and {Sch{\"o}ller}, Markus and {Lopez}, Bruno},
  booktitle = {Optical and Infrared Interferometry and Imaging VII},
  title     = {{Commissioning MATISSE: operation and performances}},
  year      = {2020},
  editor    = {{Tuthill}, Peter G. and {M{\'e}rand}, Antoine and {Sallum}, Stephanie},
  month     = dec,
  pages     = {114460L},
  series    = {Society of Photo-Optical Instrumentation Engineers (SPIE) Conference Series},
  volume    = {11446},
  adsurl    = {https://ui.adsabs.harvard.edu/abs/2020SPIE11446E..0LP},
  doi       = {10.1117/12.2562569},
  eid       = {114460L},
}

@Article{Pinte2019,
  author        = {{Pinte}, C. and {van der Plas}, G. and {M{\'e}nard}, F. and {Price}, D.~J. and {Christiaens}, V. and {Hill}, T. and {Mentiplay}, D. and {Ginski}, C. and {Choquet}, E. and {Boehler}, Y. and {Duch{\^e}ne}, G. and {Perez}, S. and {Casassus}, S.},
  journal       = {Nat. Astron.},
  title         = {{Kinematic detection of a planet carving a gap in a protoplanetary disk}},
  year          = {2019},
  month         = aug,
  pages         = {1109-1114},
  volume        = {3},
  adsurl        = {https://ui.adsabs.harvard.edu/abs/2019NatAs...3.1109P},
  archiveprefix = {arXiv},
  doi           = {10.1038/s41550-019-0852-6},
  eprint        = {1907.02538},
  primaryclass  = {astro-ph.SR},
}

@Article{Pohl2017,
  author        = {{Pohl}, A. and {Benisty}, M. and {Pinilla}, P. and {Ginski}, C. and {de Boer}, J. and {Avenhaus}, H. and {Henning}, Th. and {Zurlo}, A. and {Boccaletti}, A. and {Augereau}, J.-C. and {Birnstiel}, T. and {Dominik}, C. and {Facchini}, S. and {Fedele}, D. and {Janson}, M. and {Keppler}, M. and {Kral}, Q. and {Langlois}, M. and {Ligi}, R. and {Maire}, A.-L. and {M{\'e}nard}, F. and {Meyer}, M. and {Pinte}, C. and {Quanz}, S.~P. and {Sauvage}, J.-F. and {Sezestre}, {\'E}. and {Stolker}, T. and {Szul{\'a}gyi}, J. and {van Boekel}, R. and {van der Plas}, G. and {Villenave}, M. and {Baruffolo}, A. and {Baudoz}, P. and {Le Mignant}, D. and {Maurel}, D. and {Ramos}, J. and {Weber}, L.},
  journal       = {\apj},
  title         = {{The Circumstellar Disk HD 169142: Gas, Dust, and Planets Acting in Concert?}},
  year          = {2017},
  month         = nov,
  number        = {1},
  pages         = {52},
  volume        = {850},
  adsurl        = {https://ui.adsabs.harvard.edu/abs/2017ApJ...850...52P},
  archiveprefix = {arXiv},
  doi           = {10.3847/1538-4357/aa94c2},
  eid           = {52},
  eprint        = {1710.06485},
  primaryclass  = {astro-ph.EP},
}

@Article{Price2018,
  author        = {{Price}, Daniel J. and {Cuello}, Nicol{\'a}s and {Pinte}, Christophe and {Mentiplay}, Daniel and {Casassus}, Simon and {Christiaens}, Valentin and {Kennedy}, Grant M. and {Cuadra}, Jorge and {Sebastian Perez}, M. and {Marino}, Sebastian and {Armitage}, Philip J. and {Zurlo}, Alice and {Juhasz}, Attila and {Ragusa}, Enrico and {Laibe}, Guillaume and {Lodato}, Giuseppe},
  journal       = {\mnras},
  title         = {{Circumbinary, not transitional: on the spiral arms, cavity, shadows, fast radial flows, streamers, and horseshoe in the HD 142527 disc}},
  year          = {2018},
  month         = jun,
  number        = {1},
  pages         = {1270-1284},
  volume        = {477},
  adsurl        = {https://ui.adsabs.harvard.edu/abs/2018MNRAS.477.1270P},
  archiveprefix = {arXiv},
  doi           = {10.1093/mnras/sty647},
  eprint        = {1803.02484},
  primaryclass  = {astro-ph.SR},
}

@Software{Gommers2025,
  adsurl    = {https://ui.adsabs.harvard.edu/abs/2025zndo..15716342G},
  author    = {{Gommers}, Ralf and {Virtanen}, Pauli and {Haberland}, Matt and {Burovski}, Evgeni and {Reddy}, Tyler and {Weckesser}, Warren and {Oliphant}, Travis E. and {Nelson}, Andrew and {Cournapeau}, David and {alexbrc} and {Roy}, Pamphile and {Polat}, Ilhan and {Peterson}, Pearu and {Wilson}, Josh and {endolith} and {Mayorov}, Nikolay and {Colley}, Lucas and {van der Walt}, Stefan and {Brett}, Matthew and {Bowhay}, Jake and {Laxalde}, Denis and {Larson}, Eric and {Sakai}, Atsushi and {Millman}, Jarrod and {Steppi}, Albert and {Lars} and {peterbell10} and {Carey}, CJ and {van Mulbregt}, Paul and {eric-jones}},
  doi       = {10.5281/zenodo.15716342},
  eid       = {10.5281/zenodo.15716342},
  month     = jun,
  publisher = {Zenodo},
  title     = {{scipy/scipy: SciPy 1.16.0}},
  version   = {v1.16.0},
  year      = {2025},
}

@Article{Riello2021,
  author        = {{Riello}, M. and {De Angeli}, F. and {Evans}, D.~W. and {Montegriffo}, P. and {Carrasco}, J.~M. and {Busso}, G. and {Palaversa}, L. and {Burgess}, P.~W. and {Diener}, C. and {Davidson}, M. and {Rowell}, N. and {Fabricius}, C. and {Jordi}, C. and {Bellazzini}, M. and {Pancino}, E. and {Harrison}, D.~L. and {Cacciari}, C. and {van Leeuwen}, F. and {Hambly}, N.~C. and {Hodgkin}, S.~T. and {Osborne}, P.~J. and {Altavilla}, G. and {Barstow}, M.~A. and {Brown}, A.~G.~A. and {Castellani}, M. and {Cowell}, S. and {De Luise}, F. and {Gilmore}, G. and {Giuffrida}, G. and {Hidalgo}, S. and {Holland}, G. and {Marinoni}, S. and {Pagani}, C. and {Piersimoni}, A.~M. and {Pulone}, L. and {Ragaini}, S. and {Rainer}, M. and {Richards}, P.~J. and {Sanna}, N. and {Walton}, N.~A. and {Weiler}, M. and {Yoldas}, A.},
  journal       = {\aap},
  title         = {{Gaia Early Data Release 3. Photometric content and validation}},
  year          = {2021},
  month         = may,
  pages         = {A3},
  volume        = {649},
  adsurl        = {https://ui.adsabs.harvard.edu/abs/2021A&A...649A...3R},
  archiveprefix = {arXiv},
  doi           = {10.1051/0004-6361/202039587},
  eid           = {A3},
  eprint        = {2012.01916},
  primaryclass  = {astro-ph.IM},
}

@Article{Rufener1988,
  author   = {{Rufener}, F. and {Nicolet}, B.},
  journal  = {\aap},
  title    = {{A new determination of the Geneva photometric passbands and their absolute calibration.}},
  year     = {1988},
  month    = nov,
  pages    = {357-374},
  volume   = {206},
  adsurl   = {https://ui.adsabs.harvard.edu/abs/1988A&A...206..357R},
}
\begin{appendix}
    \onecolumn
    \section{Observations}
    \label{app:observations}
    
    \begin{table*}[ht!]
        \caption{The HD~142527 observations used in this study.}
        \bgroup
        \def\arraystretch{1.3}
        \setlength{\tabcolsep}{0.44em}
        \begin{center}
            \begin{tabular}{l l c c l l | c l c c c}
                \toprule\toprule
                \multicolumn{6}{c|}{HD~142527} & \multicolumn{5}{c}{Calibrator} \\
                \midrule
                Date and Time & Instrument & Seeing & $\tau_0$ & Array: Stations & Band & $\Delta $Time & Name & \acs{ldd} & Seeing & $\tau_0$\\
                (\acs{utc}) & & (\unit{\arcsecond}) & (\unit{\milli\second}) & & & (h:mm) & & (\unit{\milliArcsecond}) & (\unit{\arcsecond}) & (\unit{\milli\second}) \\
                \midrule
                 \DTLforeach{observations}{\sI=Column1,\sII=Column2,\sIII=Column3, \sIV=Column4, \sV=Column5, \sVI=Column6, \sVII=Column7, \sVIII=Column8, \sIX=Column9, \sX=Column10, \sXI=Column11}{\sI & \sII & \sIII & \sIV & \sV & \sVI & \sVII & \sVIII & \sIX & \sX & \sXI \DTLiflastrow{}{\tabularnewline}}\\
                \\\bottomrule
            \end{tabular}
            \tablefoot{%
                \textit{Columns}: Instrument, date and time, atmospheric conditions (seeing and coherence time $\tau_0$), array configurations and corresponding stations, spectral band, and information on the calibrator. Observations highlighted in red were not used. Letters in the `Array: Stations' column are abbreviations of the various configurations (i.e. `S' -- small, `M' -- medium, `L' -- large, and `U' -- UTs). For HD~142527, seeing and coherence time $\tau_0$ are the mean value during observations, while for the calibrator they are the value obtained at the beginning of observations.\\\noindent
                \tablefoottext{\ensuremath{\dagger}}{Archival data obtained from \acs{oidb} \citep{Lazareff2017}.} \tablefoottext{\ensuremath{\dagger\dagger}}{Reduced and calibrated data obtained from \citet{GRAVITYcollaboration2019}.}
                \tablefoottext{\ensuremath{\#}}{No calibrator information logged for this observation.} \tablefoottext{\ensuremath{\ddagger}}{Although good atmospheric conditions, artifacts or a low \acs{sn} is present.}
            }
            \label{tab:observations}
        \end{center}
        \egroup
    \end{table*}
    
    \begin{figure*}[ht!]
        \centering
        \includegraphics[width=\textwidth]{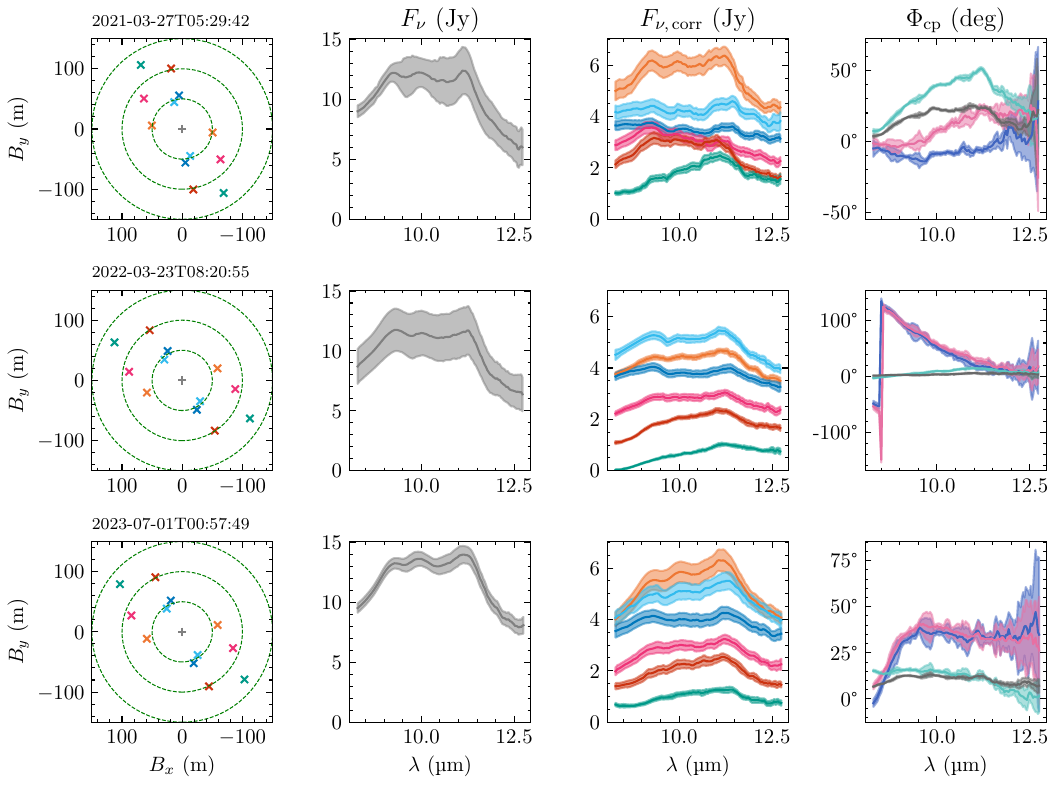}
        \caption{%
            \acs{matisse} \textit{N}-band observations of HD~142527 (Table~\ref{tab:observations}). \textit{Left to right}: Baseline distribution $(B_x,B_y)$, single dish spectra $F_\nu$, correlated fluxes $F_{\nu,\text{corr}}$, and closure phases $\Phi_{\nu,\text{cp}}$.
            }
        \label{fig:nbandData}
    \end{figure*}
    
    \begin{figure*}[p]
        \centering
        \includegraphics[width=\textwidth]{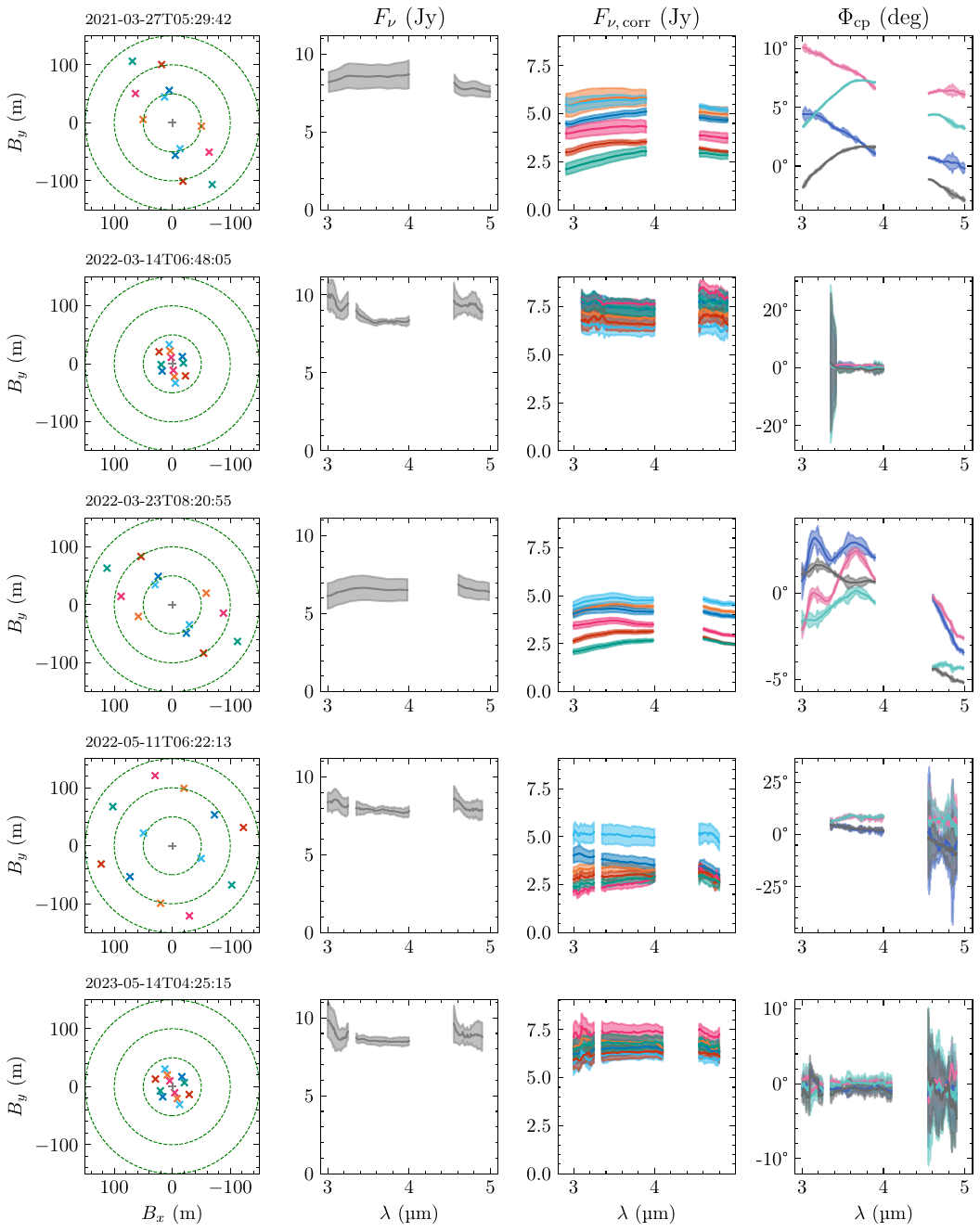} \\
        \caption{%
            Individual \textit{L}/\textit{M}-band, \acs{matisse} observations of HD~142527 (Table~\ref{tab:observations}). Layout is identical as for Fig.~\ref{fig:nbandData}. Single-dish spectra originate from chopped mode, with the other observables from non-chopped mode. Emission peaks not relevant to this study, poor-quality data, and ranges outside the band windows were flagged and are not shown.
            }
        \label{fig:lbandData}
    \end{figure*} 
    
    \begin{figure*}[ht!]
        \centering
        \includegraphics[width=\textwidth]{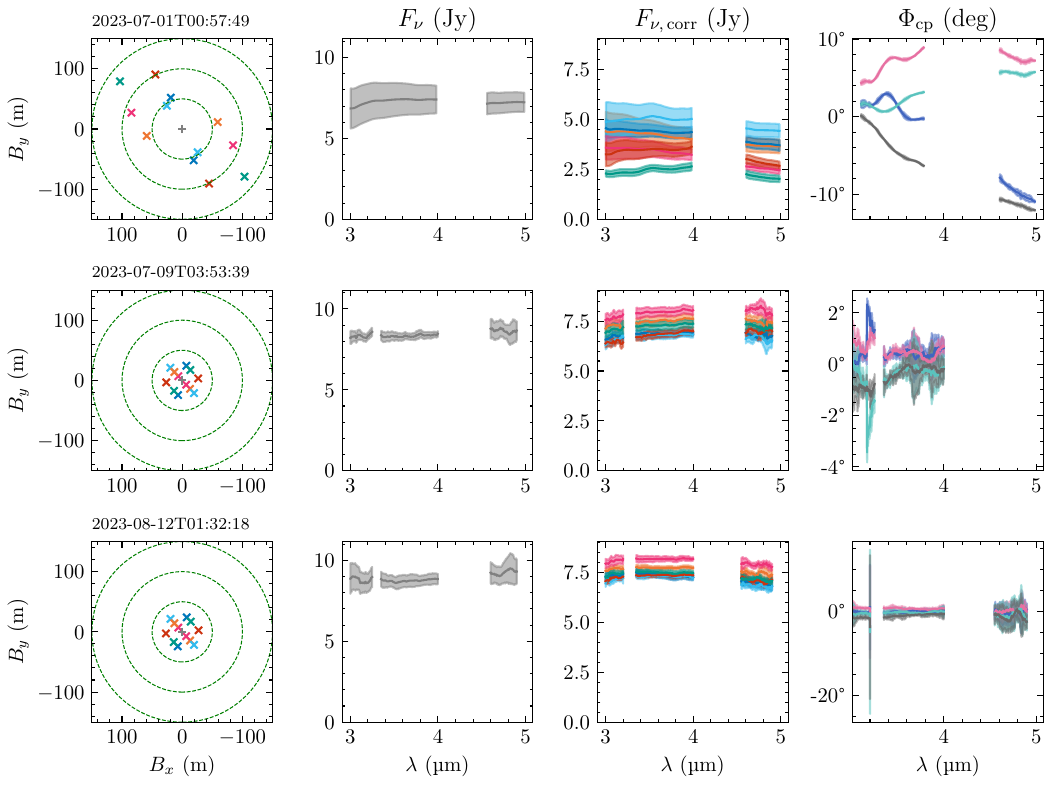} \\
        \caption{Continuation of Fig.~\ref{fig:lbandData}}
        \label{fig:lbandDataContinuation}
    \end{figure*} 
    
    \begin{figure*}[p]
        \centering
        \includegraphics[width=\textwidth]{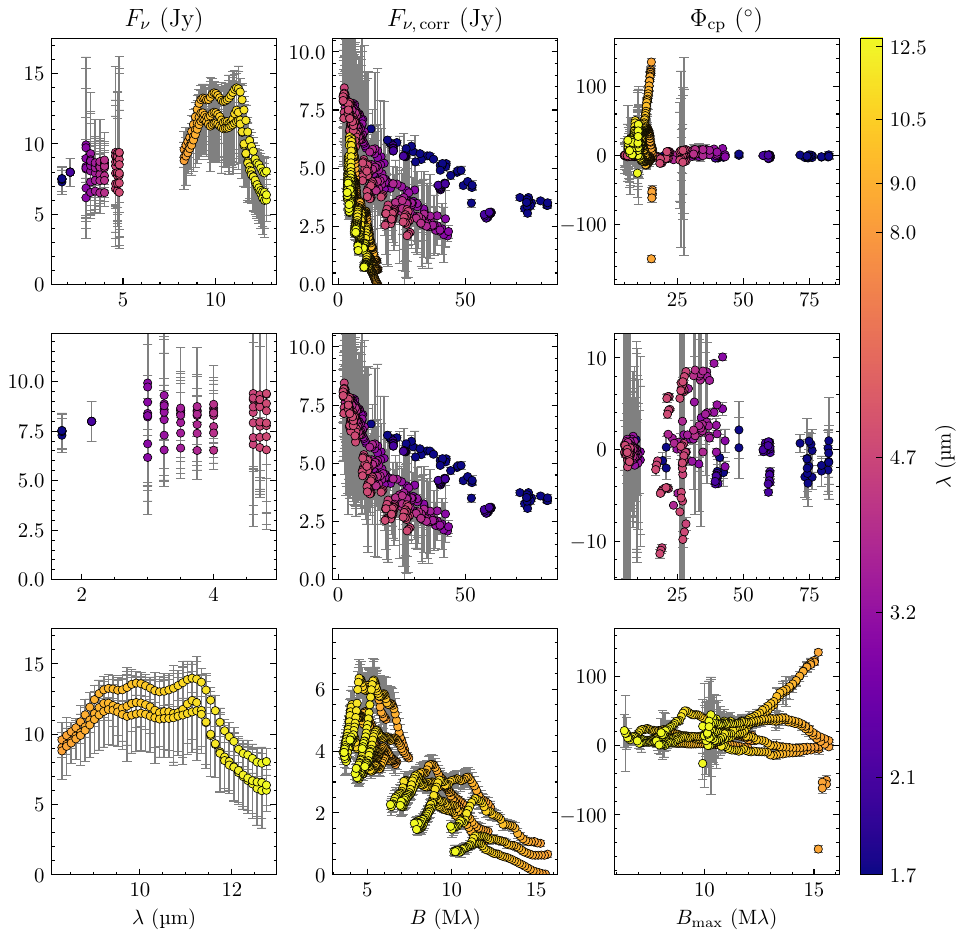}
        \caption{%
            Overview of binned data used for model-fitting. \textit{Left to right}: Single dish spectra $F_\nu$, correlated fluxes $F_{\nu,\text{corr}}$, and closure phases $\Phi_{\nu,\text{cp}}$. \textit{Top to bottom}: Data from \textit{H}-\textit{N} bands, a zoom-in to the \textit{H}-\textit{M} bands, and a zoom-in to the \textit{N} band.
        }
        \label{fig:allDataSpatial}
    \end{figure*}
    
    \begin{figure*}[ht!]
        \centering
        \includegraphics[width=\textwidth]{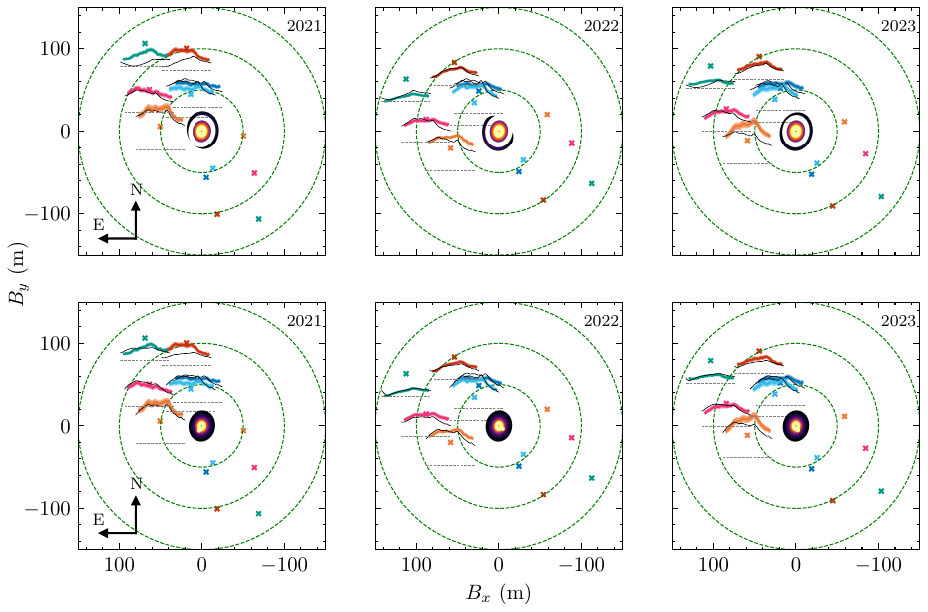}
        \caption{%
            Baseline distribution $(B_x,B_y)$ of all \textit{N}-band epochs. Each plot shows correlated fluxes $F_{\nu,\text{corr}}$ (coloured lines) in its top-left corner overlaid with the model (black lines) at baseline positions (coloured crosses) and equidistant baseline contours (green-dashed circles; \qty{50}{\metre} step-width). Additionally, zero lines of the y-axis (grey-dashed lines) correspond to the correlated flux curves. \textit{Left to right}: Three \textit{N}-band epochs (2021, 2022, and 2023). \textit{Top}: Two-zone disc model. \textit{Bottom}: One-zone disc model with a Gaussian.
        }
        \label{fig:uv}
    \end{figure*}
    
    \FloatBarrier
    \twocolumn
    \section{Model fitting}
    \label{app:bayesianFitting}
    We chose Bayesian inference, more specifically nested sampling, to fit our model to the data. Nested sampling efficiently explores the probability distribution of the parameter space (even if not well-defined; typically problematic with \ac{mcmc} methods). It estimates the evidence
    \begin{equation}
        \mathcal{Z}\equiv P(\vec{D}\vert M)=\int^\infty_{0}X(x)\odif{x}=\int\mathcal{L}(X)\odif{X},
    \end{equation}
    by treating the integral of the posterior over all parameters as an integral over the prior volume
    \begin{equation}
        X(x)=\int_{\vec{\Theta}:\mathcal{L}(\vec{\Theta})>x}\pi(\vec{\Theta})\odif{\vec{\Theta}}
    \end{equation}
    instead. Then, the iso-likelihood contour $\mathcal{L}_i=\mathcal{L}(X_i)$ associated with samples is evaluated from the prior volume and the evidence computed over `nested' shells. At each iteration $i$, the remaining evidence in the prior volume is roughly bounded by $\Delta\mathcal{Z}_i\approx\mathcal{L}_\text{max}X_i$. The difference between remaining and current evidence $\mathcal{Z}_i$,
    \begin{equation}
        \Delta\ln\mathcal{Z}_i\equiv\ln\left(\mathcal{Z}_i+\Delta\mathcal{Z}_i\right)-\ln\mathcal{Z}_i,
        \label{eq:deltaEvidenceNestedSampling}
    \end{equation}
    serves as the stopping criterion \citep{Skilling2004,Speagle2020}.
    
    We use the least-squares minimisation for the likelihood
    \begin{equation}
        \mathcal{L}(\vec{\Theta})\equiv P(\vec{D}\vert\vec{\Theta},M)\equiv\chi^2(\vec{\Theta})=\sum^{N}_{n=1}\left(\frac{\vec{y}_n-M(\vec{\Theta})}{\sigma_n}\right)^2,
    \end{equation}
    enabling the fitting of a model to data of $N$ points $y_n$ measured with (uncorrelated) errors $\sigma_n$. Estimating the fit-goodness and for model comparison, we also use the reduced form:
    \begin{equation}
        \chi^2_\text{r}(\vec{\Theta})=\frac{\mathcal{L}(\vec{\Theta})}{N-N_{\vec{\Theta}}},
    \end{equation}
    which takes the \ac{dof} $N-N_{\vec{\Theta}}$ into account \citep{Andrae2010}.
    The number of model parameters $N_{\vec{\Theta}}$ may vary between epochs (i.e. time-variable parameters) and/or for each observable.
    
    The likelihood $\mathcal{L}$ consists of contributions from the total spectra, the correlated fluxes, and the closure phases, which are computed separately and subsequently summed up:
    \begin{equation}
        \mathcal{L}_\text{tot}=w_{F_\nu}\mathcal{L}_{F_\nu}+w_{F_{\nu,\text{corr}}}\,\mathcal{L}_{F_{\nu,\text{corr}}}+w_{\Phi_{\nu,\text{cp}}}\,\mathcal{L}_{\Phi_{\nu,\text{cp}}}.
        \label{eq:leastSquares}
    \end{equation}
    However, the total reduced least squares $\chi^2_\text{r,tot}$ is generally not a simple addition as \ac{dof} for the whole model need to be taken into account.
    
    We only use the weights $w$, to achieve an equal contribution of the observables in respect to their amount of data points. Interferometric instruments at the \ac{vlti} have, per observation and frequency element $\nu$, the following distribution of data points for each observable
    \begin{flalign}
        \nonumber
        n_{\nu,F_\nu}=N\lor1,& \\
        \nonumber
        n_{\nu,F_{\nu,\text{corr}}}=\frac{N(N-1)}{2},&\qquad\text{with}\qquad\begin{pmatrix} N \\ 2 \end{pmatrix}, \\
        n_{\nu,\Phi_{\nu,\text{cp}}}=\frac{N(N-1)(N-2)}{6},&\qquad\text{with}\qquad\begin{pmatrix} N \\ 3 \end{pmatrix}.
    \end{flalign}
    Here, $n_{\nu,F_\nu}$ is the number of single-dish spectra (often averaged; i.e. $n_{\nu,F_\nu}=1$), $n_{\nu,F_{\nu,\text{corr}}}$ the number of correlated fluxes, and $n_{\nu,\Phi_{\nu,\text{cp}}}$ the number of closure phases, with $N$ being the number of telescopes.
    
    \section{Model parameters}
    \label{app:modelParameters}
    \FloatBarrier
    
    \begin{table}[ht!]
        \bgroup
        \def\arraystretch{1.25}
        \begin{centering}
            \caption{Model parameters.}
            \label{tab:modelParams}
            \begin{tabular}{l c l}
                \toprule\toprule
                Parameter & Unit & Description \\
                \midrule
                \textbf{Dust opacity model} & & \\
                \textbf{\textsc{Free}} & & \\
                $s$ & & Scaling factor\\
                $T_\text{c}$ & (\unit{\kelvin}) & Charac. temperature \\
                $\kappa_{\nu,\text{abs},k}$\tablefootmark{a} & (\unit{\square\centi\meter\per\gram})  & Absorption opacity \\
                $w_k$ & & Opacity weight \\
                \midrule
                \textbf{Stellar spectrum model} & & \\
                \textbf{\textsc{Free}} & & \\
                $\theta$ & (\unit{\milliArcsecond}) & Apparent diameter \\
                $A_\text{V}$ & (\unit{\magnitude}) & V-band extinction \\
                \textbf{\textsc{Fixed}} & & \\
                $T_\star$ & (\unit{\kelvin}) & Temperature \\
                $A_\lambda$ & (\unit{\magnitude}) & Extinction \\
                $g$ & (\unit{\metre\per\square\second}) & Surface gravity \\
                \midrule
                \textbf{Asymmetric disc} & & \\
                \textbf{\textsc{Free}} & & \\
                $R_{\text{in},n}$\tablefootmark{b} & (\unit{\astronomicalunit}) & Inner radius \\
                $R_{\text{out},n}$ & (\unit{\astronomicalunit}) & Outer radius \\
                $w_{\text{cont},n}$ & & Continuum weight \\
                $\Sigma_{0,n}$ & (\unit{\gram\per\square\centi\meter}) & $\Sigma_n$ at $R_0$ \\
                $p_n$ & & $\Sigma_n$ exponent \\
                $T_0$ & (\unit{\kelvin}) & $T$ at $R_0$ \\
                $q$ & & $T$ exponent \\
                $A_{n,m}$\tablefootmark{c} & & Modulation amp. \\
                $\phi_{n,m}$ & (\unit{\degree}) & Modulation angle \\
                \textbf{\textsc{Fixed}} & & \\
                $i_\text{in}$ & (\unit{\degree}) & Inclination angle \\
                $\theta_\text{in}$ & (\unit{\degree}) & Position angle \\
                $R_0$ & (\unit{\astronomicalunit})  & Reference radius \\
                $\kappa_{\nu,\text{abs,sil}}$ & (\unit{\square\centi\meter\per\gram})  & Silicate opacity \\
                $\kappa_{\nu,\text{abs,cont}}$ & (\unit{\square\centi\meter\per\gram})  & Continuum opacity \\
                \midrule
                \textbf{Gaussian} & & \\
                \textbf{\textsc{Free}} & & \\
                $f$ & & Scaling factor\\
                $\rho$ & (\unit{\astronomicalunit}) & Radial position \\
                $\Phi$ & (\unit{\degree}) & Position angle \\
                $\Sigma$ & (\unit{\gram\per\square\centi\meter}) & Surface density \\
                \textbf{\textsc{Fixed}} & & \\
                $a$ & (\unit{\astronomicalunit}) & \ac{fwhm} \\
                \bottomrule
            \end{tabular}
        \end{centering}
        \egroup
        \tablefoot{
            \tablefoottext{a}{Dust stoichiometries denoted by $k$.}
            \tablefoottext{b}{Disc zones denoted by $n$.}
            \tablefoottext{c}{Modulation orders denoted by $m$.}
            }
    \end{table}
    \FloatBarrier
    
    \section{Stellar spectrum and parameters}
    \label{app:stellarSpectrum}
    \FloatBarrier
    To estimate the contribution of the stellar photosphere to our observed signals, we perform model atmosphere fitting to optical photometry of HD~142527. We use the effective temperature $T_\star=\qty{6500}{\kelvin}$ derived by \citet{Fairlamb2015}, from X-shooter spectroscopy, and use their spectroscopically derived surface gravity of $\log g=3.93$ as an initial value. We calculate synthetic photometry from a reddened \texttt{PHOENIX} model spectrum by \citet{Hauschildt2025}, with the corresponding parameters adopting solar composition. Then, we tune the apparent stellar diameter $\theta$ and the \textit{V}-band extinction $A_\text{V}$, calculating synthetic photometry until reaching the best photometry fit through $\chi^2$ minimisation, while keeping $T_\star$ and $\log g$ fixed. Adopting the GAIA distance, we then obtain an estimate of the stellar radius and luminosity. We compare these values to \ac{pms}, evolutionary tracks by \citet{Choi2016} to obtain an estimate of the stellar mass and age. From the stellar radius and mass estimate, we retrieve an updated estimate of the surface gravity, and iteratively repeat the procedure using a \texttt{PHOENIX} model spectrum with the new $\log g$ value. Ultimately, we achieve a best-fit, reddened model atmosphere spectrum, and a set of stellar parameters that are mutually consistent. The procedure is described in detail below.
    
    \subsection{Method details}
    \label{app:stellarSpectrum:methodDetails}
    We adopt optical photometry in the Geneva system from \citet{Paunzen2022} and from \ac{gaiadr3} \citep{GaiaCollaboration2016a,GaiaCollaboration2023}. We also consider \acl{ir} photometry from the \ac{2mass} point-source \citep{Skrutskie2006,Cutri2003} and ALLWISE \citep{Wright2010,Mainzer2011} catalogues. However, these are ignored in the fit to the photosphere as they contain \acl{ir} excess emission from the circumstellar material. We adopt the spectro-photometric response curves $R_\lambda$ for the Geneva system from \cite{Rufener1988}, for \ac{gaiadr3} from \cite{Riello2021}, for \ac{2mass} from \citet{Cohen2003}, and for ALLWISE from \citet{Wright2010}. The photometric zero-point (i.e. $\phi_{\lambda,0}$ for the respective bands) computation is done as follows. For the Geneva system, we use the \texttt{alpha\_lyr\_mod\_002.fits} model spectrum from the CALSPEC database\footnote{Available at \url{https://www.stsci.edu/hst/instrumentation/reference-data-for-calibration-and-tools/astronomical-catalogs/calspec}.} and the $\alpha~\text{Lyr}$ magnitudes from Table~11 of \cite{Rufener1988}; for \ac{gaiadr3} we adopt the \texttt{alpha\_lyr\_mod\_002.fits} from the CALSPEC database, scaled to \qty{3.62286e-11}{\watt\per\square\metre\per\nano\metre} at \qty{550}{\nano\metre}, following \citet{Riello2021}; for \ac{2mass} we use a spectrum constant $wF_\lambda$ at \num{3.129e-13}, \num{1.133e-13}, and \qty{4.283e-14}{\watt\per\square\metre\per\micro\metre} for the \textit{J}, \textit{H}, \textit{K} bands, respectively, following \cite{Cohen2003}; and for the ALLWISE catalog we use an $\alpha~\text{Lyr}$ model by Martin Cohen, (Wright, private communication). All photometric facilities have detectors that are based on the photoelectric effect:
    \begin{equation}
      m_i=-2.5\log_{10}\left(\frac{\int_0^\infty\phi_{\lambda}R_\lambda\odif{\lambda}}{\int_0^\infty\phi_{\lambda,0}R_\lambda\odif{\lambda}}\right),
    \end{equation}
    where $m_{i}$ is the observed magnitude in photometric band $i,\phi_\lambda$ is the target spectrum expresses as photon flux (e.g. $\left[\gamma\right]=[\unit{\per\second\per\square\centi\metre\per\micro\metre}]$), and $\phi_{\lambda,0}$ is the corresponding spectrum that defines the photometric zero point of the respective band.\\
    The model photosphere spectrum $\phi_\lambda$ is calculated as follows:
    \begin{equation}
        \phi_\lambda=\Omega I_{\lambda,0}(T_\star,\log g)\cdot10^{-0.4 A_\lambda},
    \end{equation}
    where $I_{\lambda,0}$ is the disc-integrated average surface brightness of the \texttt{PHOENIX} stellar model atmosphere without foreground extinction, $\Omega=\pi\theta^2/4$ is the solid angle subtended by the star as viewed from Earth, and $A_\lambda$ is the wavelength-dependent extinction curve adopted from \cite{Indebetouw2005} and \cite{Fitzpatrick2009}, scaled so that its value at \qty{0.55}{\micro\metre} matches the fit parameter $A_\text{V}$.
    
    Note that in our photometric data, we measure the combined signal of the primary Herbig Ae star and the low-mass M-type companion, ignoring the contribution of the latter in our analysis. Given the much lower luminosity and photospheric temperature of the companion compared to the primary, its emission should affect the optical fluxes at no more than $\approx\qty{1}{\percent}$.
    
    \subsection{Fit results}
    \label{app:stellarSpectrum:fitResults}
    
    \begin{figure}[ht!]
        \centering
       \includegraphics[width=0.5\textwidth]{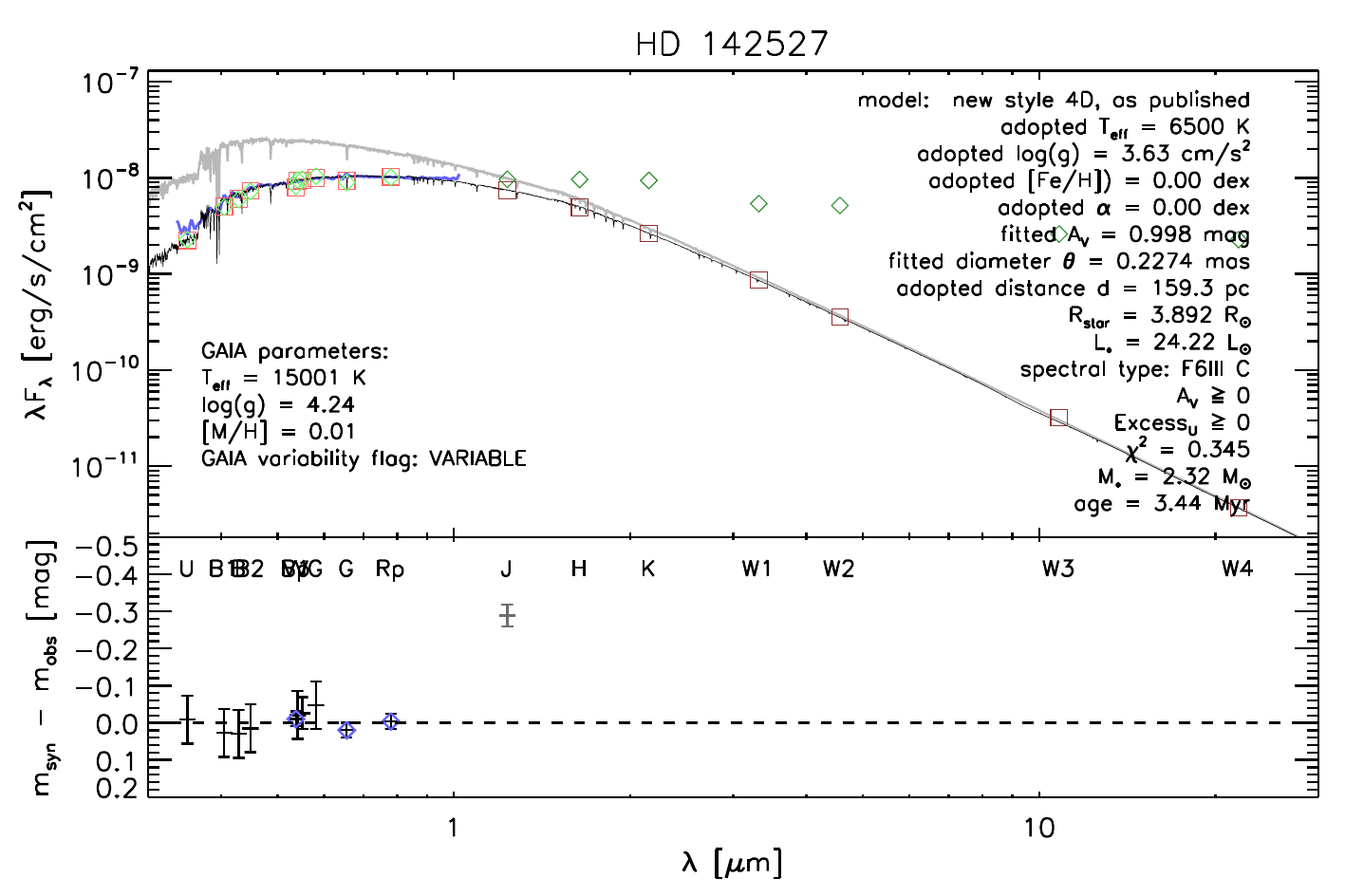}
        \caption{%
            \acs{sed} fit. \textit{Top}: Best-fit stellar atmosphere model (black) overlaid with the zero-extinction model (grey), fitted to the observed photometry (green diamond) and the low-resolution GAIA XP spectrum (blue). The observed photometry is contrasted by the synthetic photometry (red squares). \textit{Bottom}: Residuals of observed and synthetic magnitudes.
        }
        \label{fig:modelAtmosphereFit}
    \end{figure}
    
    Figure~\ref{fig:modelAtmosphereFit} shows the model atmosphere fit, with excess emission at near- and mid-\ac{ir} wavelengths, as seen from divergence in observed and synthetic photometry. No scaling has been applied to the low-resolution GAIA XP spectrum, and the excellent match in flux levels illustrates both the validity of our method and the excellent absolute calibration of the GAIA spectra.
    
    We find an adequate fit to the spectrum when using the spectroscopically derived temperature $T_\star=\qty{6500}{\kelvin}$ from \citet{Fairlamb2015} with the \textit{V}-band extinction $A_\text{V,fit}\approx\qty{1.0}{\magnitude}$ significantly different from $A_\text{V}\approx\qty{0}{\magnitude}$ of this earlier work. Consequently, we derive a substantially larger stellar radius $R_\star\approx\qty{3.9}{\radius\sun}$, compared to $R_\star\approx\qty{2.2}{\radius\sun}$ found by \cite{Fairlamb2015}, and we find a correspondingly higher luminosity $L_\star\approx\qty{24}{\lum\sun}$. We compare this to \ac{pms} evolutionary tracks yielding a mass estimate of $M_\star\approx\qty{2.3}{\mass\sun}$ and an age of $\approx\qty{3.4}{\mega\year}$ \citep[i.e. an object substantially more massive and younger than $\approx\qty{1.6}{\mass\sun}$ and $\approx\qty{8.1}{\mega\year}$, as found by][]{Fairlamb2015}. A model with $T_\star=\qty{6500}{\kelvin}$ and $A_\text{V}\approx\qty{0}{\magnitude}$, as derived by \cite{Fairlamb2015}, yields too blue a spectrum to be consistent with the observed optical photometry (grey curve; Fig.~\ref{fig:modelAtmosphereFit}).
    
    \citet{Nowak2024} find a combined mass of $\approx\qty{2.3}{\mass\sun}$ for the HD~142527~A and B system from an astrometrically measured orbit fit. Mass estimates for the companion range from $\approx$~\qtyrange{0.13}{0.34}{\mass\sun} \citep{Lacour2016,Christiaens2018}. Our combined mass estimate is somewhat higher than the one derived from the astrometric orbit.
    \FloatBarrier
    
    \section{Dust opacity model}
    \label{app:dustOpacityModel}
    \FloatBarrier
    
    \begin{table}[ht!]
        \caption{Best-fit parameters to the averaged, \textit{N}-band single-dish spectrum.}
        \begin{center}
            \bgroup
            \def\arraystretch{1.2}
            \begin{tabular}{l c c}
                 \toprule\toprule
                 Parameter & Unit & Value \\
                 \midrule
                 \DTLforeach{opacityFit}{\sI=Column1,\sII=Column2,\sIII=Column3}{\sI & \sII & \sIII \DTLiflastrow{}{\tabularnewline}}\\
                 \midrule
                 $\chi^2_\text{r}$ & & 0.03 \\
                 \bottomrule
            \end{tabular}
            \egroup
        \end{center}
        \tablefoot{%
            Over-fitting ($\chi^2_\text{r}<1$) is caused by the single, averaged dataset with many \acs{dof}. The weights for the silicate stoichiometries (i.e. excluding $w_\text{cont}$ and $w_\text{\acs{pah}}$) are normed.
        }
        \label{tab:dustOpacityFit}
    \end{table}
    
\begin{figure}[ht!]
        \centering
        \includegraphics[width=0.5\textwidth]{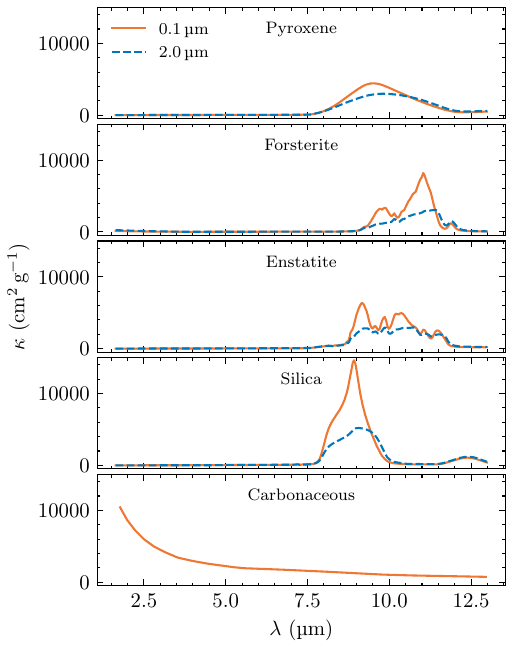}
        \caption{%
            Mass absorption coefficients of dust stoichiometries. Table~\ref{tab:dustSpecies} shows types (amorphous or crystalline) and chemical formulas. Coefficients are used in modelling the opacity contributions of HD~142527 (Sect.~\ref{sec:modelling:dustOpacityModel}).
            }
        \label{fig:dustSpecies}
    \end{figure}

    We analyse the composition of the prominent \textit{N}-band silicate emission feature of HD~142527 by fitting a model to the \textit{N}-band single-dish spectrum. The single-dish \textit{N}-band spectroscopic data are comprised of the average of the spectra from the three \ac{matisse} \ac{ut} epochs. With the methodology and dust species of \citet{VanBoekel2005}, we describe the observed spectrum as a blackbody source function $B_\nu(T)$ at characteristic temperature $T_\text{c}$ multiplied by a weighted sum of the absorption opacities and a \ac{pah} flux contribution:
    \begin{flalign}
        \nonumber
        F_{\nu,\text{model}}&=(F_{\nu,\text{sil}}+F_{\nu,\text{cont}})+w_\text{\acs{pah}}F_\text{\acs{pah}} \\
        &=s B_\nu(T_\text{c})\sum_\text{k}w_{k}\kappa_{\nu,\text{abs},k}+w_\text{\acs{pah}}F_{\nu,\text{\acs{pah}}}
        \label{eq:dustOpacityModel}
    \end{flalign}
    Here, free parameters are the dust-stoichiometry weights $w$, the \ac{pah} weight $w_\text{\acs{pah}}$, $T_\text{c}$, and the scale factor $s$. The \ac{pah} flux $F_\text{\acs{pah}}$ is an empirical \acs{pah} spectral template adopted from \citet{VanBoekel2005}. Each dust stoichiometry one weight per grain size (i.e. 0.1 and \qty{2}{\micro\metre}). The absorption opacities $\kappa_{\nu,\text{abs},k}$ are computed with \texttt{optool}\footnote{Available at \url{https://github.com/cdominik/optool}}, using the \ac{dhs} \citep{Min2005} method. Absorption opacities of the \ac{grf} method \citep{Min2007} were provided by M. Min. Fig.~\ref{fig:dustSpecies} and Table~\ref{tab:dustSpecies} show individual dust stoichiometries and structures.
    
   We chose amorphous carbonaceous grains for the continuum component, as they have no emission features in our wavelength range. Other species (e.g. iron) may present a valid alternative.
    
   Decoupling the opacities (Eq.~\ref{eq:verticalOpticalDepth}) enables a unique silicate-to-carbonaceous ratio for the disc component (Sect.~\ref{sec:modelling:modelComponents:asymTempGradDisc}). This fit method yields a reasonable, albeit imperfect, approximation of the \textit{N}-band silicate opacity (see Fig.~\ref{fig:dustOpacityFit}). We replace the silicate feature of the best-fit model with that from the observed data, using
   \begin{equation}
       \kappa_{\nu,\text{abs,sil}}=\frac{F_\nu}{s B_\nu(T_\text{c})}-\kappa_{\nu,\text{abs,cont}},
   \end{equation}
   as we do not focus on quantitative dust spectroscopy.
    
   \begin{table*}[t!]
       \caption{Dust stoichiometries used in this work.}
       \begin{center}
           \bgroup
           \def\arraystretch{1.15}
           \begin{tabular}{l l l c c c}
               \toprule\toprule
               Stoichiometric & Lattice & Chemical & Method & Grain & Reference \\
               (Grain) Type & Structure & Formula & & Sizes & \\
               \midrule
               Pyroxene & Amorphous & \ce{Mg_{x}Fe_{1-x}SiO3} & \ac{grf} & 0.1, 2.0  & (\ref{itm:dorschner}) \\
               Forsterite & Crystalline & \ce{Mg2SiO4} & \ac{grf} & 0.1, 2.0 & (\ref{itm:sogawa}) \\
               Enstatite & Crystalline & \ce{MgSiO3} & \ac{grf} & 0.1, 2.0 & (\ref{itm:jaeger}) \\
               Silica & Crystalline & \ce{SiO2} & \ac{grf} & 0.1, 2.0 & (\ref{itm:henning}) \\
               Carbonaceous & Amorphous & \ce{C} & \ac{dhs} & 0.1 & (\ref{itm:zubko}) \\
               \bottomrule
           \end{tabular}
           \egroup
       \end{center}
       \tablebib{%
           \counterlabel{dustStoichiometries}{itm:dorschner} \citet{Dorschner1995};
           \counterlabel{dustStoichiometries}{itm:sogawa} \citet{Sogawa2006};
           \counterlabel{dustStoichiometries}{itm:jaeger} \citet{Jaeger1998};
           \counterlabel{dustStoichiometries}{itm:henning} \citet{Henning1997};
           \counterlabel{dustStoichiometries}{itm:zubko} \citet{Zubko1996};
       }
       \tablefoot{%
           Stoichiometries and sizes are inspired by previous analysis from \citet{VanBoekel2005} and \citet{Juhasz2010}
       }
       \label{tab:dustSpecies}
   \end{table*}
   \FloatBarrier
    
   \section{Computation of observables}
   \label{app:computationOfObservables}
   The complex correlated flux $\mathfrak{F}_\nu$ is a linear operator, enabling the addition of those of individual model components. The single-dish spectrum, the correlated flux and the closure phase can be extracted from the complex correlated flux as shown in \citet{Buscher2015}.
   
   For this, the observed spatial frequencies $\vec{q}=(u,v)=(B_{x}/\lambda,\,B_{y}/\lambda)$ have to be de-projected to reconstruct a face-on disc by following the steps from \citet{Berger2007} and \citet{Matter2014}. We apply the counter-clockwise rotation $R(\theta)$ and the scaling matrix $S(i)$ to de-project the spatial frequencies
   \begin{flalign}
       \nonumber
       q^\prime=q^\prime(i,\theta)&=\norm{S(i)R(\theta)\vec{q}} \\
       &=\norm{
       \begin{pmatrix}
           \cos(i) & 0\\
           0 & 1
       \end{pmatrix}
       \begin{pmatrix}
           \cos(\theta) & -\sin(\theta)\\
           \sin(\theta) & \cos(\theta)
       \end{pmatrix}
       \begin{pmatrix}
           u \\
           v
       \end{pmatrix}}.
       \label{eq:deprojectedSpatialFrequencies}
   \end{flalign}
   This yields the de-projected spatial frequency $q^\prime(i,\theta)$ with the angle $\psi=\arctan\left(\frac{u^\prime}{v^\prime}\right)$ (polar coordinate representation of $\vec{q}^\prime$).
   
   To achieve the single-dish spectrum, we take the real part of the complex correlated flux at its 0th spatial frequency, and the modulus for the correlated fluxes:
   \begin{equation}
       F_\nu=\Re[\mathfrak{F}_\nu(0)],\qquad\text{and}\qquad F_{\nu,\text{corr}}=\abs{\mathfrak{F}_\nu}.
   \end{equation}
   The phase information can also be extracted from the complex correlated flux. However, turbulence in the atmosphere makes a direct observation of the absolute phases impossible, which is why the closure phases
   \begin{equation}
       \Phi_{\text{cp},ijk}=\phi_{ij}+\phi_{jk}+\phi_{ki},\qquad\text{with}\qquad\Phi_{ij}=\phi_{ij}+\epsilon_i-\epsilon_j
       \label{eq:closurePhase}
   \end{equation}
   are used. Here, $\phi_{ij}$ is the real phase between two telescopes, and $\epsilon_i$ is the atmospheric aberration per telescope. In a closure triangle, the atmospheric aberrations cancel each other, and the closure phases are computed with the spatial frequencies of the triangle
   \begin{equation}
       \vec{q}_{ijk}=\begin{pmatrix}
           1 & 0 \\
           0 & 1 \\
           1 & 1
       \end{pmatrix}
       \begin{pmatrix}
           u_i & v_i \\
           u_j & v_j
       \end{pmatrix},\quad\text{where}\quad u_k=u_i+u_j.
   \end{equation}
   Similar as before, we de-project the spatial frequencies with Eq.~\eqref{eq:deprojectedSpatialFrequencies} and then compute the closure phases using the bispectrum
   \begin{equation}
      \Phi_{\text{cp},ijk}=\arg\left(\mathfrak{F}_{ij}\mathfrak{F}_{jk}\mathfrak{F}_{ki}^*\right)=\arg\left(\mathfrak{F}_{ij}\mathfrak{F}_{jk}\mathfrak{F}_{ik}\right).
       \label{eq:phases}
   \end{equation}
    
   \section{Residuals}
   \label{app:residuals}
   
   \begin{figure*}[ht!]
       \centering
       \includegraphics[width=\textwidth]{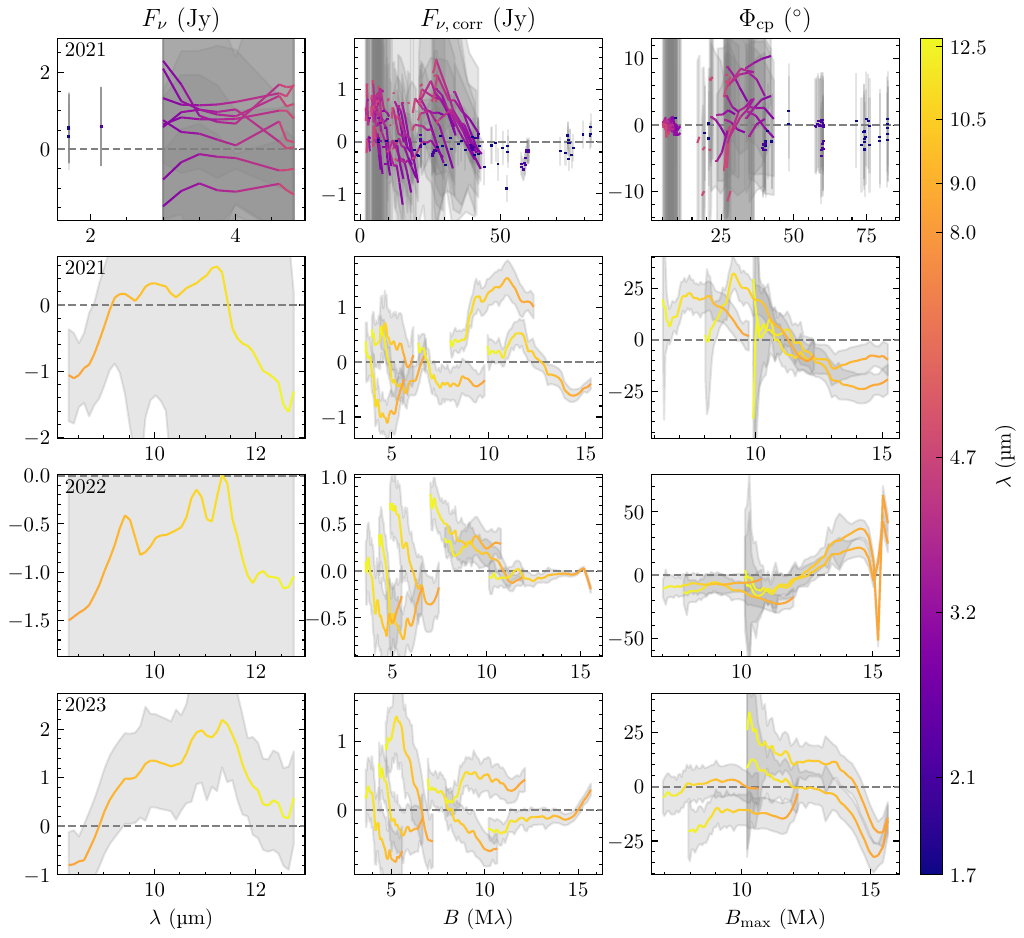}
       \caption{Residuals of Fig.~\ref{fig:bestFit} with identical layout.}
       \label{fig:bestFitResiduals}
   \end{figure*}
   
\end{appendix}
\end{document}